\documentclass[preprint2]{aastex}

\input psfig
\newcommand{\HI}{H\,{\sc i}}
\newcommand{\HII}{H\,{\sc ii}}
 
\newcommand{\kms}{~km\,s$^{-1}$}
\newcommand{\kkms}{km\,s$^{-1}$}
\newcommand{\vhel}{$v_{\rm hel}$}
\newcommand{\vopt}{$v_{\rm opt}$}
\newcommand{\vHI}{$v_{\rm HI}$}
\newcommand{\vsys}{$v_{\rm sys}$}
\newcommand{\vrot}{$v_{\rm rot}$}
\newcommand{\vLG}{$v_{\rm LG}$}
\newcommand{\wtw}{$w_{\rm 20}$}
\newcommand{\wfi}{$w_{\rm 50}$}
\newcommand{\Speak}{$S_{\rm peak}$}
\newcommand{\FHI}{$F_{\rm HI}$}
\newcommand{\MHI}{$M_{\rm HI}$}
\newcommand{\Mtot}{$M_{\rm tot}$}
\newcommand{\Msun}{~M$_{\sun}$}

\newcommand{\MMsun}{M$_{\sun}$}
\newcommand{\AB}{$A_{\rm B}$}
\newcommand{\Ho}{$H_0$}
\newcommand{\ta}{$^{\tiny a}$}
\newcommand{\tb}{$^{\tiny b}$}
\newcommand{\tc}{$^{\tiny c}$}
\newcommand{\td}{$^{\tiny d}$}
\newcommand{\te}{$^{\tiny e}$}
\newcommand{\tf}{$^{\tiny f}$}
\slugcomment{re-vised version submitted to the Astronomical Journal}
\shorttitle{The 1000 Brightest HIPASS Galaxies: \HI\ Properties}
\shortauthors{Koribalski et al.}

\begin{document}
\title{The 1000 Brightest HIPASS Galaxies: \HI\ Properties}
\author{B. S. Koribalski\altaffilmark{1}, 
L. Staveley-Smith\altaffilmark{1}, 
V. A. Kilborn\altaffilmark{1,2},
S. D. Ryder\altaffilmark{3},
R. C. Kraan-Korteweg\altaffilmark{4},
E. V. Ryan-Weber\altaffilmark{5,1},
R. D. Ekers\altaffilmark{1},
H. Jerjen\altaffilmark{6},
P. A. Henning\altaffilmark{7},
M. E. Putman\altaffilmark{8},
M. A. Zwaan\altaffilmark{5,9},
W. J. G. de~Blok\altaffilmark{10,1},
M. R. Calabretta\altaffilmark{1},
M. J. Disney\altaffilmark{10},
R. F. Minchin\altaffilmark{10},
R. Bhathal\altaffilmark{11},
P. J. Boyce\altaffilmark{10}, 
M. J. Drinkwater\altaffilmark{12},
K. C. Freeman\altaffilmark{6},
B. K. Gibson\altaffilmark{2},
A. J. Green\altaffilmark{13},
R. F. Haynes\altaffilmark{1},
S. Juraszek\altaffilmark{13},
M. J. Kesteven\altaffilmark{1},
P. M. Knezek\altaffilmark{14},
S. Mader\altaffilmark{1},
M. Marquarding\altaffilmark{1},
M. Meyer\altaffilmark{5}, 
J. R. Mould\altaffilmark{15},
T. Oosterloo\altaffilmark{16},
J. O'Brien\altaffilmark{6,1},
R. M. Price\altaffilmark{7},
E. M. Sadler\altaffilmark{13}
A. Schr\"oder\altaffilmark{17},
I. M. Stewart\altaffilmark{17},
F. Stootman\altaffilmark{11},
M. Waugh\altaffilmark{5,1},
B. E. Warren\altaffilmark{6,1},
R. L. Webster\altaffilmark{5},
and 
A. E. Wright\altaffilmark{1}}
\altaffiltext{1}{Australia Telescope National Facility, CSIRO, 
	         P.O. Box 76, Epping, NSW~1710, Australia.}
\altaffiltext{2}{Centre for Astrophysics and Supercomputing, Swinburne 
		 University of Technology, P.O. Box 218, Hawthorn, VIC~3122,
		 Australia.}
\altaffiltext{3}{Anglo-Australian Observatory, 
		 P.O. Box 296, Epping, NSW~1710, Australia.}
\altaffiltext{4}{Departamento de Astronom\'\i{a}, Universidad de Guanajuato, 
		 Apartado Postal 144, Guanajuato, Gto 36000, Mexico.}
\altaffiltext{5}{School of Physics, University of Melbourne, 
                 VIC~3010, Australia.}
\altaffiltext{6}{Research School of Astronomy \& Astrophysics, Mount Stromlo
		 Observatory, Cotter Road, Weston, ACT~2611, Australia.}
\altaffiltext{7}{Institute for Astrophysics, University of New Mexico, 
                 800 Yale Blvd, NE, Albuquerque, NM~87131, USA.}
\altaffiltext{8}{CASA, University of Colorado, Boulder, CO 80309-0389, USA}
\altaffiltext{9}{European Southern Observatory, Karl-Schwarzschild-Str. 2,
                 D-85748 Garching bei Muenchen, Germany}
\altaffiltext{10}{Cardiff School of Physics \& Astronomy, Cardiff University,
                  5 The Parade, Cardiff, CF24 3YB, U.K.}
\altaffiltext{11}{Department of Physics, University of Western Sydney 
                 Macarthur, P.O. Box 555, Campbelltown, NSW~2560, Australia.}
\altaffiltext{12}{Department of Physics, University of Queensland, 
                  QLD 4072, Australia.}
\altaffiltext{13}{School of Physics, University of Sydney, 
                 NSW~2006, Australia.}
\altaffiltext{14}{WIYN, Inc. 950 North Cherry Avenue Tucson, AZ 85726, USA.}
\altaffiltext{15}{National Optical Astronomy Observatories, P.O. Box 26732, 
                 950 North Cherry Avenue, Tucson, AZ, USA.}
\altaffiltext{16}{ASTRON, P.O. Box 2, 7990 AA Dwingeloo, The Netherlands.}
\altaffiltext{17}{Department of Physics \& Astronomy,
                 University of Leicester, Leicester LE1 7RH, U.K.}

\begin{abstract}
We present the HIPASS Bright Galaxy Catalog (BGC) which contains the 1000 
\HI-brightest galaxies in the southern sky as obtained from the \HI\ Parkes 
All-Sky Survey (HIPASS). The selection of the brightest sources is based on 
their \HI\ peak flux density (\Speak\ $\ga$ 116 mJy) as measured from the 
spatially integrated HIPASS spectrum. The derived \HI\ masses range from 
$\sim10^7$ to $4 \times 10^{10}$\Msun. While the BGC ($z < 0.03$) is complete 
in \Speak, only a subset of $\sim$500 sources can be considered complete in 
integrated \HI\ flux density (\FHI\ $\ga 25$ Jy\kms). 

The HIPASS BGC contains a total of 158 new redshifts. These belong to 91 
new sources for which no optical or infrared counterparts have previously 
been cataloged, an additional 51 galaxies for which no redshifts were 
previously known, and 16 galaxies for which the cataloged optical velocities 
disagree.
Of the 91 newly catalogued BGC sources, only four are definite \HI\ clouds: 
while three are likely Magellanic debris with velocities around 400\kms, one 
is a tidal cloud associated with the NGC~2442 galaxy group. The remaining 87 
{\em new} BGC sources, the majority of which lie in the Zone of Avoidance, 
appear to be galaxies. We identified optical counterparts to all but one of 
the 30 {\em new} galaxies at Galactic latitudes $|b| > 10\degr$. Therefore, 
the BGC yields no evidence for a population of "free-floating" intergalactic 
\HI\ clouds without associated optical counterparts.

HIPASS provides a clear view of the local large-scale structure. The dominant 
features in the sky distribution of the BGC are the Supergalactic Plane and 
the Local Void. In addition, one can clearly see the Centaurus Wall which 
connects via the Hydra and Antlia clusters to the Puppis filament. Some 
previously hardly noticed galaxy groups stand out quite distinctively in 
the \HI\ sky distribution. Several new structures are seen for the 
first time, not only behind the Milky Way.
\end{abstract}


\section{Introduction} % Section 1
\label{sec:intro}
Neutral hydrogen (\HI) is a major component of the interstellar medium (ISM)
in galaxies. The ISM provides fuel for the initial formation of molecular 
clouds and stars, and conversely acts as a reservoir for recycled gas from 
stars and supernovae. The total \HI\ content of galaxies at a given epoch 
provides important constraints on the evolution of galaxies (Pei, Fall \& 
Hauser 1999). At high redshifts, \HI\ content is commonly estimated through 
observations of the damped Lyman-$\alpha$ (DLA) absorption-line systems (Rao
\& Turnshek 2000; Storrie-Lombardi \& Wolfe 2000; Turnshek \& Rao 2002). 
These observations appear to indicate that the \HI\ density of the Universe
is substantially higher at redshift $z \approx 1-3$ compared to the $z=0$
density measured from 21-cm observations of optically-selected galaxy 
samples (e.g. Rao \& Briggs 1993). However, there are large uncertainties 
in the interpretation of DLA statistics at high redshift because of lensing 
and dust obscuration. There are also uncertainties at low redshifts because 
our knowledge of \HI\ in galaxies is heavily biased towards optically-selected 
samples. Are there, for example, many gas-rich galaxies or intergalactic clouds
missing from the nearby galaxy census as some absorption-line observations of 
low column density systems ($10^{12.5} - 10^{15.5}$ cm$^{-2}$) appear to 
indicate (Penton, Shull \& Stocke 2000; McLin et al. 2002; Ellison et al. 
2002)\,?

Within individual galaxies, \HI\ is frequently found to extend well outside 
the stellar radius (e.g. Broeils \& van Woerden 1994; Meurer et al. 1996; 
Salpeter \& Hoffman 1996; Broeils \& Rhee 1997). Between galaxies neutral 
hydrogen is found mostly in the form of tidal tails, bridges and rings 
(Hibbard \& van Gorkom 1996; Duc \& Mirabel 1997; Hibbard et al. 2001; Ryder 
et al. 2001; Koribalski \& Dickey 2004), which trace the interaction history of 
galaxies. Out of tidal \HI\ debris, which constitute part of the intergalactic 
medium, new galaxies can form. These are known as `tidal-dwarf galaxies' 
because of their similarity to classical dwarf irregulars and blue compact
dwarf galaxies (Barnes \& Hernquist 1992; Duc \& Mirabel 1998; Duc et al. 
2000; Hibbard et al. 2001; Braine et al. 2001). The closest and most prominent
interacting galaxies are, of course, the Magellanic Clouds, which are 
associated with a range of extended tidal \HI\ features, most prominently 
the Magellanic Bridge between the LMC and SMC, the Magellanic Stream 
(Mathewson, Cleary \& Murray 1974) and the recently discovered Leading Arm 
(Putman et al. 1998; Putman et al. 2003). Other \HI\ structures such as the 
Leo ring (Schneider 1989) and the tidal \HI\ remnant known as the Virgo cloud 
(Giovanelli \& Haynes 1989; Chengalur et al. 1995) have been discovered by 
accident and indicate the potential for new discoveries in blind \HI\ surveys.

The \HI\ content of normal galaxies varies strongly, depending on morphological
type, optical diameter and the environment (Roberts \& Haynes 1994; Haynes, 
Giovanelli \& Chincarini 1984; Giovanelli \& Haynes 1990; Solanes, Giovanelli 
\& Haynes 1996; Solanes et al. 2001; Verdes-Montenegro et al. 2001). Whereas 
early-type galaxies tend to be gas-poor (Duprie \& Schneider 1996; Sadler 2001;
Oosterloo et al. 2002), late-type galaxies are typically gas-rich (Matthews, 
Gallagher \& Littleton 1995; Solanes et al. 2001). \HI\ can also be easily 
detected in many low surface-brightness (LSB) and late-type dwarf galaxies, 
often barely visible in the optical (e.g., Impey \& Bothun 1997). Therefore, 
blind \HI\ surveys will reveal classes of galaxies that are under-represented 
in optical surveys. Since galaxy clustering is known to decrease from early-
to late-type galaxies (Giuricin et al. 2001) this will affect our overall view 
of the local large-scale structure.

The galaxy distribution in the nearby Universe reveals many coherent structures
(Geller \& Huchra 1989; Fairall et al. 1990; Fairall 1998), such as the 
Supergalactic Plane (SGP), the Fornax and Centaurus Walls, as well as the 
Local Void (to name the most prominent in the southern sky). Since our 
knowledge about the local large-scale structure has, so far, mostly been 
determined by extensive optical and infrared surveys of the sky (e.g. 
Lauberts 1982, Jarrett et al. 2003), not much is known about the connectivity 
of structures across the artificial gap created by the extinction of dust in 
our own Galaxy. This can affect up to $\sim$25\% of the sky (e.g. 
Kraan-Korteweg \& Lahav 2000) and obscures our view of important regions such 
as the putative Great Attractor (Lynden-Bell et al. 1988). However, 21-cm 
observations are unaffected by dust extinction, so are easily able to 
complement optical and infrared galaxy surveys and substantially improve the 
census of galaxies and the measurement of the baryon content of the local 
Universe. 

\subsection{\HI\ Surveys}
\label{sec:HIintro}
Many \HI\ surveys have been targeted at cataloged galaxies, galaxy groups,
clusters and selected regions --- for a summary see Salzer \& Haynes (1996) 
or Zwaan et al. (1997). The biggest targeted \HI\ surveys to-date are those 
by (a) Fisher \& Tully (1981a) who obtained \HI\ spectra for 1171 galaxies
(out of 1787 targets) with the NRAO 91-m and 43-m telescopes and the Bonn 
100-m telescope, (b) Schneider et al. (1990) who obtained \HI\ spectra for
762 dwarf and other LSB galaxies with the Arecibo telescope, (c) Mathewson,
Ford, \& Buchhorn (1992; hereafter MFB92) who obtained \HI\ spectra for 551 
galaxies with the Parkes 
telescope, (d) Haynes et al. (1997) who obtained \HI\ spectra for about 500 
galaxies in 27 Abell clusters visible from Arecibo, (e) Theureau et al. (1998) 
who present 2112 \HI\ spectra of cataloged spiral galaxies obtained with the 
Nan\c{c}ay radiotelescope, and (f) Haynes et al. (1999) who obtained \HI\ 
spectra for 881 galaxies, over 500 of which were detected with the 91-m Green 
Bank telescope.

Overall, there are currently about three times more \HI\ measurements in the 
northern hemisphere ($\ga$10300) than in the south ($\ga$3500), as determined
from the Lyon/Meudon Extragalactic Database (LEDA). At declinations south of 
--45\degr\ ($\sim$30\% of the southern sky) only 560 \HI\ measurements are 
available. The most prominent (targeted) southern \HI\ catalog is that by MFB92
who published \HI\ profiles for 551 galaxies of type Sb to Sd. Earlier, Reif et 
al. (1982) and Longmore et al. (1982) successfully obtained 196 and 100 \HI\ 
profiles, respectively, with the Parkes telescope, and Fouqu\'e et al. (1990a) 
published \HI\ profiles of 242 southern late-type galaxies in the declination 
range --38\degr\ to --17\degr\ obtained with the Nan\c{c}ay telescope. 

One of the earliest {\em blind} \HI\ surveys was that by Kerr \& Henning 
(1987) who detected 16 previously uncataloged galaxies in the Zone of 
Avoidance (ZOA) and considered the 
possibility of future whole sky surveys with multibeam instruments (see also
Henning 1995). Schneider (1996) gives a summary of other blind \HI\ survey 
programs. Until recently, the Arecibo Dual-Beam Survey (ADBS) and the deeper 
Arecibo \HI\ strip survey (AHISS, see Zwaan et al. 1997) were the largest 
blind 21-cm surveys (see Table~\ref{tab:hisurveys}) carried out, covering 
areas of 430 and 65 square degrees, respectively. With the advent of a 21-cm 
multibeam system at the 64-m Parkes telescope (Staveley-Smith et al. 1996) 
as well as new observing and data reduction software (Barnes et al. 2001), 
much larger and deeper surveys are now possible. For example, Henning et al.
(2000) did a shallow survey of 1800 square degrees of sky in the southern 
Zone of Avoidance, finding 110 \HI\ sources, 67 of which had no previously 
published optical counterpart. \\

This paper is organized as follows:
the observations and data reduction are described in Section~2, followed 
by a detailed description of the `HIPASS Bright Galaxy Catalog' in Section~3. 
The \HI\ properties from the BGC are discussed in Section~4, followed by a 
study of the large-scale structure in Section~5. Our conclusions are given 
in Section~6.

In related papers:
Ryan-Weber et al. (2002) study 138 galaxies from the BGC which had no redshift 
measurements prior to the commencement of the \HI\ multibeam surveys. Zwaan et 
al. (2003) analyze the BGC completeness and derive a new \HI\ mass function.

\section{HIPASS Observations \& Data Reduction} % Section 2
\label{sec:obs}

The observations utilized the 21-cm multibeam receiver (Staveley-Smith et al.
1996) installed at the prime focus of the Parkes\footnote{The Parkes telescope 
  is part of the Australia Telescope which is funded by the Commonwealth 
  of Australia for operation as a National Facility managed by CSIRO.}
64-m radio telescope. The receiver has thirteen beams, each with two linear 
polarizations, and system temperatures of around 20 K. The correlator has
a bandwidth of 64 MHz divided into 1024 channels, covering a velocity range 
of $-1200 < cz < +12700$\kms, at a channel spacing of 13.2\kms. The velocity 
resolution, after Tukey 25\% smoothing (Barnes et al. 2001) is 18\kms. Data 
for HIPASS was gathered by scanning the telescope across the sky in 8\degr\ 
strips in Declination ($\delta$). Five sets of independent scans were made 
of each region, resulting in a final sensitivity of $\sim40$ mJy\,beam$^{-1}$
(3$\sigma$) per channel. Assuming a velocity width of 100\kms\ the theoretical
5$\sigma$ \HI\ column density of HIPASS is $3 \times 10^{18}$~cm$^{-2}$ (only
applicable for very extended objects filling the beam). 
Observations were conducted from 1997 to 2000, and the whole of the southern 
sky ($\delta < +2\degr$) was covered. Note that the HIPASS BGC (see Section~3) 
only includes sources with $\delta < 0\degr$. Observations for the northern 
extension to HIPASS ($+2\degr < \delta < +25\degr$) and for the Jodrell Bank 
HIJASS survey ($\delta > +22\degr$) will be reported elsewhere (see, e.g., 
Boyce et al. 2001, Lang et al. 2003).

The data was bandpass corrected, calibrated and Doppler-tracked using the 
{\sc aips++} program {\sc LiveData} and then gridded with {\sc Gridzilla}
into 388 slightly overlapping cubes of size $8\degr \times 8\degr$, and 
pixel size 4\arcmin. The mean Parkes beam is 14\farcm3, but the gridding 
process slightly degrades the angular resolution. Moreover, the beam varies 
slightly with source strength and extent (see Barnes et al. 2001). For the 
\HI\ parameterization (see Section~\ref{sec:hipara}) we use a beam size of
15\farcm5. All velocities are in the usual `optical' convention ($v = cz$) 
and heliocentric reference frame. 
% Skipped. The radial velocity axis in the HIPASS cubes is given in 
%   the "radio definition" using the heliocentric restframe. 
%   \vhel\ = ($c \times v_{\rm radio}$)/($c - v_{\rm radio}$).
Continuum ripple, proportional in amplitude to the detected 20-cm continuum
emission, was removed using a `scaled template method', though some residual 
ripple at the position of strong radio continuum sources (e.g. the nearby 
Centaurus\,A galaxy, see Section~\ref{sec:hiabs}) still remains. HIPASS
spectra, smoothed spatially to a pixel size of 8\arcmin, are available at
  {\tt ~~www.atnf.csiro.au/research/multibeam/}. The \HI\ spectra of all 
1000 HIPASS BGC sources are displayed in Fig.~\ref{fig:hispectra} (see 
Section~\ref{sec:hipassbgc}).

\subsection{Flux Calibration}
\label{sec:fhical}
The absolute flux calibration of HIPASS (see Barnes et al. 2001) is based 
on the 20-cm flux density of the extragalactic radio sources PKS\,1934--638 
(14.9 Jy at 1420 MHz) and Hydra\,A (40.6 Jy at 1395 MHz). Whereas 
PKS\,1934--638 is a compact source, unresolved with the Parkes telescope, 
Hydra\,A is an extended radio galaxy (diameter $\sim$8\arcmin, Taylor et al. 
1990; Lane et al. 2004) with a total flux density of 43.2 Jy at 1395 MHz 
(Baars et al. 1977). To account for beam dilution the flux density of 
Hydra\,A is reduced by $\sim$6\% (J. Reynolds, priv. comm.) for the HIPASS 
calibration. 
% NOTE. Hydra A flux densities: 
%   FHI = 42.50  +- (5%) Jy  @ 1420 MHz (Baars et al. 1977, Tab 5).
%          -> C = 1.08       @ 1420 MHz (J. Reynolds, priv. comm.)
%   FHI = 42.78  +- (5%) Jy  @ 1410 MHz (Baars et al. 1977, Tab 5).
%   FHI = 42.83  +- (5%) Jy  @ 1408 MHz (Baars et al. 1977, Tab 5).
%   FHI = 43.05  +- (5%) Jy  @ 1400 MHz (Baars et al. 1977, Tab 5).
%   FHI = 43.20  +- (5%) Jy  @ 1395 MHz (Baars et al. 1977, Tab 5).
%          -> C = 1.0640 => 40.6 Jy (see above)
%   FHI = 42.65  +- 0.11 Jy  @ 1408 MHz (Ott et al. 1993).
%   FHI = 43     +- 1    Jy  @ 1415 MHz (Lane, priv. comm.)

For comparison, MFB92 for their absolute flux calibration
use Hydra\,A with a flux density of 43.5 Jy at 1410 MHz (no correction for
beam dilution was applied) and the galaxy IC\,4824 with an integrated \HI\ 
flux density of 20.0 Jy\kms. We measure \FHI\ = $17.0 \pm 2.7$ Jy\kms\ for 
IC\,4824 (HIPASS J1913--62), slightly lower than the value used by MFB92
and consistent with the difference in absolute flux calibration. 
See Section~\ref{sec:pgcflux} for a comparison of the BGC \FHI\ measurements 
with other \HI\ surveys.

\section{The HIPASS Bright Galaxy Catalog} % Section 3
\label{sec:hipassbgc}
A catalog of the 1000 \HI-brightest galaxies was compiled using HIPASS data
from the whole southern sky ($\delta < 0\degr$). Throughout this paper we refer
to this catalog as the HIPASS Bright Galaxy Catalog (BGC). The galaxy finding, 
selection criteria, optical identification and \HI\ parameterization are 
described in detail in the following subsections. The \HI\ spectra of all 1000 
BGC sources are displayed in Fig.~\ref{fig:hispectra} and the \HI\ properties 
together with their optical (or in some cases infrared) identifications, where 
available, are given in Table~\ref{tab:bgctable}. For a description of the 
table columns see Section~\ref{sec:hitable}. The full versions of 
Fig.~\ref{fig:hispectra} and Table~\ref{tab:bgctable} are in the electronic 
edition of the Journal, while the printed edition contains only a sample. The 
HIPASS BGC and additional information is available at
{\tt ~~www.atnf.csiro.au/research/multibeam/}.

The HIPASS BGC overlaps with recently published \HI\ catalogs of particular 
regions of the sky based on the same set of data: Banks et al. (1999; the 
nearby Centaurus\,A Group); Putman et al. (2002; HVC catalog, including some 
nearby galaxies); Kilborn et al. (2002; South Celestial Cap) and Waugh et al. 
(2002; Fornax region). Although the \HI\ parameters of a HIPASS source can 
vary slightly between catalogs, depending on the chosen fitting parameters, 
the HIPASS name of each source (e.g. HIPASS J1712--64) is maintained for 
consistency and cross-identification purposes. Following standard nomenclature,
letters (`A', `B', etc.) are appended to a name where the positions are similar
enough that two or more sources would have the same name.

The HIPASS BGC also overlaps with the HIPASS Catalogue (HICAT: Meyer et al. 
2004, Zwaan et al. 2004) which contains 4315 \HI\ sources with systemic 
velocities between $\sim$300 and 12700\kms\ (see Table~\ref{tab:hisurveys}). 
Optical identification of the HICAT sources is under way. Note that HICAT 
and the BGC are based on the same survey data, but were compiled and 
parameterized independently.

\subsection{Galaxy Finding}
\label{sec:finding}
An automatic galaxy-finding algorithm, {\sc MultiFind} (Kilborn 2001; 
Kilborn et al. 2002), was run on the 388 edge-masked HIPASS cubes covering 
the whole southern sky, over all 1024 channels (i.e. --1200 to +12700\kms),
with an \HI\ peak flux density cutoff of 60 mJy\,beam$^{-1}$. The resulting 
source list is very large because the algorithm is designed to 
detect all signals above a certain strength when present over two or more 
consecutive channels. Beside \HI\ emission from our own Galaxy and many other
galaxies, it also contains Galactic recombination lines and HVCs as well as 
artifacts such as interference and residual continuum ripple. 
% Skipped.
% It is important to set the peak flux cutoff level well below the desired 
% detection limit, in order not to miss galaxies with narrow Gaussian or 
% double-horn \HI\ profiles (which are rejected if appearing in only one 
% channel above the cutoff level). 
After applying the selection criteria (see below), each \HI\ spectrum was
inspected by eye to eliminate any artifacts as well as Galactic signals. In 
many cases the corresponding HIPASS cube was inspected to reliably classify
the signal based on the surrounding pixels and neighbouring channels. For 
example, Galactic recombinations lines were rejected on the basis that they
appear at several, well-defined frequencies (see Section~\ref{sec:velcover}). 
Residual interference, where present, was rejected as it is usually narrow, 
intermittent and scattered over many degrees.

\subsection{Selection Criteria}
\label{sec:selection}
From the extensive source list we extracted a galaxy candidate list with 
systemic velocity \vsys\ = --1200 to 8000\kms\ (excluding $\pm$350\kms), and 
Galactic latitude $|b| > 3$\degr. The range $|v| < 350$\kms\ was omitted in the
selection because of confusion with HVCs (see Putman et al. 2002) which makes 
it difficult to identify galaxies at those velocities. The latitude range of 
$|b| < 3$\degr\ was omitted due to contamination by extensive continuum ripple 
from Galactic sources and therefore particularly difficult for an automatic 
galaxy finder. This was possible because this area had already been searched 
extensively by eye using the Parkes \HI\ ZOA shallow survey (HIZSS; Henning 
et al. 2000) which was conducted contemporaneously with HIPASS. To overcome 
the velocity and latitude gaps we added to our galaxy candidate list the 
positions of all known galaxies with systemic velocities \vsys\ $<$ 350\kms\ 
(using the NASA/IPAC Extragalactic Database, NED) as well as galaxies from 
the HIZSS ($|b| < 5$\degr). All galaxies are then parameterized using HIPASS 
data.

\subsection{\HI\ Parameterization}
\label{sec:hipara}
The \HI\ parameterization is carried out semi-automatically using several 
{\sc miriad} programs. For each source, the \HI\ emission profile was 
inspected using preliminary positions and velocities as obtained from {\sc 
MultiFind} or, in case of the added sources (see above), from the HIZSS and 
NED. If necessary, the velocities were adjusted, and an integrated \HI\ 
intensity map produced (program {\sc moment}). This, in turn, was used to 
estimate the \HI\ centroid using a Gaussian fit (program {\sc imfit}). The 
fitting area (a polygon) was set manually according to the source size. 
Based on the measured centroid, a weighted spectrum was then formed.
To this spectrum, a first-order baseline was fit and subtracted (see 
Fig.~\ref{fig:hispectra}). The \HI\ peak and integrated flux densities, 
the systemic velocity and the velocity widths were then measured (program 
{\sc mbspect}). 

Most galaxies are point sources in HIPASS; i.e. their intrinsic size is much
smaller than the 15\farcm5 angular resolution of the gridded data. For these
sources we considered data in a region of size $20\arcmin \times 20\arcmin$, 
centered on each source, and weighted each pixel according to the expected 
beam response. To find extended sources, we also directly summed the flux in 
a region of size $28\arcmin \times 28\arcmin$. Sources with at least 10\% 
more flux were flagged as potentially extended, and the cube was manually 
inspected. Where the increased flux was due to confusion with a neighbouring 
galaxy, the point source model was used and the galaxy marked as confused 
(c). Truly extended galaxies (marked with "e") were re-parameterized using 
{\sc mbspect} with a more appropriate box size (see Section~\ref{sec:hiext}). 

% NOTE: Total number 8384 candidates (5927 multiples)
Altogether, more than 8000 spectra ($\sim$70\% of which were multiple entries) 
were inspected. Galactic signals (e.g. recombination lines) as well as 
artifacts (interference, continuum ripple, etc.) were removed or cataloged 
separately (e.g. HVCs; see Putman et al. 2002). Of the 1400 galaxies finally 
parameterized, 1000 have \HI\ peak flux densities $\ga$ 116 mJy. Assuming
a velocity width of 100\kms\ this corresponds to a column density limit of
$\sim$10$^{19}$ cm$^{-2}$ (for objects filling the beam) and an \HI\ mass 
limit of $2 \times 10^6 \times D^2$\Msun, where $D$ is the galaxy distance 
in Mpc.
% NOTE. FHI = 0.116 Jy x 100 km/s x 0.7 = 8.12 Jy km/s

The Large Magellanic Cloud (LMC) and the Small Magellanic Cloud (SMC) obey the 
BGC selection criteria. However, their huge angular sizes make it difficult to 
parameterize their \HI\ properties in the same way as the BGC sources. Their 
\HI\ parameters have therefore been obtained from separate Parkes observations 
and are not included in Table~2 (or any of the discussions). Both the standard 
HIPASS reduction technique, and the modified {\sc minmed5} method used by 
Putman et al. (2003) substantially underestimate the integrated \HI\ flux 
density in the Magellanic Clouds. Staveley-Smith et al. (2003) measure an 
integrated \HI\ flux density of $8.1 \times 10^5$ Jy\kms\ for the LMC. Adopting
a distance of 50 kpc this gives an \HI\ mass of $4.8 \times 10^8$\Msun. 
Stanimirovic et al. (1999) measure an integrated \HI\ flux density of $4.5 
\times 10^5$ Jy\kms\ for the SMC. Adopting a distance of 60 kpc this gives 
an \HI\ mass of $3.8 \times 10^8$\Msun. We refer to Westerlund (1997) for a 
detailed discussion of the Magellanic Cloud distances.

\subsection{The \HI\ Data}
\label{sec:hitable}
Fig.~\ref{fig:hispectra} shows the integrated \HI\ spectra for all 1000 BGC 
sources (only two example pages are given in the printed version of the 
journal; all \HI\ spectra are given in the electronic version).
Marked in each HIPASS spectrum are several measured \HI\ parameters (the peak 
flux density, 50\% and 20\% velocity widths) as well as the velocity range 
over which the \HI\ emission profile is parameterized. The velocity range
displayed outside the profile was used for the first order baseline fit.
The width-maximized 50\% and 20\% points are highlighted with a circle and
the width-minimized points are highlighted with a cross (in most cases these
fall on the same positions due to the chosen velocity range).

The \HI\ properties obtained for all 1000 HIPASS BGC sources are listed in
Table~\ref{tab:bgctable}. The columns are as follows:  \\
{\em ---Col.\,(1)} HIPASS name, \\
{\em ---Cols.\,(2+3)} fitted \HI\ position in RA, DEC (J2000), \\
{\em ---Cols.\,(4+5)} Galactic longitude, $l$, and latitude, $b$, 
          in degrees, \\
{\em ---Col.\,(6)} most likely optical (or infrared) identification(s)
          (see Section~\ref{sec:optids}), \\
{\em ---Col.\,(7+8)} the \HI\ peak flux density, \Speak, and its uncertainty, 
          $\sigma$(\Speak), in Jy, (spatially integrated, where appropriate) \\
{\em ---Col.\,(9+10)} the integrated (spatially and spectrally) \HI\ flux 
          density, \FHI, and its uncertainty, $\sigma$(\FHI), in Jy\kms, \\
{\em ---Col.\,(11+12)} the \HI\ systemic velocity, \vsys, measured at the 
          midpoint of the 50\% level of the peak flux density, and its 
          uncertainty, $\sigma$(\vsys), in \kkms, (all velocities are in the 
          heliocentric velocity frame using the optical convention), \\
{\em ---Col.\,(13)} velocity line widths, \wfi, in \kkms, measured at the 50\% 
          level of the peak flux density (the uncertainty, $\sigma$(\wfi), 
          is about $2 \times \sigma$(\vsys), see Col.\,12),  \\
{\em ---Col.\,(14)} velocity line widths, \wtw, in \kkms, measured at the 20\% 
          level of the peak flux density (the uncertainty, $\sigma$(\wtw), 
          is about $3 \times \sigma$(\vsys), see Col.\,12), \\
{\em ---Col.\,(15)} the Local Group velocity, \vLG, in \kkms, calculated using
          $v_{\rm LG} = v_{\rm sys} + 300 \sin l \cos b$, and \\
{\em ---Col.\,(16)} the logarithm of the \HI\ mass, \MHI, calculated using 
          \MHI\ [\MMsun] = $2.36 \times 10^5~D^2$~\FHI, where \FHI\ is the 
          integrated \HI\ flux density in Jy\kms, and $D$ the distance in Mpc.
          Distances are derived using $D$ = \vLG\ / \Ho, except for nine  
          galaxies (see Section~\ref{sec:distances} and 
          Table~\ref{tab:closest}) for which we use independent distances. 
          The latter are marked with a star (*) in this column. 
          We adopt a Hubble constant of \Ho\ = 75\kms\,Mpc$^{-1}$.  \\
{\em ---Col.\,(17)} gives flags noting extended (e) and confused (c) sources, 
galaxies with no previous velocity measurement (:) or potentially incorrect 
velocity measurements (w), baseline ripple (r), and Hanning smoothing (h). 
Galaxy pairs and groups are also marked as such.
 
Note that we use a first order baseline fit for all but three BGC sources
(HIPASS J0037--22, J0930--35, and J1324--42). For the latter a severe baseline 
ripple was present, and had to be subtracted out by high-order polynomial fits.
Hanning smoothing was applied to the spectrum of one galaxy, HIPASS J0403--01, 
(to reduce spectral ringing) before the velocity range of the \HI\ emission 
could be determined. 

\subsubsection{Uncertainties in the measured HI properties}
\label{sec:hierror}
In the following we analyze the uncertainties in the measured \HI\ properties,
some of which can be determined more accurately than others. Uncertainties in 
the peak and integrated flux densities, $\sigma$(\Speak) and $\sigma$(\FHI),
respectively, and the systemic velocity, $\sigma$(\vsys), are given for each 
BGC source (see Cols.~8, 10 and 12 in Table~\ref{tab:bgctable}). Other 
uncertainties can be calculated from these values or are given more generally.

The positional accuracy of a HIPASS source can be calculated as the gridded 
beam (15\farcm5) divided by the \HI\ signal-to-noise ratio. Since BGC sources 
generally have signal-to-noise ratios $\ga$9, the positional uncertainty is 
expected to be smaller than 
1\farcm7. By comparing the \HI\ positions given in Table~\ref{tab:bgctable} 
with those of the corresponding cataloged, optical or infrared counterparts 
(see Section~\ref{sec:optids}) we find a standard deviation in RA, Dec of 
1\farcm3. 

The uncertainty in the peak flux density, $\sigma$(\Speak), of a HIPASS source 
is generally dominated by the r.m.s. noise in the corresponding \HI\ spectrum. 
We measure an r.m.s. of $\sim13$ mJy for most of the survey. Increased noise 
levels are measured in the \HI\ spectra of extended sources and in regions of 
high 20-cm continuum flux density (see Section~\ref{sec:obs}). In addition,
$\sigma$(\Speak) increases slightly with rising peak flux density (see Barnes 
et al. 2001); we estimate this as 5\% of \Speak, resulting in 
$\sigma$(\Speak)$^2$ = rms$^2$ + (0.05 \Speak)$^2$. 
% NOTE: We adopt: rms = crms + 5 mJy (crms = clipped r.m.s.) 
During the \HI\ parametrisation of all BGC sources we measured the {\em 
clipped} r.m.s. (see Fig.~\ref{fig:rmshist}), which is lower than the standard 
r.m.s. noise ($\sigma_{\rm rms}$). The clipped r.m.s. is calculated after five 
iterations of 2$\sigma$ clipping which is necessary to obtain a robust 
baseline fit in the line-free region of the HIPASS spectra. 

There are various methods to estimate the uncertainty in the integrated flux 
density (see Reif et al. 1982, Schneider et al. 1990, Fouqu\'e et al. 1990a).
We use $\sigma(F_{\rm HI}) = 4 \times SN^{-1} 
                             (S_{\rm peak}~F_{\rm HI}~\delta v)^{1/2}$, 
modified from the slightly more conservative estimate by Fouqu\'e et al. 
(1990a). Here, $SN$ is the ratio of \Speak\ to $\sigma$(\Speak), and 
$\delta v$ = 18\kms\ is the velocity resolution of HIPASS. The mean \FHI\ 
uncertainty in the BGC is 15\% \FHI ($\sigma$ = 7\%), although this is 
likely an overestimate (see Section~\ref{sec:pgcflux}).

The uncertainty in the systemic velocity is approximately $\sigma$(\vsys) = 
$3 \times SN^{-1}$ ($P~\delta v$)$^{1/2}$. Again, we slightly modified the 
estimate by Fouqu\'e et al. (1990a; see also Schneider et al. 1986, 1990). 
Here, $P$ = 0.5 $\times$ (\wtw -- \wfi) is a measure of the steepness of the 
profile edges. We find median values of $P$ = 12\kms\ and $\sigma$(\vsys) = 
4\kms.

Following Schneider et al. (1986) the uncertainties in the 50\% and 20\% 
velocity line widths are $\sigma$(\wfi) = $2 \times \sigma$(\vsys) and 
$\sigma$(\wtw) = $3 \times \sigma$(\vsys), resulting in median values of 
8\kms\ and 12\kms, respectively. The velocity widths listed in 
Table~\ref{tab:bgctable} have not been corrected for resolution and internal 
velocity dispersion (see, e.g., Tully \& Fouqu\'e 1985, Bottinelli et al. 1990,
Fouqu\'e et al. 1990b).
 
\subsubsection{Distances}
\label{sec:distances}
For most BGC sources approximate distances are derived using $D$ = 
\vLG\ / \Ho, where $v_{\rm LG} = v_{\rm sys} + 300 \sin l \cos b$ is the 
Local Group velocity and \Ho\ = 75\kms\,Mpc$^{-1}$. Although independent 
distances are currently available for about 45 galaxies in the BGC (see
Mateo 1998, Willick \& Batra 2001, Karachentsev et al. 2003, and references
therein) we adopted these only for the closest galaxies ($D < 2.2$ Mpc, see 
Table~\ref{tab:closest}). We note that all but one of the galaxies with 
\vLG\ $<$ 131\kms\ have independent distance estimates; the exception is 
the newly catalogued galaxy HIZSS\,003 (HIPASS J0700--04) which has a Local 
Group velocity of 115\kms. 
Fig.~\ref{fig:lv} shows distances versus \HI\ Local Group velocities
for all galaxies in the BGC with $D \la 5$ Mpc. For error bars on the 
independent distances (typically 10\%) consult the above references. 

\subsection{Optical Identification}
\label{sec:optids}
The search for optical (or infrared) counterparts was conducted using the
NASA/IPAC Extragalactic Database (NED). For the identification we used 
both positional and velocity information. A search radius of 6\arcmin\ was 
sufficient in most cases to find a cataloged galaxy, but because of the 
15\farcm5 gridded beam, sources within that sky area and with similar 
velocities may contribute and confuse the \HI\ emission spectrum and flux 
density measurement. The measured \HI\ systemic velocity was used to identify 
the most likely optical counterpart(s), whenever previously cataloged 
velocities were available in NED. For further analysis, in particular to 
obtain the optical properties of galaxies in the BGC (presented in a 
forthcoming paper) we also used the Lyon/Meudon Extragalactic Database (LEDA).
% NOTE: once a matching galaxy was found in NED, we did not inspect
%       the DSS to check if there are other potential candidates.
Figs.~\ref{fig:scatter} and \ref{fig:radecsep} show the separations between 
the fitted \HI\ position, from HIPASS, and the optical positions for all 853 
galaxies with single optical identifications in LEDA, i.e. galaxy pairs and
groups as well as the newly cataloged galaxies and \HI\ clouds are not 
included here.
% NOTE: 44 pairs + 11 groups + 91 new objects + ESO174-G?001 = 147 
The difference between the measured \HI\ systemic velocity of HIPASS BGC 
members and the optical velocities of their LEDA counterparts is shown in 
Fig.~\ref{fig:velsep}.
The scatter diagram (Fig.~\ref{fig:scatter}) reveals no systematic offset of 
the positions at any peak flux density level. The standard deviation in RA 
and Dec is $\sigma_{\rm RA}$ = 1\farcm0 and $\sigma_{\rm DEC}$ = 0\farcm8, 
respectively.
Galaxy positions from HIPASS are therefore generally accurate to within a 
few arcminutes, with the actual value depending on the \HI\ peak flux density 
and source extent (Barnes et al. 2001). Large offsets between the \HI\ and 
the optical positions usually occur when multiple galaxies contribute to the 
signal or when the \HI\ distribution is asymmetric or peculiar. It is likely
that occasionally an optical galaxy has been mis-identified as the counterpart
to a BGC source, which might result in a relative large position offset. For
example, 
we found one source, HIPASS J0622--07 (\vsys\ = 754\kms), for which the 
proposed optical counterpart, CGMW1-0080, at a projected distance of 5\farcm6 
is not the correct identification (see Fig.~\ref{fig:hipass0622}). 
\HI\ follow-up observations with the Australia Telescope Compact Array (ATCA) 
revealed a previously uncataloged galaxy, less than 1\arcmin\ from the HIPASS 
position and easily visible in the Digitized Sky Survey (the galaxy center is 
close to a star; see also Whiting, Hau \& Irwin 2002).
% HIPASS J0622--07 = WHI B0619-07 @ RA,DEC(J2000) = 06:22:13.8,-07:50:23

For 91 sources, NED had no catalog entry, mainly because of Galactic 
extinction (see Section~\ref{sec:newgal}). Additionally, for 51 galaxies, 
no velocity information was recorded in NED (these are flagged with `:' 
in Table~\ref{tab:bgctable}). For 16 galaxies, the first-listed NED velocity 
differs significantly, mostly a result of some incorrect optical measurements; 
these are labeled `w' (for details see Appendix~\ref{app:wrong}). At least 
44 BGC sources correspond to galaxy pairs and 11 to compact galaxy groups.
More than 68 galaxies are confused (`c') by a neighbouring galaxy. As a 
result their \HI\ flux densities maybe slightly overestimated.

\subsection{\HI\ absorption}
\label{sec:hiabs}
The \HI\ spectra of the galaxies NGC~253 (HIPASS J0047--25), NGC~3256 (HIPASS 
J1028--43), NGC~4945 (HIPASS J1305--49), Circinus (HIPASS J1413--65), NGC~5128 
(Centaurus\,A, HIPASS J1324--42) and possibly others are affected by \HI\ 
absorption against their bright radio nuclei. The integrated \HI\ flux density 
and \HI\ mass of these galaxies is therefore underestimated. Furthermore, their 
strong radio continuum sources cause baseline ripple resulting in further 
uncertainty in the measured parameters. 

The \HI\ spectrum at the position of the nearby radio galaxy NGC~5128 
represents the most difficult case. Due to the large extent and high flux 
density of its 20-cm radio continuum emission, the resulting 
(artificial) baseline ripple as well as the presence of \HI\ absorption and 
extended \HI\ emission over a large velocity range, the \HI\ properties of
NGC~5128 are difficult to measure. The fit to the \HI\ position in HIPASS
is very uncertain and offset by 8\farcm9 from the optical position (see 
Table~\ref{tab:bgctable}). We used a 9th order baseline fit to the original 
\HI\ spectrum within a carefully selected velocity range to obtain the \HI\ 
properties. The NGC~5128 \HI\ spectrum shown by Gardner \& Whiteoak (1976) 
was used to set the appropriate velocity range for the \HI\ emission. The 
fitting results have large uncertainties. Using the same method as for the
other BGC sources, we measure an integrated \HI\ flux density of only $91.8
\pm 13.2$ Jy\kms, which is clearly an underestimate when compared with the 
\HI\ measurements (\FHI\ = 122 Jy\kms) by van Gorkom et al. (1990).
% NOTE. van Gorkom et al. 1990: MHI=7.2x10^8 Msun, D=5Mpc => FHI = 122 Jy km/s
%     Schiminovich et al. 1994: MHI=4.5x10^8 Msun (disk), D=3.5Mpc => 156
%                                  +1.5x10^8 Msun (shells)
%                                  =6.0x10^8 Msun => FHI = 207.5 Jy km/s
Schiminovich et al. (1994) in addition find \HI\ emission associated with 
the diffuse shells of NGC~5128 and measure \FHI\ = 208 Jy\kms. The extended 
emission is also visible in the HIPASS data. Using a Gaussian fit we find an 
\HI\ diameter of 47\farcm5 ($\pm$7\farcm3) $\times$ 26\farcm1 ($\pm$4\farcm0)
at $PA$ = 29\degr\ and \FHI\ $\approx$ 186 Jy\kms\ (\MHI\ = $9 \times 
10^8$\Msun).
% NOTE: IMFIT output.
% Peak value:                  36.05
% Total integrated flux:       186.3
% Right Ascension:                13:25:43.1
% Declination:                    -42:52:53.7
% Major axis (arcsec):        2852 +/-436
% Minor axis (arcsec):        1567 +/-240
% Position angle (degrees):      28.81 +/- 9.52
% Deconvolved Major, minor axes (arcsec): 2696 1261
% Deconvolved Position angle (degrees):      28.8
The Local Group velocity of NGC~5128 is 338\kms, resulting in a distance of 
4.5 Mpc. Recent estimates imply a closer distance of 3.5 Mpc (Hui et al. 1993).

We note that some galaxies with bright radio nuclei may be easily detected in 
\HI\ absorption, but not in \HI\ emission. Such galaxies would be missing from 
any \HI\ peak-flux limited catalogs, despite their substantial \HI\ content. 
An example is the edge-on disk galaxy NGC~5793 (HIPASS J1459--16A) which is not
part of the BGC. It has an extremely luminous radio nucleus and is observed in
\HI\ absorption at velocities from 3420 to 3590\kms\ (see also Pihlstroem et 
al. 2000). A very weak emission feature at 3200\kms\ matches the blue-shifted 
H$_2$O maser emission of NGC~5793 measured by Hagiwara et al. (1997). \HI\ 
emission is also detected at $\sim$2900\kms\ (HIPASS J1459--16B) and is either 
associated with the elliptical galaxy NGC~5796 or an uncataloged neighbour 
to the East of NGC~5973.

\subsection{Completeness and Reliability}
\label{sec:complete}
The selection of sources for the BGC is based on their \HI\ peak flux density 
(\Speak\ $\ga$ 116 mJy). Since HIPASS has a uniform noise level of typically 
13 mJy, BGC \HI\ spectra generally have high signal-to-noise ratios (see 
Fig.~1). Identification and parameterization are both therefore expected to be 
highly reliable. Similarly, completeness is expected to be very high (see also 
Zwaan et al. 2003) with the exception of some areas of increased noise. These 
can arise from: residual narrow-band interference; strong Galactic hydrogen 
recombination lines (see Section~\ref{sec:velcover}); scanning sidelobes north 
and south of bright extended \HI\ sources (see
Barnes et al. 2001); and residual baseline ripple at the position of strong 
radio continuum sources. As it affects all velocities, the latter is probably 
the most important, with the areas most affected being: (1) the Galactic Plane 
($|b| < 1\degr$), particularly the spiral arm regions where the 20-cm radio
continuum emission is very high; (2) the locations of individual strong radio 
continuum sources, some of which are the nuclei of gas-rich galaxies (e.g.,
NGC~5128, see Section~\ref{sec:hiabs}). Within the BGC velocity range, only
 $\sim$2.5\% of all southern HIPASS spectra have an r.m.s. noise greater than 
23 mJy, i.e. at this noise level \HI\ sources in the BGC are still detected 
with \Speak\ $> 5\sigma$. This gives a useful upper limit to the incompleteness 
of the BGC. For comparison, Fig.~\ref{fig:rmshist} shows a histogram of the 
clipped r.m.s. as measured in the spatially and spectrally integrated \HI\ 
spectra of all BGC sources.

% NOTE.
% In a static Euclidean geometry (constant space density, no evolution, ...)
% the slope of the differential (integral) source count is -2.5 (-1.5).
% integral: N (>S), differential: N (S to S+dS), 

The \HI\ peak flux density distribution (Fig.~\ref{fig:peakhist}) is well 
described by an Euclidean power law ($N \propto$ \Speak$^{-2.5}$) suggesting
that the BGC is complete. The best-fit linear regression to the log-log version
of the histogram gives a slope of $-2.57 \pm 0.12$ and an offset of $0.040 \pm
0.073$. The fit results vary with bin size and the number of bins, but are well
% NOTE. Using a bin width of 20 mJy, the best fit was obtained for 16+-1 bins.
reflected by the given error bars as long as each bin has a sufficient number 
of sources. Note that \Speak\ varies with velocity resolution and is affected 
by the noise of the \HI\ spectrum. 

The velocity-integrated \HI\ flux density of a galaxy, \FHI, is a much more 
useful physical measurement than \Speak, because it relates more directly to 
its \HI\ mass: \MHI\ $\propto$ \FHI\ $D^2$. The \FHI\ distribution 
(Fig.~\ref{fig:fhihist}) can be described by the same power law ($N \propto$ 
\FHI$^{-2.5}$) as the \Speak\ distribution but only for \HI\ sources with 
\FHI\ $\ga 25$ Jy\kms. This limit gives a highly complete subsample of 
$\sim$500 BGC sources. Whereas the BGC completeness limit is quite well 
defined the slope of the distribution is rather uncertain (and not well 
reflected in the given error bar). The best-fit linear regression to the 
log-log version of the histogram gives a slope of $-2.47 \pm 0.13$ and an 
offset of $5.54 \pm 0.23$.
% NOTE. Starting at 25 Jy km/s and using a bin width of 4 Jy km/s, only 
%       14 bins have more than 5 sources. 
% programs: scount.f scount3.f
We note that HICAT contains twenty \HI\ sources with \FHI\ $>$ 25.0 Jy\kms\ 
that are not listed in the BGC. These have \Speak\ $<$ 116 mJy (the BGC
\Speak\ cutoff) and 20\% velocity widths greater than $\sim$400\kms. This
suggests that the \FHI-limited BGC sub-sample is about 95\% complete.

In any \HI\ peak flux limited sample, like the BGC, sources at a given
integrated \HI\ flux density are detected more easily the narrower their 
velocity widths are (see Fig.~\ref{fig:w50fhi}). This favours the detection 
of sources that are either intrinsically narrow-lined or galaxies viewed 
close to face-on. On the other hand, sources with a velocity width smaller 
than the HIPASS velocity resolution of 18\kms\ and low signal to noise are 
very difficult to detect because the spectrum is likely to resemble noise 
or interference. The peak flux selection therefore means that the lowest 
integrated \HI\ flux densities are found in galaxies with rather narrow 
profiles (typically \wfi\ $<$ 40\kms), as shown in Fig.~\ref{fig:w50fhi}. 
From the work by Ryan-Weber et al. (2002) we know that many of the 
narrow-lined sources in the BGC are late-type dwarf galaxies, either 
previously uncatalogued or without prior velocity information. For example, 
we determined new velocities for 8 out of the 10 galaxies with the lowest 
\FHI\ values; these galaxies are generally under-represented in optical 
surveys because of their small diameters and low surface-brightness. The 
optical properties for all BGC sources, such as morphological types, 
inclination angles, and luminosities, will be presented in a later paper.

\subsubsection{Velocity coverage}
\label{sec:velcover}
Because of confusion with Galactic gas, the BGC includes galaxies with
$|v_{\rm sys}| < 350$\kms\ only where these objects were previously known 
(and meet the BGC selection criteria). As such, the BGC may not be complete 
in this velocity range. 

Putman et al. (2002) cataloged \HI\ objects with $|v_{\rm LSR}| < 500$\kms\ 
(mostly HVCs), excluding only $|v_{\rm LSR}| < 90$\kms, and identified 40 
galaxies. Among them are two newly cataloged galaxies: HIPASS J0746--28 
(HIZSS\,021, Henning et al. 2000) and HIPASS J1337--39 in the Centaurus\,A 
Group (see also Banks et al. 1999). Objects were classified as galaxies 
based on either being cataloged in LEDA or NED or being visible in the 
Digitized Sky Survey (DSS). Since this classification was based on optical 
% NOTE: the DSS was searched at the position of CHVCs and :HVCs
data, some previously uncataloged galaxies (especially in the ZOA) may have 
been mis-classified as HVCs. The BGC contains 37 \HI\ objects with \vsys\
$< 500$\kms, six of which are newly cataloged objects including three \HI\ 
clouds (see Table~\ref{tab:hiclouds}). That means there may be undiscovered 
low-velocity or Local Group galaxies missing from the BGC with $|v_{\rm sys}| 
< 350$\kms.

We did not search for objects with $v > 8000$\kms. This limit, rather than 
the 12700\kms\ limit of HIPASS, was set for practical reasons: the residual 
continuum ripple is worse at high velocities and occasional emissions from the 
GPS L3 beacon at velocities of $\sim$8200\kms\ cause many false detections in 
{\sc MultiFind}. Whilst visual inspection of the data cubes easily eliminates 
such candidates, the extra work of searching this, and higher velocities, was 
not considered worthwhile given the extremely low probability of detection (a 
face-on galaxy with a velocity width of 50\kms\ would need to have an \HI\ mass
more than $2\times 10^{10}$ M$_{\sun}$ to be included in the BGC). The HIPASS 
Catalog (Meyer et al. 2004) contains $\sim$100 galaxies with \HI\ systemic 
velocities larger than 8000\kms, all of which lie below the BGC peak flux 
cutoff.
% NOTE. There is only one source with Speak > 100 mJy and v > 8000 km/s.
% It is HIPASS J1422--17 with Speak = 106 mJy, FHI = 8.3 Jy km/s at vhel = 9169 
% km/s (vLG = 9028 km/s => MHI = 2.8 x 10^10 Msun). In NED it is known as 
% CSRG0775 (v = 9169 km/s, FHI = 4.9 Jy km/s, SB(r)cd).

The Galactic hydrogen recombination lines H166$\alpha$, H210$\beta$ and 
H167$\alpha$ occur within the BGC velocity range at frequencies which, if 
interpreted as a velocity, correspond to --911, 4340, 4507\kms\ (for an example
see Fig.~\ref{fig:recomb}). Recombination lines are narrow, and generally
restricted to regions of strong thermal radio emission in the Galactic Plane. 
So narrow velocity-width galaxies behind the Galactic Plane can be difficult 
to identify at some velocities. However, because of their distinctive 
spectrum, \HII\ regions and other regions with recombination line emission 
are easily removed from the BGC.

\section{Results \& Discussions} % Section 4
\label{sec:results}
In the following we briefly discuss the newly cataloged galaxies and \HI\
clouds (Section~\ref{sec:newgal}) as well as the galaxies with the largest
angular sizes (Section~\ref{sec:hiext}). Section~\ref{sec:hiprop} is devoted 
to the \HI\ properties of the BGC sources. In Section~\ref{sec:pgcflux} we 
compare the integrated \HI\ flux density measurements in the HIPASS BGC to 
those in the literature, where available. 

\subsection{Newly Cataloged Galaxies \& \HI\ Clouds}
\label{sec:newgal}
For 91 BGC members NED has no catalog entry; 87 of these are most likely
galaxies (see below). The remaining four new sources appear to be \HI\ 
clouds without optical counterparts. Three of these lie outside the 
Zone of Avoidance: HIPASS J1712--64 (Kilborn et al. 2000), J1718--59 
(Koribalski 2001) and J0731--69 (Ryder et al. 2001); the fourth one is 
J1616--55 (Staveley-Smith et al. 1998). Their \HI\ parameters are listed 
in Table~\ref{tab:hiclouds}.

Whereas HIPASS J0731--69 has been identified as a potential tidal cloud near 
NGC~2442, the others are at much lower velocities (\vsys\ $<$ 500\kms).
HIPASS J1718--59 ($\sim$3\degr\ in length) lies only a few degrees away from
J1712--64, but at a slightly lower systemic velocity. Together with J1616--55 
and other, much weaker clouds at similar velocities, these objects could be 
tidal debris related to the Magellanic Clouds and the Leading Arm (Putman et 
al. 1998). The failure of deep observations by Lewis et al. (2002) to detect 
any stars in J1712--64 tends to confirm this theory.

% Skipped.
%We also detected HI\,1225+01 (the `Virgo Cloud'; HIPASS J1227+01), originally
%discovered by Giovanelli \& Haynes (1989), which will be part of the northern
%extension ($\delta$ = 0 to 25\degr) to the HIPASS Bright Galaxy Catalog.
%Although, in contrast to the \HI\ clouds mentioned above, the NE component of 
%the Virgo cloud has an optical counterpart (see Chengalur et al. 1995).

Of the 87 likely new galaxies, 57 ($\sim$70\%) lie at Galactic latitudes 
below 10\degr. These constitute a substantial fraction of the total of 154 
BGC members in this region. Fig.~\ref{fig:bhist} shows a histogram of the 
galaxy latitudes. There are 37 BGC objects with $|b| < 5$\degr, of which 32 
are Henning et al. (2000) HIZSS galaxies (counted as `new' for the purposes
of this paper). The five additional galaxies are HIPASS J1526--51 (\vsys\ = 
605\kms; also known as HIZOA J1526--51, see Juraszek et al. 2000), J1441--62 
(672\kms), J1758--31 (3316\kms), J1812--21 (1533\kms), and J1851--09 
(5485\kms); no optical counterparts have as yet been identified for these. 
They were most likely missed in the HIZSS because of their very narrow \HI\ 
profiles (\wfi\ $\approx$ 40--50\kms; except HIPASS J1851--09 for which we 
measure \wfi\ = 90\kms).
% NOTE. W50 =  52,  39,  41,   50,  90  km/s
%       FHI = 4.7, 6.0, 5.5, 11.2, 9.8  Jy km/s

Although most of the newly cataloged galaxies lie in or close to the Zone of 
Avoidance, some are near individual bright stars. Narrow \HI\ velocity widths 
are typical for most of the newly cataloged galaxies outside the ZOA. This, 
together with their morphology suggests they are mostly dwarf irregular 
galaxies. Of the 21 new galaxies with $|b| > 15$\degr, only one (HIPASS 
J0546--68) has no obvious optical counterpart as it lies behind the Large 
Magellanic Cloud. For a detailed analysis of all the newly cataloged galaxies 
see Ryan-Weber et al. (2002).

\subsection{Extended BGC Sources}
\label{sec:hiext}
We find at least 24 BGC sources to be extended when compared to the Parkes 
beam of 15\farcm5; these are listed in Table~\ref{tab:ext}. Although they 
have the largest angular sizes, mostly due to their proximity, only a few of 
them are physically large. BGC sources with very large angular sizes, like
the Circinus galaxy (HIPASS J1413--65), may have their \HI\ flux density 
underestimated due to strong negative sidelobes, an artifact of the bandpass 
calibration (see Barnes et al. 2001). In Table~\ref{tab:ext} we give their 
deconvolved 
Gaussian \HI\ diameters as well as some of their \HI\ properties, which are 
also listed in Table~\ref{tab:bgctable}. The diameters were determined from 
a Gaussian fit to the integrated \HI\ intensity distribution and represent, 
in nearly all cases, a good fit to the HIPASS data. For most HIPASS sources 
in Table~\ref{tab:ext} we give the fitted \HI\ major and minor axes diameters 
as well as the position angles, $PA$. For sources which are resolved in one 
direction only, we list the fitted \HI\ major axis diameter. Most of the 
extended sources in the BGC are nearby, well studied spiral galaxies, the 
largest of which is NGC~6744 (HIPASS J1909--63a). In addition, there are (at 
least) four extended galaxy pairs/groups and two \HI\ clouds. Note that the 
extended \HI\ envelope of a galaxy can greatly exceed the fitted Gaussian 
\HI\ diameter. All extended sources are briefly described in 
Appendix~\ref{app:extended}. 

\subsection{\HI\ Properties}
\label{sec:hiprop}

\subsubsection{Peak and Integrated HI Flux Densities}
\label{sec:hiflux}
The ten \HI-brightest galaxies in the BGC are, in order of their \HI\ peak flux
densities (\Speak), NGC~6822, NGC~300, NGC~55, NGC~3109, Circinus, NGC~5236 
(M\,83), the Wolf-Lundmark Melotte (WLM), NGC~3621, NGC~6744, and NGC~247. The 
list of galaxies with the highest integrated \HI\ flux densities (\FHI) is 
nearly identical (instead of the WLM it includes NGC~253). All are well-known,
% NOTE: in order of \FHI: NGC~6822, NGC~55, NGC~300, M\,83, Circinus, 
%       NGC~3109, NGC~6744, NGC~3621, NGC~253 and NGC~247.
nearby spiral galaxies which, except for the WLM, are extended with respect 
to the Parkes beam (see Tables~\ref{tab:closest} \& \ref{tab:ext}). For 
comparison, the optically brightest galaxies (according to their blue 
magnitude as listed 
in LEDA), are NGC~5128, NGC~5236, NGC~253, NGC~300, NGC~4594, ESO356-G004 
(Fornax dwarf spheroidal), NGC~6744, NGC~4945, NGC~6822, and NGC~1068. While 
there are several sources in common, three of the optically brightest 
galaxies (NGC~4594, ESO356-G004, and NGC~1068) did not make it into the 
BGC, emphasizing again that there are many optically bright (elliptical) 
galaxies with little \HI\ gas.
Histograms of the \HI\ peak and integrated flux densities of the BGC sources
are shown in Figs.~\ref{fig:peakhist} and \ref{fig:fhihist}, respectively, 
and are discussed in Section~\ref{sec:complete}.

\subsubsection{HI Velocity Distribution}
\label{sec:hivelo}
Fig.~\ref{fig:velhist1} shows the BGC \HI\ Local Group velocity histogram for 
all 1000 sources as well as for the complete \FHI-limited sub-sample (\FHI\ 
$>$ 25 Jy\kms, see Section~\ref{sec:complete}). We overlaid the selection 
functions as derived by Zwaan et al. (2003, see their Fig.~17) based on the 
BGC \HI\ mass function. Several overdensities are clearly visible indicating 
significant structure in the nearby galaxy distribution (see 
Section~\ref{sec:lss}).

For comparison we show in Fig.~\ref{fig:velhist2} the \HI\ systemic velocity 
histograms from the targeted samples of Mathewson et al. (1992) and Theureau 
et al. (1998). While some local structure is clearly visible, it differs 
significantly from that seen in the BGC due to selection effects. Note that 
MFB92 target Sb--d galaxies with inclination angles $i > 40$\degr\ and 
Galactic latitudes $|b| > 11$\degr, selected mainly from optical catalogs, 
% NOTE. Their selection criteria are: type Sb--d, $i > 40$\degr, 
%        opt. diameter $>$ 1\farcm7, $|b| > 11$\degr, mostly v<7000 km/s
while Theureau et al. (1998) target more distant Sa--Sdm galaxies. 
The differences between the BGC velocity histogram and that of the optical 
comparison sample obtained from LEDA are discussed in Section~\ref{sec:lss}.

% \paragraph{The Nearest Galaxy Groups.}
Galaxies with low systemic velocities are of particular interest as they are 
very close and can be studied in detail (see also Section~\ref{sec:lss}). The 
BGC contains about 50 (160) galaxies within a distance of 5 (10) Mpc many of 
which are located in the Local Group, the Sculptor Group or the Centaurus\,A 
Group. There are many other nearby groups, e.g. the NGC~3056 group 
(10$^{\rm h}$, --30\degr, \vsys $\sim$ 950\kms, overlapping with LGG\,180) 
with around ten \HI-rich members, but none as prominent as the Sculptor or 
Centaurus\,A Groups.

% NOTE: The southern Local Group galaxies are:
%       WLM, Cetus dSph, SMC, LMC, ESO245-G007, Fornax dSph, Carina, 
%       Sextans dSph, ESO594-G004, NGC6822, DDO210, and Tucana.

In addition to the Magellanic Clouds, we detected four {\em Local Group}
galaxies in the HIPASS BGC (see Table~\ref{tab:closest}) out of the 36 
listed by van den Bergh (2000): these are the dwarf irregular galaxies 
NGC~6822 (DDO\,209; see Section~\ref{sec:hiext}), DDO\,210 (Aquarius Dwarf), 
WLM (DDO\,221) and ESO594-G004 (SagDIG, see Young \& Lo 1997). The remaining 
six southern Local Group members are gas-poor dwarf spheroidal galaxies. 
Further away, in the outskirts of the Local Group, we detected the dwarf 
irregular galaxies IC\,5152, Sextans\,A and NGC~3109 (see 
Appendix~\ref{app:extended} and \ref{app:lgroup}). No optical identification 
has so far been obtained for the galaxy HIZSS\,003 ($b$ = 0\fdg1) which lies 
at a distance of $\sim$1.5 Mpc.

Of the 35 potential {\it Sculptor Group} members in the velocity range from 
zero to 800\kms\ we detected 25 (20 of which are part of the BGC), six were not 
detected and four objects (SC\,02, SC\,18, SC\,24, and SC\,42) were discarded 
because the neutral hydrogen gas at the published velocities (Cot\'e et al. 
1997) appears to be part of an HVC complex. 

Banks et al. (1999) searched for galaxies in the {\it Centaurus\,A Group}
using HIPASS data. They detected \HI\ in 30 galaxies, nine of which were 
identified as new group members. Four galaxies were previously uncataloged: 
HIPASS J1321--31, J1337--39, J1348--37 and J1351--47; the first two are also 
in the BGC. In total, the BGC contains at least 21 Centaurus\,A Group members 
(incl. NGC~5128, see Section~\ref{sec:hiabs}).

\subsubsection{HI Mass Distribution}
\label{sec:himass}
Fig.~\ref{fig:mhihist} shows the BGC \HI\ mass distribution for all 1000
sources as well as for the complete \FHI-limited sub-sample. The derived \HI\ 
masses range from $2 \times 10^6$\Msun\ to $4 \times 10^{10}$\Msun. The BGC 
sources with the highest \HI\ masses are described in 
Appendix~\ref{app:himassive}. We find a median \MHI\ of $2.9 \times 10^9$\Msun\
(log \MHI\ = 9.46), similar to the \HI\ mass of the Milky Way, for all BGC 
sources, and a slightly higher value of $3.9 \times 10^9$\Msun\ (log \MHI\ = 
9.59) for the complete sub-sample.
% NOTES.
% MHI Stats: N = 1000, mean(logMHI) = 9.35, sigma = 0.65
%            N =  496, mean(logMHI) = 9.54, sigma = 0.55
The BGC \HI\ mass function (Zwaan et al. 2003) is characterized by a Schechter 
function with a slope of $\alpha = -1.30 \pm 0.08$ and a "knee" of 
log ($M_{\rm HI}^{*}$/\Msun) = $9.79 \pm 0.06$. 

Figs.~\ref{fig:logdmhi} and \ref{fig:logw50mhi} show the \MHI\ distribution of 
all BGC sources with respect to their distance, $D$, and velocity width, \wfi. 
In Fig.~\ref{fig:logdmhi}, which illustrates the depth of our \HI\ sample, we 
have indicated the approximate detection limit of the BGC (\FHI\ $\approx$ 5 
Jy\kms) as well as the \FHI\ completeness limit (\FHI\ = 25 Jy\kms). Note that 
the \HI\ peak flux density cutoff, \Speak\ = 116 mJy, of the BGC is similar to 
the 5$\sigma$ detection limit of the \HI\ survey by Fisher \& Tully (1981a). 
Fig.~\ref{fig:logdmhi} is closely matched to figure 1 of Briggs (1997) who 
compared the depths of several \HI-rich galaxy samples. 
Fig.~\ref{fig:logw50mhi} shows the selection effects due to the peak flux
 limited nature of our sample as well as a strong correlation between the 
\HI\ velocity width and the \HI\ mass of galaxies. Note that we display the 
measured velocity widths uncorrected for galaxy inclination. At a given 
\HI\ flux density (\HI\ mass), the BGC is biased towards sources with low 
velocity width, and, vice versa, at a given velocity width it is biased 
towards galaxies with high \FHI\ (\MHI); see also Fig.~\ref{fig:w50fhi}.

Only 3 (38) sources in the BGC have \HI\ masses below 10$^7$ (10$^8$) \Msun, 
of these 1 (10) were previously uncataloged. The galaxies with the lowest 
\HI\ masses are DDO\,210 (Aquarius dwarf), HIPASS J1247--77 (see Kilborn et 
al. 2002), ESO594-G004 (SagDIG), ESO349-G031 (SDIG), NGC~5237, HIZSS\,003, 
IC~3104, UGCA~015 (DDO\,006), ESO383-G087 and HIPASS J1337--39 (Banks et al.
1999). They are typically nearby late-type dwarf galaxies, characterized by 
very narrow \HI\ velocity widths and \vsys\ $<$ 500\kms.

\subsubsection{HI Velocity Width Distribution}
\label{sec:hiwidth}
Fig.~\ref{fig:whist} shows a histograms of the 50\% and 20\% \HI\ velocity 
widths, \wfi\ and \wtw, respectively. While \wfi\ is usually a very robust 
measurement for the BGC sources, \wtw\ is more easily affected by noise and/or 
confusion by companions within the Parkes beam. We find a mean width of
$<w_{\rm 50}>$ = 173\kms\ ($\sigma$ = 106\kms) and 
$<w_{\rm 20}>$ = 210\kms\ ($\sigma$ = 116\kms). The median values for \wfi\ 
and \wtw\ are $\sim$20\kms\ lower: 155\kms\ and 191\kms, respectively. The 
BGC sources with the largest measured 50\% and 20\% velocity widths are 
described in Appendix~\ref{app:hiwidths}. For comparison we also show the 
distribution of the measured velocity widths in the complete \FHI-limited 
sub-sample of the BGC. Their mean values are 
$<w_{\rm 50}>_{\rm sub}$ = 228\kms\ ($\sigma$ = 108\kms) and 
$<w_{\rm 20}>_{\rm sub}$ = 269\kms\ ($\sigma$ = 119\kms), significantly higher
than for all BGC sources. The median values are 212\kms\ (\wfi) and 253\kms\
(\wtw). Henning (1995) obtained mean 50\% line widths for five different galaxy 
samples, ranging from $<w_{\rm 50}>_{\rm LSB}$ = 155\kms\ ($\sigma$ = 82\kms) 
for the LSB sample to $<w_{\rm 50}>_{\rm IRAS}$ = 288\kms\ ($\sigma$ = 124\kms) 
for the IRAS sample, clearly showing the effects of source selection. While 
the \FHI-limited sub-sample has an intermediate mean width, suggesting it is 
drawn from a population of large spirals and dwarf galaxies, the remaining
sources in the BGC are mainly drawn from a population of late-type galaxies.

\subsubsection{The Shape of HI Profiles}
\label{sec:hishape}
The ratio of the 20\% and 50\% velocity widths (see Fig.~\ref{fig:w20w50})
gives an indication of the \HI\ profile shape, in particular the steepness of
the profile edges. The steepest edges are typically found in regular spiral 
galaxies with extended, flat rotation curves, resulting in well-defined double 
horn profiles (e.g. HIPASS J0833--22 = NGC~2613, \wtw\ / \wfi\ = 1.03), while 
irregular dwarf galaxies tend to show Gaussian profiles (e.g. HIPASS J0919--12 
% NOTE: the theoretical value for a Gaussian profile is: w20/w50 = 1.52
= DDO\,060, \wtw\ / \wfi\ = 1.53). The median \wfi\ / \wtw\ ratio in the BGC 
is 1.18. The \HI\ profiles of galaxy pairs/groups and confused galaxies 
generally show much higher ratios. Values of \wtw\ / \wfi\ $>$ 4 were found 
in the irregular galaxy UGC~08127, and the groups HIPASS J2206--31 and 
J0605--14 (see Ryan-Weber et al. 2002). In these cases a narrow \HI\ component 
is accompanied by a broad \HI\ spectrum.

The \HI\ spectrum of a single galaxy is determined by its \HI\ distribution, 
velocity field and inclination. While a symmetric \HI\ spectrum usually 
indicates a symmetric, regular rotating galaxy, an asymmetric \HI\ spectrum 
can be due to a variety of internal and external factors (e.g., asymmetric 
\HI\ distribution, irregular rotation pattern, warps, neighboring \HI\ 
sources) which are best investigated using interferometric \HI\ data (see, 
e.g., Baldwin et al. 1980, Swaters et al. 2002). Richter \& Sancisi (1994) 
as well as Haynes et al. (1998) find that (at least) half of the single-dish 
\HI\ spectra obtained for large samples of galaxies show significant 
asymmetries. Following Haynes et al. we use the fractional velocity 
difference criterion, $|v_{\rm sys} - v_{\rm mom1}| / w_{\rm 50} > 0.02$, 
to estimate the approximate number of asymmetric \HI\ profiles in the BGC.
Here \vsys\ is the \HI\ systemic velocity, measured at the midpoint of the 
50\% level of the peak flux density (see Table~2), and $v_{\rm mom1}$ is 
the first moment. We find the \HI\ spectra of half the BGC sources to be 
asymmetric, in agreement with previous studies. For about a quarter of 
BGC \HI\ spectra we calculate a fractional velocity difference larger than 
0.04. Fig.~1 shows a large range of spectral shapes, ranging from symmetric 
over mildly asymmetric to severely distorted. We find that the difference 
criterion is not appropriate for BGC sources with narrow velocity widths.
Nevertheless, after excluding \HI\ sources with \wfi\ $>$ 90\kms, we again 
find about half the remaining \HI\ sources to be asymmetric.
There are 70 BGC sources with \wtw\ / \wfi\ $>$ 1.8, among them many galaxy 
pairs and groups, all of which fulfill the above asymmetry criterion.
% NOTE. There are 742 BGC sources with w50 > 90.0 km/s)
% 530 (361) sources have  asym > 0.02
% 281 (163) sources have  asym > 0.04

\subsection{Comparison of \HI\ fluxes with other surveys}
\label{sec:pgcflux} 
Here we compare the \FHI\ measurements in the HIPASS BGC with those obtained
by Mathewson et al. (1992) and Theureau et al. (1998) as well as with the 
21-cm magnitudes given in LEDA, which includes the first two \HI\ samples. 

The BGC and the targeted sample by MFB92 have 228 galaxies in common, although
some galaxies are confused by neighouring sources. Fig.~\ref{fig:mfbfhi} shows
the percentage difference between $F_{\rm HI,BGC}$ and $F_{\rm HI,MFB92}$. The 
median offset from zero is --9\%, consistent with the difference in absolute 
flux calibration (see Section~\ref{sec:fhical}).
% NOTE. MFB92 selection criteria: 
%       type=Sb-d, i>40degr, |b|>11degr, also S/N>=5, deltav = 7\kms.

Theureau et al. (1998), who observed over 1200 southern galaxies in \HI\ with 
the Nan\c{c}ay telescope (HPBW = $3\farcm6 \times 22\arcmin$ at $\delta$ = 
0\degr), use numerous "calibrating galaxies", three of which are also in the 
BGC. For NGC~1518, NGC~1637, and NGC~6814 they measure $F_{\rm HI,T98}$ = 
% NOTE. These are HIPASS J0406--21, J0441--02, and J1942--10
50.66, 38.43, and 21.03 Jy\kms, respectively, whereas we measure 
$F_{\rm HI,BGC}$ = $72.0 \pm 5.7, 72.4 \pm 5.9$, and $37.3 \pm 4.0$ Jy\kms\ 
(see Table~\ref{tab:bgctable}), $\sim$40--70\% larger than their values. Our 
\FHI\ measurements agree with those listed by Huchtmeier \& Richter (1989). 
As the \HI\ envelopes of galaxies are typically a factor two larger than 
their optical diameters (see, e.g., Salpeter \& Hoffman 1996), it is likely 
that only a fraction of the galaxy \HI\ extent was detected within the 
Nan\c{c}ay telescope beam.
% NOTES.
% NGC1637: Roberts, M.S., Hogg, D.E., Schulman 2001, Soccoro meeting (VLA data)
%          measure D_HI/D_25 = 3.0
% NGC6814: Liszt, H.S., \& Dickey, J.M. 1995, AJ 110, 998 (VLA data)
%          measure D_HI/D_0  = 2.0, D_HI = 7' = 23 kpc (D=20.7 Mpc)
%          FHI = 29.6 Jy\kms\ (+10%), D25 = 3' x 2.8' => D_HI/D25 = 2.3

We found LEDA 21-cm magnitudes, $m_{21}$, for 692 galaxies in the BGC, where 
$m_{21} = 17.4 - 2.5\log (F_{\rm HI})$. The mean of the $m_{21}$ uncertainties
quoted in LEDA is 0.22 ($\sigma$ = 0.08) or 22\% \FHI. For galaxies with 
multiple \FHI\ measurements, a general LEDA search outputs a single, preferred 
value. The percentage difference between $F_{\rm HI,BGC}$ and $F_{\rm HI,LEDA}$
is shown in Fig.~\ref{fig:pgcflux}. The median of the distribution is +0.8\%, 
showing good agreement with other \HI\ data for the majority of galaxies. For 
nearly 80\% of the BGC sources the \FHI\ difference is less than 25\%. If we 
ignore any sources which differ by more than 40\% in \FHI, the mean difference 
of the remaining 610 sources is --0.06\% ($\sigma$ = 15.2\%). 
% NOTE. 82 sources differ by more than 80\%

The largest \FHI\ difference occurs for the edge-on spiral galaxy MCG-02-14-003
(HIPASS J0511--09) which has an \HI\ flux density of 30.1 Jy\kms\ (BGC), in 
agreement with MFB92 who measure \FHI\ = 30.65 Jy\kms. The 
value of \FHI\ = 2.4 Jy\kms\ stated by Theureux et al. (1998) is likely to be 
a typographical error as their \HI\ spectrum is similar to ours. Discrepancies 
were already noted for HIPASS J0622--07 (for which CGMW1-0080 is not the 
correct optical counterpart; see Section~\ref{sec:optids}) and HIPASS J1324--42
(NGC~5128, for which the standard fitting routine severely underestimates the 
total \HI\ flux density). We found the published \HI\ flux densities for the 
galaxies with the largest discrepancies, ESO001-G006 (MFB92), 
ESO274-G016 and ESO375-G071 (Fisher \& Tully 1981) severely underestimated.
For the latter two galaxies our \FHI\ measurements agree roughly with those
by Longmore et al. (1982).
% NOTE. Longmore's absolute flux calibration is based on PKS 0521--36 and
%       PKS 1934--638, but no 20-cm flux densities are quoted.
%       Reif et al. (1982) quote PKS 0521--36 (17.5 Jy) and 
%       PKS 1934--638 (16.4 Jy). 
%       => 14.9 Jy / 16.4 Jy = 0.9
The \FHI\ value quoted by Theureux et al. (1998) for the galaxy UGC04358 (see 
the left-most point on Fig.~\ref{fig:pgcflux}), is also a typographical error.

\section{Large-Scale Structure} % Section 5
\label{sec:lss}
Here, we analyze the large-scale structures revealed by the HIPASS Bright 
Galaxy Catalog (BGC) and compare to the 1000 {\em optically}-brightest
galaxies in the southern sky.

\subsection{The optical comparison sample}
The optical galaxy sample was defined using the Lyon/Meudon Extragalactic
Database (LEDA), selecting the brightest galaxies based on their apparent 
total blue magnitudes ($B_{\rm T}$). There are 1000 southern galaxies with 
$B_{\rm T} \la 13\fm26$: 38 of these have no velocities recorded in LEDA, 
five have velocities beyond the BGC limit of 8000\kms. The HIPASS BGC and the 
LEDA comparison sample have only $\sim$400 galaxies in common. For $\sim$600 
galaxies in the optical sample, 21-cm fluxes are listed in LEDA, indicating 
(1) a large fraction of gas-poor galaxies (see Koribalski 2002) and/or (2) 
a lack of (cataloged) \HI\ measurements, probably both. Either way, it is 
clear that the \HI- and optically-selected samples differ significantly in 
their properties. 

Note also that the low angular resolution of single dish \HI\ galaxy samples 
compared to optically selected galaxy samples results in numerous unresolved 
galaxy pairs and groups for the HIPASS BGC, while the LEDA sample is mostly 
composed of individual galaxies.
% NOTE: There are $\sim$20 sources in the LEDA sample which have no name 
%       (neither type or vlg) and might not be galaxies. Quasars ? 

\subsection{Galaxy distribution on the sky}
Fig.~\ref{fig:BGCdis} shows the sky distribution of the 1000 \HI-brightest 
galaxies (the HIPASS BGC) and Fig.~\ref{fig:LEDAdis} that of the optical 
comparison sample. The sky distributions of the two samples  show a broad 
similarity, but strongly differ in some details.

The first concerns the Zone of Avoidance (ZOA): the \HI\ galaxies in the BGC 
give us 
the first view of the local Universe in the southern hemisphere uninhibited by 
the foreground stars and dust from our own Galaxy. Numerous known large-scale 
structures can be fully traced across the sky, the most prominent being the 
Supergalactic Plane (see Section~\ref{sec:sgp}), the second one certainly the 
Centaurus Wall which connects via the Hydra and Antlia clusters to the Puppis 
filament crossing the ZOA in two locations (two of the dinosaur's toes as
pointed out by Lynden-Bell 1994). The structures also allow a much improved 
delineation of voids as for example in the case of the Local Void (see 
Section~\ref{sec:lvoid}). Amazingly enough it is not only behind the Milky Way 
that new structures can be mapped for the first time. Some previously hardly 
noticed galaxy groups stand out quite distinctively in the \HI\ sky 
distributions. These are discussed in further detail in the description of 
the various velocity slices (see Fig.~\ref{fig:BGCvel}).

The second striking point concerns the clustering properties of the two 
samples: although both follow overall the same structures, the \HI\ 
distribution appears much more homogeneous and less clumpy than the optical 
one. This is due to the fact that (a) the optical sample contains many more 
elliptical and early-type galaxies than the HIPASS BGC which is dominated 
by late-type spiral galaxies (see Koribalski 2002), and 
(b) early-type galaxies mark the bottoms of the potential wells of the high 
density peaks in the Universe (from rich groups to clusters) whereas the
spiral or irregular galaxies --- although also found in clusters and rich 
groups --- are the main building blocks of poor groups of galaxies and 
particularly the filamentary cellular-like structures that seem to connect 
groups and clusters. A study of the statistical properties of both samples 
is under way.

A more subtle reason for the discrepancy in the distribution lies in the 
respective velocity histograms (see Fig.~\ref{fig:velhist3}). The peak of the 
velocity distribution of the BGC lies at a very low \vhel\ $\sim$ 1600\kms\ 
(width $\sim$600\kms), preceded by a secondary maximum of \vhel\ $\sim$ 
900\kms\ (width $\sim$300\kms). This results in a sharp drop-off for \vhel\ 
$\ga$ 2000\kms\ and little sensitivity to structures beyond 3000\kms. With 
less than 50 galaxies with \vhel\ $>$ 4000\kms, the BGC has no power to make 
predictions about the Great Attractor (GA) region and its overdensity with 
respect to the remaining part of the southern sky. The mean velocity in the 
BGC is $\sim$1900\kms\ ($\sigma \sim 1000$\kms). The optical comparison sample 
has a considerably higher mean of $\sim$2300\kms\ ($\sigma \sim 1300$\kms), 
with peaks around 1600\kms\ and 2600\kms, and a slight excess, compared to 
the BGC, of sources around $\sim$4200\kms. The center of the GA overdensity 
remains obscured because of the Milky Way. 
% NOTES: BGC  vsys mean = 1914, sigma = 1035
%        LEDA vsys mean = 2283, sigma = 1301

An interesting, though not intuitive, result of this difference in the
velocity distribution is the fact that the BGC samples the very nearby
Universe much deeper than a similarly defined optical sample: the \HI\ galaxy 
sample contains nearly 1.5 times more galaxies within \vhel\ $<$ 2000\kms\ 
than its optical counterpart even though the latter contains early-type 
galaxies and gas-poor dwarfs as well. And while the overall appearance of 
the BGC is less clumpy, it actually outlines structures in the very {\em 
nearby} Universe with much greater precision and homogeneity, and fewer gaps 
(see e.g. the Supergalactic Plane in the top panel of Fig.~\ref{fig:BGCvel}).

In the following the most prominent structures are described in a sequence 
of increasing velocity intervals.

\subsection{The nearest velocity slice ($v < 1400$\kms)}
The top panel of Fig.~\ref{fig:BGCvel} shows the distribution of the galaxies 
with velocities below 1400\kms. The most prominent features are the 
Supergalactic Plane (SGP), the Local Void next to the Puppis filament, 
the near side of the Fornax Wall and the Volans Void (see also 
Table~\ref{tab:lss}).

\subsubsection{The Supergalactic Plane} 
\label{sec:sgp}
The Supergalactic Plane (SGP) contains the greatest concentration of nearby 
groups and clusters of galaxies in the local Universe (see Lahav et al. 2000, 
and references therein). It is partly a projection of the Local Supercluster 
with the Virgo cluster near its center and the Local Group and the Sculptor 
Group at its outer boundary and partly a projection of the more distant 
Centaurus Wall. The SGP was defined by de Vaucouleurs in the 1950's based on 
a study of the Shapley-Ames Catalog of the brightest 1246 galaxies in the 
% NOTE: REF is Shapley & Ames 1932
whole sky. He noted its flattened pancake-like structure and defined a 
coordinate system centered on this structure. The Supergalactic equator (see 
de Vaucouleurs et al. 1991 for its definition) is shown as the horizontal 
line in Figs.~\ref{fig:BGCdis}, \ref{fig:LEDAdis} and \ref{fig:BGCvel}. We 
can clearly see that the nearest galaxies (top panel) follow this line 
closely. Interestingly enough it has been found in the last two decades 
that the concentration of galaxies towards the SGP continues out to larger 
distances and that the two dominant structures, the GA and Perseus-Pisces 
superclusters, also lie along this equator.

The SGP is very evident in galaxies from the BGC which form a distinct and 
continuous filamentary band --- without a gap due to the ZOA --- across the 
whole southern sky ending on the righthand side in the Southern Virgo 
Extension ($\alpha,\delta$(J2000) = 12\fh5, --5\degr, \vhel\ $\sim$ 1100\kms) 
which is quite a dominant agglomeration in this slice. Beside this clumping, 
the width of the band is no more than $\sim$10\degr. Deviations of up to 
10\degr\ -- 15\degr\ from the supergalactic equator are apparent.  

\subsubsection{The Puppis filament}
There is a suggestion in this sky projection that the SGP might be 2-pronged 
with a second filament which folds back across the ZOA in Puppis (8$^{\rm h}$, 
--30\degr), seemingly encircling the Fornax Wall described below.
The latter circle, which contains the here fully revealed Puppis 
filament, rises in significance due to two previously unrecognized galaxy 
groups in Antlia (centered at $\sim$900\kms\ and $\alpha,\delta$(J2000) =
$9^{\rm h}\,50^{\rm m}$, --30\degr\ and $10^{\rm h}\,40^{\rm m}$, --37\degr, 
respectively) which we suggest to be named the Antlia~G1 and Antlia~G2 
groups. They contain spiral galaxies with \HI\ masses of the order of 
10$^8$\Msun\ and are hardly visible in optical surveys. Antlia~G1 might be 
related to the Antlia Cloud although the grouping visible in the BGC does 
not coincide with the members listed in Tully (1982). The extension of the 
Puppis filament through these two groups seems to connect with the SGP at 
about $13^{\rm h}\,30^{\rm m}$, --30\degr. It is conceivable that the Puppis 
filament also connects to the slightly higher velocity Southern Virgo 
Extension.
 
\subsubsection{The Fornax Wall}
Another important feature in the nearby Universe is the Fornax Wall. It
consists of the nearby Dorado Cloud and is quite extensive on the sky. 
It is made of three loose galaxy groups centered on NGC~1672 
($\alpha,\delta$(J2000) = $4^{\rm h}\,45^{\rm m}$, --59\degr), NGC~1566 
($4^{\rm h}\,20^{\rm m}$, --55\degr) and NGC~1433 ($3^{\rm h}\,40^{\rm m}$,
--47\degr) and extending in velocity space from 900 to about 1200\kms, to the
compact Fornax cluster ($3^{\rm h}\,40^{\rm m}$, --36\degr, $\sim$1300\kms)
towards the slightly higher-velocity Eridanus Cloud ($\sim$1800\kms)
centered on NGC~1332 ($3^{\rm h}\,24^{\rm m}$, --22\degr). Fornax and 
Eridanus are still visible in the middle panel. 

\subsubsection{The Local Void}
\label{sec:lvoid}
The HIPASS Bright Galaxy Catalog also is much more stringent in defining the 
void sizes. The dominant void in the nearby Universe clearly is the Local Void 
which can be seen as the empty region above the SGP in the top and middle 
panels of Fig.~\ref{fig:BGCvel}.

The existence of the ``Local Void'' was first pointed out by Tully \& Fisher
(1987). They studied the sky distribution of galaxies within $cz$ = 3000\kms\
and discovered that relatively few galaxies lie in the region between $l$ = 
0 -- 90\degr, $b$ = --30\degr\ to +30\degr. Because the great portion of the 
Local Void is located behind the Milky Way, its size and distance have not 
been definitely measured. The Local Void might be relevant to the origin of 
the ``Local Anomaly'' motion (see, e.g., Faber \& Burstein 1988). Nakanishi 
et al. (1997) find the center of the Local Void to be located at $l, b$ = 
60\degr, --15\degr, and \vhel\ = 2500\kms, and the size is $\sim$2500\kms\ 
along the direction toward the center. Whereas Fairall (1998) gives the 
center of the Local Void to be located at $l, b$ = 18\degr, 6\degr\ 
($\alpha,\delta$(J2000) = 18$^{\rm h}$, --10\degr), and \vhel\ = 1500\kms, 
emphasizing herewith the difficulty in assessing its true dimensions.

The BGC enables us to redefine the extent of the Local Void as approximately 
$l \sim -10$\degr\ to +20\degr, $b \sim -20$\degr\ to +20\degr\ and 
\vhel\ $\la 1700$\kms\ (see Fig.~\ref{fig:lvoid}). For $b \sim -20$\degr\ 
to 0\degr\ the void appears to extend up to $l$ = 30\degr\ -- 40\degr. 
Fig.~\ref{fig:BGCvel} reveals its size to be clearly smaller in the \HI\ 
sky compared to the optical, mainly because the latter is strongly affected 
by Galactic extinction. The HIPASS BGC reveals the outer boundary of the 
Local Void (opposite the SGP in Fig.~\ref{fig:BGCvel}) which is hidden in 
optical surveys due to the broad ZOA in that region ($\pm$25\degr) caused 
by the Galactic Bulge (see Fig.~1 in Kraan-Korteweg \& Lahav 2000). We find 
nine previously uncataloged galaxies in the range $l$ = 0 to +40\degr\ and
$|b| < 20$\degr, six of which lie at velocities of 1500 -- 1700\kms.

\subsubsection{The Volans Void}
Similarly to the Local Void, the Volans Void (see ``Atlas Section'' in Fairall
1998) centered at $\alpha,\delta$(J2000) = $7^{\rm h}\,00^{\rm m}$, --70\degr, 
$\sim$800\kms\ is quite extensive in the optical (seen most clearly in the top 
right panel, below the SGP) but seems to be cut in half with galaxies 
identified in the BGC (top left panel).

\subsection{The middle velocity slice ($2200 > v > 1400$\kms)}
The middle panel of Fig.~\ref{fig:BGCvel} shows galaxies with velocities 
between 1400 and 2200\kms. In this range the BGC still probes deeper than 
the optical sample and reveals clear filamentary features while the optical 
galaxy distribution already loses the SGP as well as the Puppis filament. 
Besides the clumpings of Fornax, Eridanus and the Southern Virgo Extension 
visible in both catalogs, the BGC finds two further galaxy groups which are 
not at all visible in the optical. They are found close to the Southern 
Virgo Extension at $\alpha,\delta$(J2000) = $12^{\rm h}\,00^{\rm m}$, 
--20\degr\ and $13^{\rm h}\,15^{\rm m}$, --20\degr, respectively, both at 
a mean velocity of $\sim$1750\kms\ with quite narrow velocity dispersions. 

The first of the two groups has previously been identified as the NGC~4038 
Group in the Crater Cloud (Tully 1988, Tully \& Fisher 1987). Its brightest 
elliptical member (not detected in HIPASS) is NGC~3957 with $B_{\rm T} = 
13\fm03$ at a velocity of 1687\kms. It has over a dozen members in the BGC 
with \HI\ masses ranging between $1 - 10 \times 10^9$\Msun.

The second galaxy group has not before been recognized as such, probably 
because it does lie exactly on the SGP. Being hardly prominent in the optical 
it must have been viewed as an element of the SGP rather than a galaxy group. 
This group contains close to a dozen spirals detected in the BGC with similar 
\HI\ masses as for the NGC~4038 Group. Its brightest elliptical is the 
gas-rich galaxy NGC~5084 (HIPASS J1320--21), $B_{\rm T} = 11\fm55$, at a 
velocity of 1864\kms\ and we suggest that this new group be named after this 
galaxy, i.e. the NGC~5084 Group (see Table~\ref{tab:lss}).

\subsection{The highest velocity slice ($v > 2200$\kms)}
In the bottom panel, which shows galaxies with velocities above 2200\kms, the 
optical sample is more strongly populated than the 
BGC. Even so, the BGC seems to trace the structures in a smoother manner than 
the optical sample. This is primarily due to the fact that the main filament 
described below in further detail crosses the ZOA twice and hence is lost in 
optical surveys. Secondly, the filamentary structure in the optical sample 
is identified through a sequence of clumps, respectively of galaxy clusters, 
with few galaxies in between.

Again we have one dominant filament in this velocity slice which can be 
followed over a major fraction of the southern sky. It can be traced from the 
Indus and Pavo clusters ($\alpha,\delta$(J2000) = $21^{\rm h}\,00^{\rm m}$, 
--47\degr\ and $18^{\rm h}\,50^{\rm m}$, --63\degr, respectively) crossing 
the ZOA to the Centaurus cluster ($12^{\rm h}\,50^{\rm m}$, --40\degr) in a 
linear structure, called the Centaurus Wall by Fairall (1998), at a slight 
angle with respect to the SGP. From there it bends over to the Hydra 
($10^{\rm h}\,40^{\rm m}$, --25\degr) and Antlia ($10^{\rm h}\,30^{\rm m}$, 
--35\degr) clusters and folds back across the ZOA through Puppis where yet 
another spiral-rich new galaxy group is uncovered.

This galaxy group at $7^{\rm h}\,50^{\rm m}$, --35\degr, named here Puppis~G2 
lies at $\sim$2800\kms\ and contains only newly identified galaxies (HIZSS 
plus new BGC galaxies). However, this agglomeration might point to a fairly 
massive group because at this velocity distance only the most \HI-massive 
galaxies are revealed in the BGC (they indeed have \HI\ masses between $3 - 
10 \times 10^9$\Msun) and with an optical extinction $A_B$ of over 4~mag at 
this location, even the deep optical ZOA surveys are unlikely to uncover many 
of the possible early-type members of this group. This whole filament seems
to encircle the Eridanus Void, a large void that extends out to about 
5000\kms\ (Fairall 1998).

It should be stressed again that the optical galaxy comparison sample probes 
the nearby Universe to much higher velocities which results in the fact that 
we may recognize high velocity features in the optical velocity slice to which 
the BGC is not sensitive anymore. This is quite notable for the Great Attractor
region (centered on $\alpha,\delta$(J2000) = $15^{\rm h}\,00^{\rm m}$, 
--60\degr, 4500\kms) where the optical sample does show a preponderance of 
galaxies in the general direction of the Great Attractor overdensity --- with 
a huge gap of its central part though, due to the ZOA. As stated above, the 
Great Attractor is not being traced by the BGC.

\section{Conclusions}
\label{sec:conclusions}
The HIPASS Bright Galaxy Catalog is a subset of HIPASS and contains the 1000 
\HI-brightest sources in the southern sky ($\delta < 0\degr$) based on their 
peak flux density (\Speak\ $\ga$ 116 mJy) as measured from the spatially
integrated \HI\ spectrum. We obtained previously catalogued optical (or 
infrared) counterparts for 909 of the 1000 BGC sources. The HIPASS BGC is the 
second largest sample of galaxies from a blind \HI\ survey to date, covering 
a volume of $\sim2 \times 10^6$ Mpc$^3$, only surpassed by the HIPASS Catalog 
itself. While the BGC is complete in peak flux density, only a subset of 
$\sim$500 BGC members can be considered complete in integrated \HI\ flux 
density (\FHI\ $\ga 25$ Jy\kms). 

The lowest detectable \HI\ flux density for relatively narrow line-width 
sources in the BGC (galaxies as well as \HI\ clouds) is given by \FHI\ = 
$0.7 \times$ \Speak\ $\times$ \wfi\ (see Fig.~\ref{fig:w50fhi}) corresponding 
to an \HI\ mass of $2 \times 10^4$ \wfi\ $D^2$ (see Fig.~\ref{fig:logdmhi}). 
This means \HI\ sources with \wfi\ = 50\kms\ and \MHI\ $\ga 10^8$ (10$^9$)
\Msun\ can be detected out to 10 (30) Mpc. 

Of the 91 newly catalogued BGC sources, 87 appear to be galaxies. A large 
fraction of these (37) lie at Galactic latitudes $|b| < 5\degr$; all but 
five were previously uncovered by the \HI\ Zone of Avoidance shallow survey 
(HIZSS; Henning et al. 2000). For all but one of the 30 {\em new} galaxies 
at $|b| > 10\degr$ optical counterparts were discovered in the Digitized Sky
Surveys (see Ryan-Weber et al. 2002). For a further 51 galaxies in the BGC no 
redshifts were previously known. In addition, we identify 16 galaxies (i.e., 
$\sim$2\% of the BGC) where the cataloged optical velocity is potentially 
incorrect, resulting in a total of 158 new redshifts. 

There are four \HI\ clouds in the BGC, three of which are most likely 
Magellanic debris (HIPASS J1712--64, J1718--59, and J1616--55) with systemic 
velocities in the range 390--460\kms. Further, detailed investigations are
under way. We discovered one intergalactic \HI\ cloud, HIPASS J0731--69, which 
is associated with the galaxy NGC~2442 (Ryder et al. 2001). The cloud, which 
lies at a projected distance of 38\arcmin\ (176 kpc) from NGC~2442 ($D$ = 
15.9 Mpc) has an \HI\ mass of nearly 10$^9$\Msun. We could have detected 
similar \HI\ clouds up to a distance of 30 Mpc, with a separation of at least 
20\arcmin\ from an associated \HI-rich galaxy. This detection suggests that 
their space density is about 1/1000th that of galaxies with the same \MHI\ 
(see also Briggs 1990). Other known tidal \HI\ clouds such as those near 
NGC~3263 (HIPASS J1029--44b) and NGC~7582 (HIPASS J2318--42), see 
Appendix~\ref{app:hiwidths}, are contained in the BGC as part of their 
associated galaxy or galaxy group.

Therefore, no definite free-floating \HI\ clouds without stars have been 
discovered in the search volume. Following Fisher \& Tully (1981b) we derive an
\HI\ density of $\rho_{\rm HIclouds} < 5.4 \times 10^{5}$\Msun\,Mpc$^{-3}$ or 
$<3.7 \times 10^{-35}$\,g\,cm$^{-3}$ for \HI\ clouds with \wfi\ = 100\kms\ and 
\MHI\ = $10^7 - 10^{10}$\Msun. We compare the limit for $\rho_{\rm HIclouds}$ 
with the critical density for the closure of the Universe, $\rho_{\rm c}$, the 
baryonic mass density, $\rho_{\rm b}$, and the \HI\ mass density of galaxies, 
$\rho_{\rm HIG}$, (see Zwaan et al. 2003):
% NOTE. rho_HI = 6.1 x 10^7 Msun/Mpc^3 = 4.1 x 10^-33 g/cm^3
$\rho_{HIclouds}$/$\rho_{\rm c} < 3.5 \times 10^{-6}$,
$\rho_{HIclouds}$/$\rho_{\rm b} < 9.9 \times 10^{-5}$, 
$\rho_{HIclouds}$/$\rho_{\rm HIG} < 8.9 \times 10^{-3}$. 

Since radio wavelengths suffer little absorption in the Galactic Plane, HIPASS
provides a clear view of the local large-scale structure. The dominant 
features in the sky distribution of the BGC are the Supergalactic Plane and 
the Local Void. In addition, one can clearly see the Centaurus Wall which 
connects via the Hydra and Antlia clusters to the Puppis filament. Some 
previously hardly noticed galaxy groups stand out quite distinctively in 
the \HI\ sky distribution. Several new structures can be mapped for the 
first time, not only behind the Milky Way.

\section*{Acknowledgments}
\begin{itemize}
\item We are grateful to the staff at the ATNF Parkes and Narrabri 
      observatories for assistance with HIPASS and follow-up observations. 
\item We thank Jay Ekers for her help with the HIPASS cube making and 
      galaxy finding.
\item This research has made use of the NASA / IPAC Extragalactic Database 
      (NED) which is operated by the Jet Propulsion Laboratory, California 
      Institute of Technology, under contract with the National Aeronautics 
      and Space Administration. We also acknowledge extensive use of the 
      Lyon/Meudon Extragalactic Database (LEDA;
      see {\tt http://leda.univ-lyon1.fr}). 
\item Digitized Sky Survey (DSS) material (UKST / ROE / AAO / STScI) is
      acknowledged.
\end{itemize}

\appendix
\twocolumn

\section{Galaxies with discrepant velocities}
\label{app:wrong}
The HIPASS BGC contains 16 galaxies where the measured \HI\ velocity disagrees
with the first-listed velocity of the likely optical counterpart given in NED 
(at the time of the optical identification, see Section~\ref{sec:optids}). In 
Table~\ref{tab:wrong} we list the systemic velocities obtained from HIPASS 
(Col.~2) as well as previous optical and \HI\ velocity measurements collected
from the literature. At the discrepant velocities no \HI\ emission was detected
in HIPASS unless stated otherwise.

ATCA \HI\ follow-up observations have confirmed the optical positions and 
\HI\ velocities obtained for HIPASS J0312--21 (ESO547-G019) and HIPASS 
J1051--34 (ESO376-G022, see Fig~\ref{fig:eso376-g022}). 
In those cases where the HIPASS and optical velocities cannot 
be reconciled (e.g. ESO172-G004, ESO173-G016) we conclude that either the 
published optical velocity is in error or, in a few cases, the field may 
contain two galaxies: e.g., one uncataloged \HI-rich LSB galaxy and one 
cataloged \HI-poor galaxy. In two cases (ESO221-G006, ESO138-G009) previously 
published \HI\ velocities are substantially offset from the HIPASS velocity.
For the galaxy "pair" ESO252-IG001 ATCA \HI\ observations have confirmed that
HIPASS J0457--42 is the western, more face-on galaxy (PA = 45\degr). The 
optical velocities (see Table~\ref{tab:wrong}) probably refer to the more 
distant, nearly edge-on, eastern galaxy, which was also detected in HIPASS 
and contains a small amount of \HI\ gas at velocities of 2800--3000\kms. 
Further ATCA \HI\ follow-up observations of BGC sources are under way to 
provide the correct identifications.

\section{Individual Galaxies in the BGC}

\subsection{Extended galaxies}
\label{app:extended}
Here we give a brief description of all extended sources in the BGC 
(see Table~\ref{tab:ext}):
{\scriptsize
\begin{itemize}
\item The Sculptor Group is so close that the largest of its members are 
  resolved with the Parkes telescope: NGC~55 (HIPASS J0015--39), NGC~247
  (HIPASS J0047--20), NGC~253 (HIPASS J0047--25), NGC~300 (HIPASS J0054--37) 
  and NGC~7793 (HIPASS J2357--32). The integrated \HI\ flux densities 
  obtained from HIPASS are generally consistent with the single dish \HI\ 
  flux densities given in the literature (see Huchtmeier \& Richter 1989), 
  although for NGC~55 and NGC~253 they vary by a factor of two, and are 
  always higher (as expected) than previous VLA measurements (see Puche \& 
  Carignan 1991, and references therein). The blue-shifted velocities of 
  some Sculptor Group galaxies are confused by Galactic \HI\ emission and 
  \HI\ clouds in the Magellanic Stream (see also Putman et al. 2003), which 
  slightly increases the measurement uncertainties. For the starburst galaxy 
  NGC~253 we measure an integrated \HI\ flux density of $693 \pm 42$ Jy\kms; 
  see Koribalski, Whiteoak \& Houghton (1995) for a detailed study.
\item NGC~1291 (HIPASS J0317--41) is a large, nearly face-on, early-type 
  spiral galaxy of type SBa. Mebold et al. (1979) and Reif et al. (1982) 
  measured an integrated \HI\ flux density of $72.5 \pm 5.9$ Jy\kms, slightly 
  lower than our measurement of \FHI\ = $90 \pm 12$ Jy\kms. Using VLA \HI\ 
  data van Driel, Rots \& van Woerden (1988) show that the \HI\ gas is 
  concentrated in an annulus (diameter $\ga$10\arcmin), extending well 
  beyond the optical bar. 
\item NGC~1313 (HIPASS J0317--66) is a late-type barred spiral galaxy which 
  has been studied in detail by Ryder et al. (1995) who find a large \HI\ disk 
  extending roughly $18\arcmin \times 10\arcmin$ (24 kpc $\times$ 13 kpc). We 
  measure an integrated \HI\ flux density of $463 \pm 33$ Jy\kms\ in agreement 
  with \FHI\ = 484 Jy\kms\ obtained by Ryder et al. (1995).
% NOTE: Ryder et al. have combined ATCA plus Parkes narrow-band data
\item HIPASS J0403--43 is an interacting galaxy pair consisting of the 
  lenticular galaxy NGC~1512 and the elliptical galaxy NGC~1510 which are
  separated by 5\farcm0. We measure an integrated \HI\ flux density of 
  $259 \pm 17$ Jy\kms\ and a deconvolved \HI\ diameter of $17\farcm4 
  \times 10\farcm7$ (48 kpc $\times$ 30 kpc). The \HI\ emission is centered 
  on NGC~1512. Hawarden et al. (1979) measure \FHI\ = $232 \pm 20$ Jy\kms\ 
  (same as Reif et al. 1982) and a deconvolved \HI\ diameter of $17\arcmin 
  \times 15\arcmin$ (47 kpc $\times$ 41 kpc).
\item NGC~1533 (HIPASS J0409--56) is a nearby, early-type galaxy with two 
  small companions, IC\,2038/9. We measure an integrated \HI\ flux density of 
  $68 \pm 8$ Jy\kms, slightly smaller than \FHI\ = $87.5 \pm 6.5$ Jy\kms\ by 
  Reif et al. (1982). Recent ATCA \HI\ observations obtained by Ryan-Weber, 
  Webster \& Staveley-Smith (2003) reveal an asymmetric \HI\ ring surrounding 
  the optical galaxy with an extent of $\sim$15\arcmin. Only a small amount 
  of \HI\ gas is associated with the galaxy IC\,2038, none with IC\,2039. 
\item NGC~1532 (HIPASS J0411--32) is a large spiral galaxy interacting with 
  its much smaller, elliptical companion NGC~1531. VLA \HI\ measurements of 
  NGC~1531/2 by Sandage \& Fomalont (1993) reveal a total \HI\ extent of 
  $\sim$20\arcmin\ ($\sim$70 kpc) and an integrated \HI\ flux density of 
  $228 \pm 15$ Jy\kms, similar to our value of $249 \pm 15$ Jy\kms. For a 
  study of the CO(1--0) and CO(2--1) content of NGC~1532 see Horellou et al. 
  (1995).
\item NGC~2915 (HIPASS J0926--76) is a weak blue compact dwarf galaxy with 
  a very extended \HI\
  disk revealing a short central bar and extended spiral arms extending well
  beyond the stellar distribution. Using the AT Compact Array Meurer et al. 
  (1996) obtained an integrated \HI\ flux density of 145 Jy\kms\ and an \HI\ 
  diameter of nearly 20\arcmin, which corresponds to five times its Holmberg 
  radius. Our measurement of the integrated flux density, \FHI\ = $108 \pm 
  14$ Jy\kms, is significantly lower, and a Gaussian fit to the \HI\ 
  distribution from HIPASS gives a diameter of only $\sim$9\arcmin\ (7 kpc). 
  Both are underestimates due to strong negative sidelobes. 
% NOTE. Meurer et al. use D = 5.3+-1.6 Mpc, Ho = 50, stellar measurement
\item NGC~3109 (DDO\,236, HIPASS J1003--26A) is a Magellanic-type spiral 
  galaxy. With a distance
  of 1.25 Mpc (Mateo 1998) it lies at the outskirts of the Local Group. We 
  measure an \HI\ diameter of $23\arcmin \times 6\arcmin$ (8 kpc $\times$ 2 
  kpc) and an integrated \HI\ flux density of \FHI\ = $1148 \pm 97$ Jy\kms. 
  The resulting \HI\ mass of $4.2 \times 10^8$\Msun\ agrees well with that 
  by Barnes \& de Blok (2001). Using the 21-cm multibeam system on the Parkes 
  telescope and narrow-band filters Barnes \& de Blok (2001) acquired a deep 
  \HI\ image of NGC~3109 and its neighbour, the Antlia dwarf galaxy. They 
  measure an \HI\ extent of $\sim$18\arcmin\ for NGC~3109. 
\item NGC~3621 (HIPASS J1118--32) is a nearby spiral galaxy with an optical 
  diameter of about
  $12\arcmin \times 7\arcmin$. Our integrated \HI\ flux density measurement of 
  \FHI\ = $884 \pm 56$ Jy\kms\ compares well with the previous estimate of 
  846 Jy\kms\ (see Huchtmeier \& Richter 1989). We determine an \HI\ diameter 
  of $19\arcmin \times 4\arcmin$ (34 kpc $\times$ 7 kpc). A deep ATCA \HI\ 
  image obtained by Ryan-Weber (priv. comm.) shows a total \HI\ extent of 
  $32\arcmin \times 11\arcmin$ (58 kpc $\times$ 20 kpc). The cepheid distance 
  of NGC~3621 is 6.6 Mpc (Willick \& Batra 2001), compared to 6.2 Mpc adopted 
  here.
\item NGC~4945 (HIPASS J1305--49) is an edge-on starburst galaxy with an \HI\ 
  extent of $\sim$20\arcmin\ ($\sim$25 kpc). We measure an integrated \HI\ 
  flux density of \FHI\ = $319 \pm 21$ Jy\kms. The ATCA measurement of \FHI\ 
  = 70 Jy\kms\ by Ott et al. (2001) is severely affected by \HI\ absorption. 
  Mathewson \& Ford (1996) measure \FHI\ = 269 Jy\kms.
\item For a detailed discussion of NGC~5128 (Centaurus\,A, HIPASS J1324--42) 
  see Section~\ref{sec:hiabs}. 
\item NGC~5236 (M\,83, HIPASS J0317--41) is a nearly face-on, gas-rich spiral 
  galaxy located within a loose group of galaxies (see Fig.~\ref{fig:m83}). 
  We measure an integrated \HI\ flux density of $1630 \pm 96$ Jy\kms\ and a 
  deconvolved Gaussian \HI\ diameter of $31\farcm2 \times 21\farcm4$ (40 kpc 
  $\times$ 28 kpc). Huchtmeier \& Bohnenstengel (1981) measure a maximum \HI\ 
  extent of $95\arcmin \times 76\arcmin$. Park et al. (2001) obtained a large 
  \HI\ mosaic with the ATCA and Parkes showing an extent of at least 60\arcmin.
  The dwarf companions NGC~5253 (HIPASS J1339--31A) and NGC~5264 (HIPASS 
  J1341--29) lie at projected distances of 113\arcmin\ and 60\arcmin\ from 
  NGC~5236, respectively (see also Banks et al. 1999 and Koribalski 2002).
\item The Circinus galaxy (HIPASS J1413--65) is a large bright galaxy at low 
  Galactic latitude ($b$ = --4\degr); for a detailed study see Freeman et al. 
  (1977) and Jones et al. (1999). We find the \HI\ diameter to be about twice 
  its Holmberg diameter. Freeman et al. (1977) and Henning et al. (2000) find 
  integrated \HI\ flux densities of 1907 and 1867 Jy\kms, respectively, 
  significantly larger than our value of \FHI\ = $1451 \pm 98$ Jy\kms\ which 
  is an underestimate due to strong negative sidelobes.
\item HIPASS J1532--56 (HIZSS\,097, HIZOA J1532--56) was discovered by 
  Staveley-Smith et al. (1998) as one of the most extended ($\le$200 kpc)
  sources in the \HI\ survey of the Zone of Avoidance (see also Henning et 
  al. 2000, Juraszek et al. 2000). ATCA \HI\ observations by Staveley-Smith 
  et al. (1998) reveal an irregular gas distribution and complex velocity 
  field suggesting this source is an interacting galaxy group, similar to 
  the M\,81/M\,82/NGC~3077 group (Yun, Ho \& Lo 1994). At a location close 
  to zero degrees latitude the Galactic extinction is so large that no 
  optical or infrared identifications have been possible. Our measurement 
  of the integrated \HI\ flux density of J1532--56 (\FHI\ = $64 \pm 15$ 
  Jy\kms) is up to a factor of two higher than previous values. 
% NOTE. FHI = 64.2 (HIPASS), 34.8 (SJK), 28.2 (HIZOA), 58.5 (HIZSS)
%       possibly due to baseline uncertainties and the large \HI\ diameter.
\item HIPASS J1616--55 (HIZSS\,102, HIZOA J1616--55) was also discovered by
  Staveley-Smith et al. (1998) who measure a total \HI\ extent of 80\arcmin. 
  ATCA \HI\ observations reveal numerous clumps, but no optical counterpart
  has been found. While this could be a pair of interacting low-surface 
  brightness galaxies, as suggested by Staveley-Smith et al., we believe it 
  is one of several extended \HI\ clouds representing tidal debris related 
  to the Magellanic Clouds and the Leading Arm (Putman et al. 1998). Our 
  measurement of the integrated \HI\ flux density of J1616--55 (\FHI\ = 
  $24 \pm 7$ Jy\kms) is similar to previous values.
% NOTE. FHI = 23.7 (HIPASS), 27.6 (SJK), 31.4 (HIZOA), 30.3 (HIZSS)
\item HIPASS J1718--59 was discovered by Koribalski (2001) and has an \HI\ 
  extent of $\sim$3\degr. No optical counterpart has been identified and we 
  believe this to be another extended \HI\ cloud. It lies only a few degrees 
  away from HIPASS J1712--64 and at a similar velocity to HIPASS J1616--55.
  The \HI\ properties of the clouds are also listed in 
  Table~\ref{tab:hiclouds}.
\item NGC~6744 (HIPASS J1909--63a) is one of the largest galaxies in the 
  southern sky. Using ATCA data Ryder, Walsh \& Malin (1999) measure a 
  total \HI\ extent of $\sim$30\arcmin, about 1.5 times the size of the 
  stellar disk. The two late-type dwarf companions, NGC~6744A (located 
  within the \HI\ disk of NGC~6744) and ESO104-G044 (HIPASS J1911--64, at 
  a projected distance of 24\farcm2 from NGC~6744) also contain \HI\ gas. 
  We derive an \HI\ mass of $2.2 \times 10^{10}$\Msun, making NGC~6744 
  one of the most massive galaxies within a distance of $\sim$25 Mpc. 
% NOTE: M101 (vsys = 240 km/s) is a very extended and massive galaxy
%       MHI = 2.4 10^10 Msun (Huchtmeier & Witzel 1979) assuming D=7.2 Mpc
\item HIPASS J1943--06 is a nearby, possibly interacting galaxy pair (\FHI\ 
  = $45 \pm 7$ Jy\kms). It consists of the two galaxies, MCG-01-50-001 and 
  NGC~6821 (MCG-01-50-002), which have nearly identical systemic velocities 
  and are separated by 18\farcm9. The HIPASS data show a common \HI\ 
  envelope, but interferometric \HI\ observations are needed to study this 
  gas-rich galaxy pair in detail. By fitting two Gaussians to the integrated 
  \HI\ distribution we obtain \FHI\ = 20.4 and 24.2 Jy\kms\ for MCG-01-50-001 
  and NGC~6821, respectively.
% NOTE. Imfit with two Gaussians gives two point sources with:
% (1) 19:43:16.1, -06:57:09.5  FHI=20.4  size=916.8 x 899.7, -0.55
%   = 19:43:12.2, -06:56:25  = MCG-01-50-001 
%     vHI = 1498+-3 km/s, size=1.2x0.3, type=IBm?sp, FHI=19.5+-1.1 Jy km/s
% (2) 19:44:17.3, -06:49:15.5  FHI=24.2 size=1060.1 x 924.1, 60.94
%   = 19:44:24.2, -06:50:02  = NGC~6821 (MCG-01-50-002)
%     vHI = 1525+-8 km/s, size=1.2x1.0, type=SB(s)d, FHI=19.6+-??? Jy km/s
\item NGC~6822 (HIPASS J1944--14) is, with a distance of 490 kpc (Mateo 1998), 
  the most nearby dwarf irregular galaxy beyond the satellites of the Milky 
  Way. Our measurement of \FHI\ = $2525 \pm 250$ Jy\kms\ is slightly 
  contaminated by Galactic emission. Using extensive ATCA observations de 
  Blok \& Walter (2000) measure \FHI\ = $2200 \pm 100$ Jy\kms\ and a total 
  \HI\ extent of 40\arcmin.
\item HIPASS J2318--42 is a compact interacting galaxy group consisting of 
  the gas-rich spiral galaxies NGC~7582, NGC~7590 and NGC~7599. These form, 
  together with NGC~7552 (= HIPASS J2316--42), the Grus Quartet. All four 
  galaxies have large amounts of \HI\ gas. ATCA \HI\ observations show tidal 
  tails emanating from NGC~7852 towards NGC~7890/9 and NGC~7552 (see 
  Koribalski 1996).
\end{itemize}
}
\subsection{Some Local Group galaxies} 
\label{app:lgroup}
Here we describe some of the Local Group galaxies detected in the BGC 
(NGC~3109 and NGC~6822 have already been described in the previous Section).

{\scriptsize
\begin{itemize}
\item The Wolf-Lundmark Melotte (DDO\,221, HIPASS J0001--15) is a Local Group 
  dwarf irregular galaxy at a distance of $\sim$900 kpc (Mateo 1998, Dolphin 
  2000, Rejkuba et al. 2000). Despite its large optical diameter of $12\arcmin 
  \times 4\arcmin$ it appears unresolved in HIPASS. However, using Effelsberg 
  \HI\ data Huchtmeier, Seiradakis \& Materne (1981) find a total \HI\ extent 
  of 45\arcmin\ (12 kpc) and an \HI\ mass of $5.3 \times 10^7$\Msun. We 
  obtained a lower value of \MHI\ = $4.7 \times 10^7$\Msun. Note that the WLM 
  is, in projection, located just north-east of the Magellanic Stream.
\item Sextans\,A (HIPASS J1010--04) is another Local Group dwarf irregular 
  galaxy at a distance 
  of 1.44 Mpc (Mateo 1998). Huchtmeier, Seiradakis, \& Materne (1981) reported 
  an \HI\ diameter of 54\arcmin\ ($\sim$23 kpc), 5.8 times the Holmberg 
  diameter, and an \HI\ mass of $1.3 \times 10^8$\Msun. We derive a lower 
  \HI\ mass of $8.2 \times 10^7$\Msun. Wilcots \& Hunter (2002) obtained an 
  \HI\ mosaic of Sextans\,A using the VLA and measured an \HI\ diameter of 
  only 18\arcmin\ ($\sim$8 kpc) and \MHI\ = $8 \times 10^7$\Msun\ (see also 
  Skillman et al. 1988). Sextans\,A appears unresolved in HIPASS. ---
  The linear separation between Sextans\,A and NGC~3109 (see above) is only
  $\sim$500 kpc. Together with Antlia and Sextans\,B this small group may 
  be the closest external clustering of galaxies (van den Bergh 1999).
\item IC\,5152 (HIPASS J2202--51) is a dwarf irregular galaxy at a distance 
  of 1.59 Mpc (Mateo 1998) with an optical diameter of $\sim$5\arcmin. Our 
  value of \FHI\ = $97 \pm 10$ Jy\kms\ agrees with that by Huchtmeier \& 
  Richter (1986) and results in an \HI\ mass of $5.8 \times 10^7$\Msun. 
  Zijlstra \& Minniti (1999) determine a similar distance of 1.7 Mpc.
\item HIPASS J0700--04 (HIZSS\,003, Henning et al. 2000) is the closest newly 
  cataloged galaxy in the BGC with \vLG\ = 115\kms. Recent observations by 
  Massey, Henning \& Kraan-Korteweg (2003) show it to be a regularly rotating, 
  dwarf irregular galaxy with an \HI\ extent of 6\arcmin.
% NOTE: HIZSS\,003 is also known as Dw217.8+0.0.
\end{itemize}
}

\subsection{Galaxies with the highest HI mass}
\label{app:himassive}
The BGC sources with the highest \HI\ masses are ESO390-G004 and NGC~5291,
followed by the galaxy pair NGC~6935/7 and the barred spiral galaxy 
ESO136-G016.

{\scriptsize
\begin{itemize}
\item ESO390-G004 (HIPASS J1620--36) lies close to the Zone of Avoidance 
  ($b$ = 9\fdg8, \AB\ = 3.1 mag); no optical velocity is available. Theureau 
  et al. (1998) measure a systemic velocity of $4448 \pm 6$\kms, in agreement 
  with our measurement, but an integrated \HI\ flux density of only $23.7 \pm 
  2.0$ Jy\kms\ using the Nan\c{c}ay telescope. Although the HIPASS spectrum 
  is affected by some baseline ripple, our measurement of \FHI\ = $47 \pm 6$ 
  Jy\kms\ is reliable. Optically the galaxy shows a bright core plus some 
  extended emission. We estimate an \HI\ mass of $3.7 \times 10^{10}$\Msun. 
  Preliminary ATCA \HI\ snapshot observations show that there is at least one 
  uncataloged companion, $\sim$2\arcmin\ SE of ESO390-G004, which contributes 
  to the \FHI\ measurement. The \FHI\ discrepancy between the Parkes and 
  Nan\c{c}ay measurements may be explained by the different beam sizes of 
  the telescopes. 
\item The peculiar system NGC~5291 (HIPASS J1347--30), located in the western 
  outskirts of the cluster Abell~3574, contains the lenticular galaxy NGC~5291 
  and a close companion, the so-called `Seashell' galaxy, as well as numerous 
  bright knots (see e.g. Duc \& Mirabel 1998). One of these knots is classified
  as the \HII\ galaxy CTS1032 (\vopt\ = $4350 \pm 99$\kms, Pena, Ruiz \& Maza 
  1991). Malphrus et al. (1997) find an unusually high amount of \HI\ gas 
  distributed along a fragmented ring of diameter 9\farcm9 (NS) $\times$
  5\farcm6 (EW), covering NGC~5291, the `Seashell' galaxy, CTS1032, and at 
  least another 10 knots. The 20\% velocity width of the NGC~5291 system is 
  the largest measured in the BGC (\wfi\ = 637\kms, \wtw\ = 757\kms). Longmore 
  et al. (1979) measured a single dish \HI\ mass of $5 \times 10^{10}$\Msun, 
  assuming $D$ = 58 Mpc. Our measurement of \MHI\ = $3.6 \times 10^{10}$\Msun\ 
  ($D$ = 56 Mpc) is similar.
\item HIPASS J2038--52 corresponds to the galaxy pair NGC~6935/7 (separation
  4\farcm5). ATCA \HI\ snapshot observations show another small galaxy possibly
  contributing to the \HI\ emission. This new galaxy lies $\sim$6\arcmin\ to
  the NNW of the early-type galaxy NGC~6935. The \HI\ emission appears to be
  dominated by the spiral galaxy NGC~6937 for which Huchtmeier \& Richter
  (1989) quote an integrated flux density of $31.8 \pm 4.1$ Jy\kms, close to
  our measurement of 33.1 Jy\kms. The \HI\ mass of the system is $3.1 \times
  10^{10}$\Msun.
\item ESO136-G016 (HIPASS J1603--60) is a distant edge-on, barred spiral 
  galaxy. At a latitude of $b$ = --6\fdg3 and \AB\ = 1.4 mag, it appears very 
  faint in the DSS\,1, but is clearly visible on red DSS\,2 images. Its 
  optical diameter is about 3\farcm5, no optical velocity is available. Our 
  measurement of \FHI\ = $25.5 \pm 3.7$ Jy\kms\ agrees well with the previous 
  estimate of $27.40 \pm 4.1$ Jy\kms\ (Huchtmeier \& Richter 1989). We derive 
  an \HI\ mass of $3.0 \times 10^{10}$\Msun. ATCA \HI\ observations are needed 
  to study the neighbourhood and gas kinematics in detail. The red DSS\,2 
  image shows a potential companion a few arcminutes to the north-west.
\end{itemize}
}

\subsection{Galaxies with the largest HI velocity widths}
\label{app:hiwidths}
The BGC sources with the largest measured 50\% velocity widths are
NGC~5084, NGC~5291 (see above) and NGC~2613.

{\scriptsize
\begin{itemize}
\item NGC~5084 (HIPASS J1320--21) is a massive and unusual lenticular galaxy 
  (see Gottesman \& Hawarden 1986). It has a radius of 8\farcm2 ($\sim$50 kpc) 
  and is close to 
  edge-on ($i >$ 86\degr). We measure \FHI\ = $101.5 \pm 6.7$ Jy\kms, \wfi\ 
  = 645\kms, and \wtw\ = 668\kms. Using \vrot\ = \wtw\ / 2 we estimate 
  a total dynamical mass of \Mtot\ = $1.3 \times 10^{12}$\Msun, and a 
  mass-to-light ratio $\ga$65. At a slightly higher velocity and a projected 
  distance of 15\arcmin\ lies the companion galaxy ESO576-G040 (HIPASS 
  J1320--22), also a member of the BGC. The total dynamical mass is derived 
  using \Mtot\ [\MMsun] = $2.31 \times 10^5 r_{\rm kpc} v_{\rm rot}^2$, 
  where \vrot\ is the inclination corrected rotation velocity of the galaxy 
  in \kkms\ and $r_{\rm kpc}$ is the galaxy radius.
% NOTE. The optical radius is slightly larger than the HI radius (see GH86)
%       D = 1551/75 = 20.68 Mpc
\item NGC~2613 (HIPASS J0833--22) is a bright Sb galaxy with an optical radius 
  of 3\farcm6 and an inclination of 76\degr. We measure an integrated \HI\ 
  flux density of $59.4 \pm 4.6$ Jy\kms, \wfi\ = 602\kms, and \wtw\ = 620\kms. 
  Using WSRT data, Bottema (1989) 
  measured a much lower \HI\ flux of 29.4 Jy\kms\ and derived a rotation 
  velocity of 315\kms. Chaves \& Irwin (2001) used the VLA to image NGC~2613 
  and its companion, ESO495-G017, and measured integrated \HI\ flux densities 
  of 55.2 Jy\kms\ and 1.6 Jy\kms, respectively. Using their \HI\ radius of 
  $\sim$5\arcmin\ ($\sim$27 kpc) we derive a total dynamical mass of \Mtot\ 
  $\sim 6 \times 10^{11}$\Msun\ for NGC~2613. The small companion galaxy, 
  ESO495-G017, lies at a projected distance of 7\farcm3.
\end{itemize}
}

The largest 20\% velocity widths were measured in NGC~5291 and NGC~5084,
followed by NGC~5183/4, ESO320-G026 and NGC~3263.

{\scriptsize
\begin{itemize}
\item The spiral galaxies NGC~5183/4 (HIPASS J1330--01; \FHI\ = $29.2 \pm 3.0$ 
  Jy\kms) are separated by only 3\farcm7. We measure velocity widths of \wfi\ 
  = 346\kms\ and \wtw\ = 666\kms. Huchtmeier \& Richter (1989) give integrated 
  \HI\ flux densities of 17.3 and 10.9 Jy\kms, respectively. The individual 
  galaxies have velocity widths of only $\sim$360\kms\ (de Vaucouleurs et al. 
  1991).
\item The Sb galaxy ESO320-G026 (HIPASS J1149--38) is confused by the smaller 
  spiral ESO320-G024, which lies at a projected distance of 6\farcm2. We 
  measure \FHI\ = $37.2 \pm 4.3$ Jy\kms, \wfi\ = 329\kms, and \wtw\ = 641\kms.
  Neither galaxy has previous \HI\ measurements.
\item NGC~3263 (HIPASS J1029--44b) is a peculiar galaxy and part of a group 
  of \HI\ rich galaxies (Koribalski et al. 2004, in prep.). Best known in 
  this loose group is the galaxy merger NGC~3256 (HIPASS J1027--43) which 
  exhibits a spectacular set of symmetric tidal arms both in the optical and 
  in \HI. A large \HI\ gas cloud to the west of NGC~3263 (English, Koribalski
  \& Freeman 2003) as well as the galaxy NGC~3261 (HIPASS J1029--44a) also 
  contribute to the integrated \HI\ flux density. We measure \FHI\ = $79.1 
  \pm 5.6$ Jy\kms, \wfi\ = 423\kms, and \wtw\ = 630\kms.
\end{itemize}
}

None of these galaxies compares with the S0/Sa galaxy UGC~12591 which appears
to have one of the largest rotational velocity (\wfi\ = 957\kms) of any known 
disk system (Giovanelli et al. 1986). We note that LEDA lists velocity widths 
larger than 700\kms\ (800\kms) for 17 (6) galaxy systems.

\newpage
\clearpage

% Figures 

\begin{figure*}[h]  % Figure 1
 % \centering{\psfig{figure=Koribalski.fig1.panel01.eps,width=16cm}}
\caption{\label{fig:hispectra}
  \HI\ spectra of all 1000 members of the HIPASS Bright Galaxy Catalog (BGC).
  The HIPASS name together with the most likely optical identification (in
  brackets) is given at the top of each spectrum. Marked are also the measured
  \HI\ peak flux density as well as the 50\% and 20\% velocity widths. The 
  vertical lines indicate the velocity range used for the first order baseline 
  fit (outside) and the \HI\ line emission analysis (inside).  --- The full
  version of this figure is in the electronic edition of the Journal.}
\end{figure*}

\setcounter{figure}{0}

\begin{figure*}[h]  % Figure 1 continued
 % \centering{\psfig{figure=Koribalski.fig1.panel02.eps,width=16cm}}
\caption{ continued.}
\end{figure*}

\newpage
\clearpage

\begin{figure} % Figure 2
 \centering{\psfig{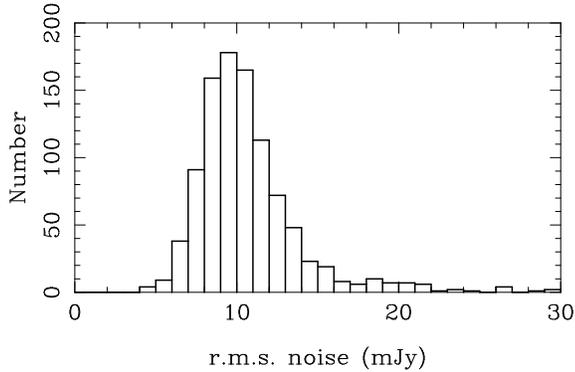}}
\caption{\label{fig:rmshist}
  Histogram of the clipped r.m.s. of the HIPASS BGC spectra, as determined
  from the {\sc mbspect} fit of the spatially and spectrally integrated \HI\ 
  spectra. Not displayed here are 24 galaxies with r.m.s.  $>$30 mJy; these 
  are mostly extended sources and some galaxies in the Zone of Avoidance.} 
\end{figure}

\begin{figure}  % Figure 3
 \centering{\psfig{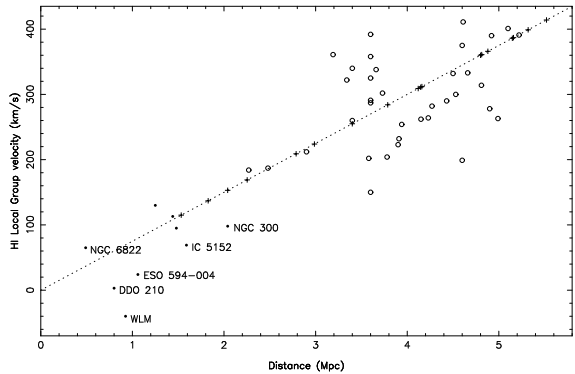}}
\caption{\label{fig:lv}
 Distances versus \HI\ Local Group velocities for the closest galaxies in 
 the BGC. Open and filled circles indicate galaxies for which independent 
 distances are available in the literature (see Section~\ref{sec:distances}), 
 while the "+" symbols denote the Hubble distance ($D$ = \vLG\ / \Ho) for 
 those galaxies for which no independent distances are currently available. 
 The filled circles show those nine galaxies for which the independent 
 distances were adopted in the BGC. The largest deviations occur where one 
 distance is given for members of a galaxy group (e.g., part of the Cen\,A 
 group at 3.6 Mpc, see Karachentsev et al. 2003). }
\end{figure}

\begin{figure}  % Figure 4
 \centering{\psfig{figure=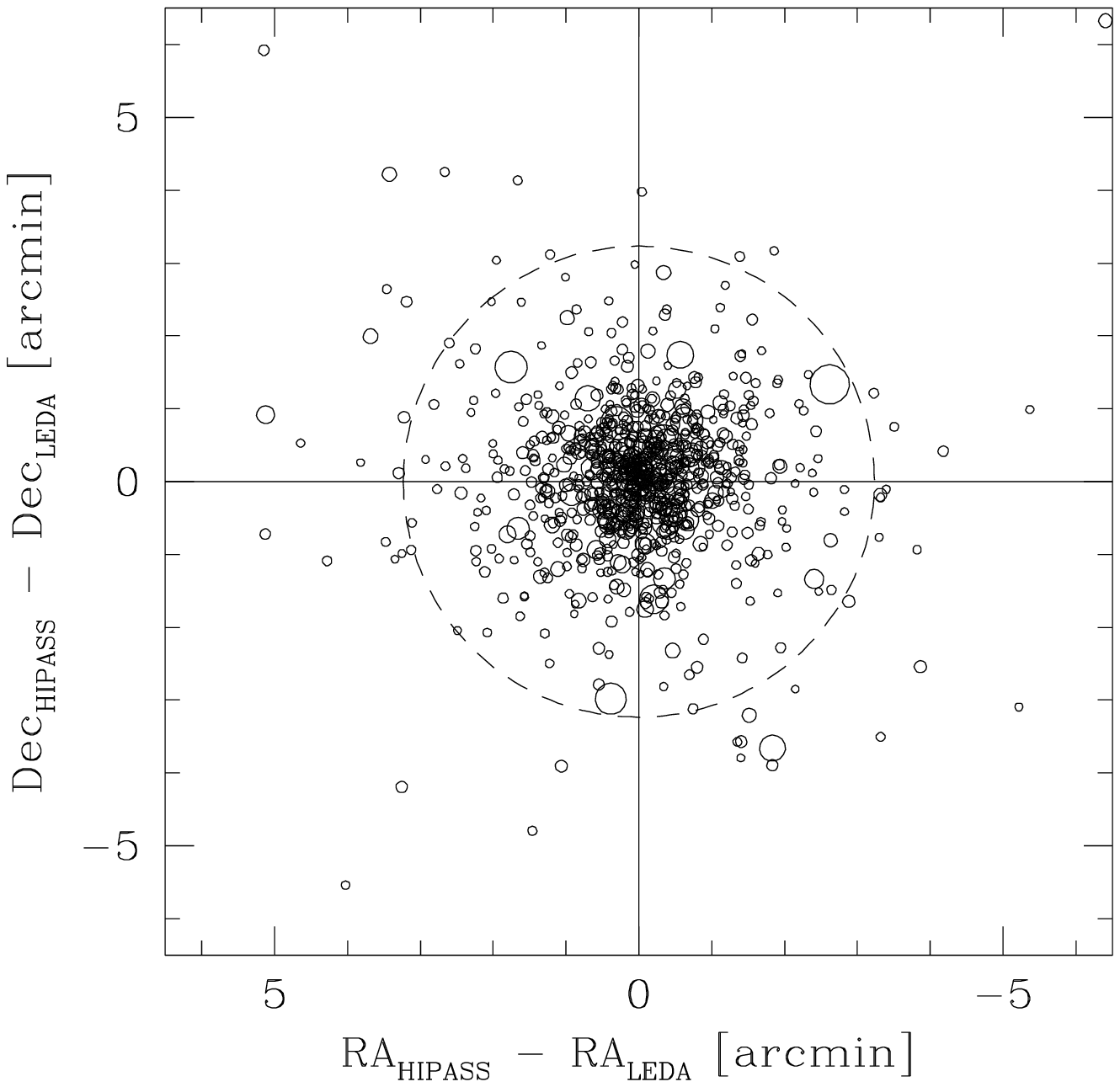,width=10cm}}
\caption{\label{fig:scatter}
  The difference between the \HI\ and optical positions of 853 HIPASS BGC 
  objects, using optical positions from LEDA. The remaining 147 BGC members 
  had either no or had multiple optical identifications in LEDA. The dashed 
  circle marks the radius containing 95\% of the galaxies at 3\farcm2. The 
  symbol size is proportional to the logarithm of the \HI\ peak flux density.}
\end{figure}

\clearpage

\begin{figure}  % Figure 5
 \centering{\psfig{figure=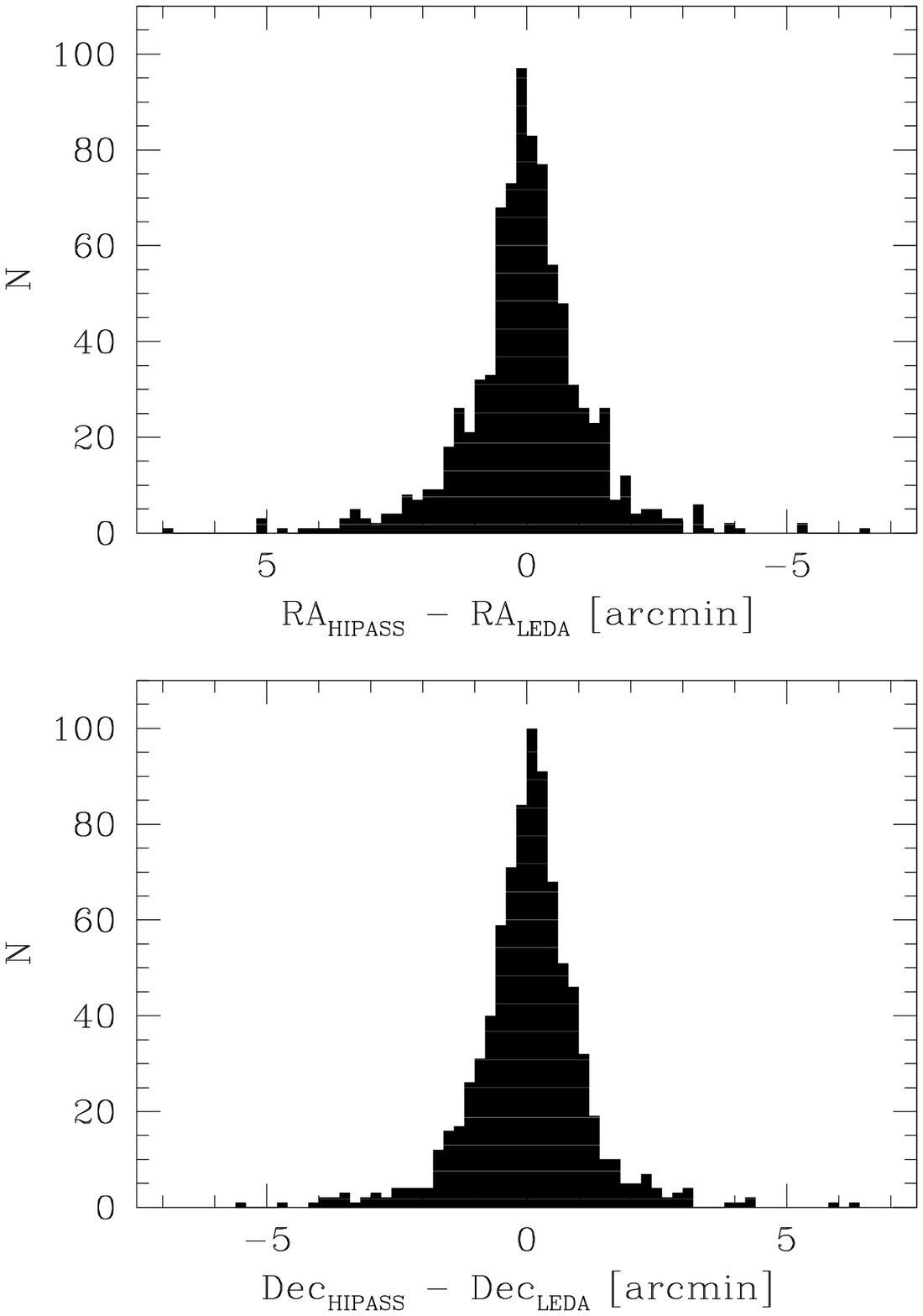,width=10cm}}
\caption{\label{fig:radecsep}
  Histograms of the difference between the \HI\ and optical Right Ascension 
  (top) and Declination (bottom) values for 853 HIPASS BGC objects.}
\end{figure}

\begin{figure}  % Figure 6
 \centering{\psfig{figure=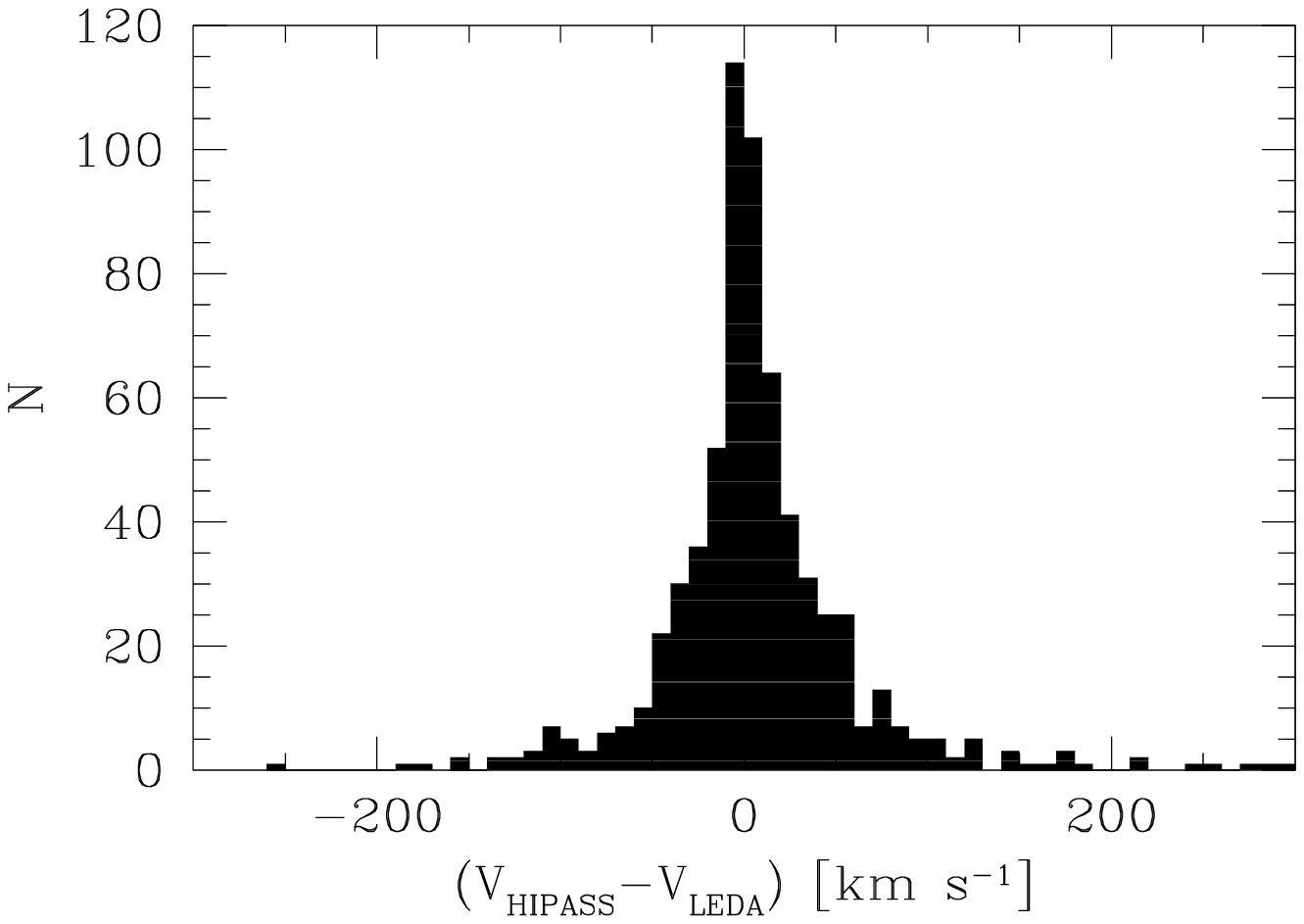,width=10cm}}
\caption{\label{fig:velsep}
  Histogram of the difference between the \HI\ and optical systematic 
  velocities for 672 HIPASS BGC objects. For 95\% of these galaxies the 
  optical velocities are within 170\kms\ of the HIPASS velocities. There 
  are 20 galaxies with differences larger than 300\kms, these lie outside 
  the velocity range displayed here.}
\end{figure}

\begin{figure} % Figure 7
% \centering{\psfig{figure=Koribalski.fig7.eps,width=7.5cm}}
\caption{\label{fig:hipass0622} Optical DSS image of the field around
   the nearby galaxy HIPASS J0622--07. The small galaxy CGMW1-0080 was
   thought to be the optical counterpart of the \HI\ source, but ATCA
   \HI\ imaging showed that the big galaxy towards the center of the
   field is the actual counterpart. CGMW1-0080 is likely a background
   galaxy.}
\end{figure}

\begin{figure} % Figure 8
 \centering{\psfig{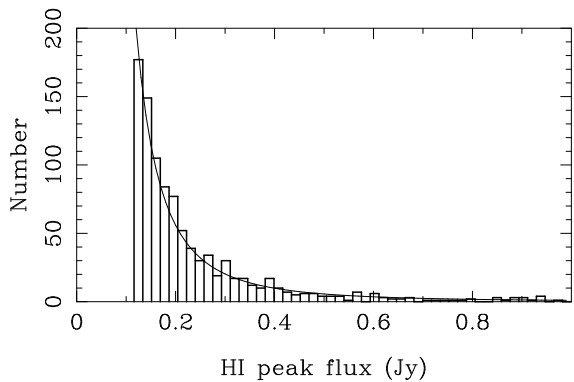}}
\caption{\label{fig:peakhist}
  Histogram of the measured \HI\ peak flux densities in the HIPASS BGC (the 
  bin size is 20 mJy). The peak flux cutoff at \Speak\ $\ga$ 116 mJy results 
  from selecting the 1000 \HI-brightest galaxies. Not displayed here are 38 
  galaxies with \HI\ peak flux densities $>$1 Jy. The overlaid curve is 
  described by \Speak$^{-2.5}$, indicating a complete catalog.}
\end{figure}

\clearpage

% Skipped.
%\begin{figure} % Figure 
%\begin{tabular}{c}
% \centering{\psfig{figure=Koribalski.fig7.ps,width=7.5cm,angle=-90}}
%\end{tabular}
%\caption{\label{fig:circ}
  %Integrated \HI\ distribution in the Circinus galaxy (HIPASS J1413--65) as 
  %obtained from the HIPASS data. The contour levels are --16, --8, --4, 4, 
  %8, 16, 32, 64, 128 and 256 Jy\,beam$^{-1}$\kms. We measure an integrated 
  %\HI\ flux density of at least 1450 Jy\kms. The negative sidelobes to the 
  %North and South are an artifact of the bandpass calibration. 
  %For high-resolution ATCA \HI\ images of the Circinus galaxy see Jones et 
  %al. (1999).} 
%\end{figure}

\begin{figure} % Figure 9
 \centering{\psfig{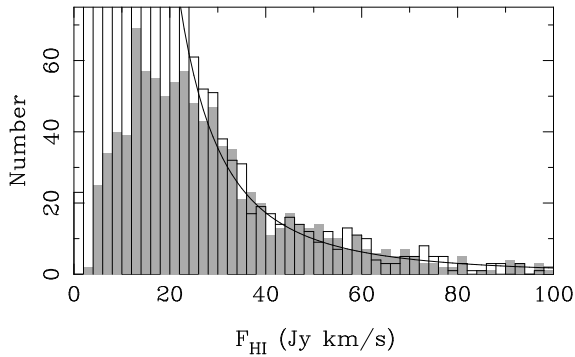}}
\caption{\label{fig:fhihist}
  Histogram of the integrated \HI\ flux densities (\FHI) in the HIPASS BGC
  (grey) and the HIPASS Catalogue (Meyer et al. 2004). The overlaid curve, 
  which is proportional to \FHI$^{-2.5}$, indicates completeness only for 
  BGC sources with \FHI\ $\ga 25$ Jy\kms. Note that the HIPASS Catalogue 
  and the BGC are based on the same survey data, but were compiled and 
  parameterized independently. --- About 60 sources with \FHI\ $\ga$ 
  100 Jy\kms\ are not displayed.}
% NOTE. 60 BGC and 56 HICAT sources have FHI>100 Jy\kms.
\end{figure}

\begin{figure} % Figure 10
 \centering{\psfig{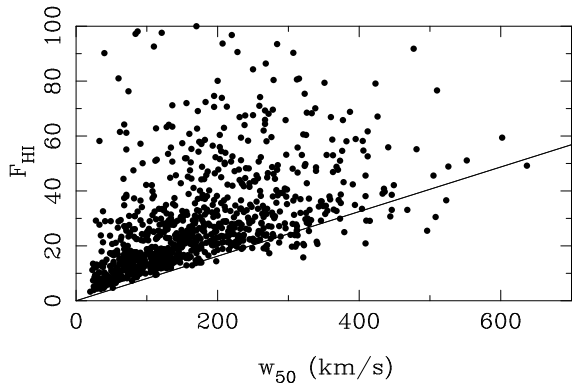}}
\caption{\label{fig:w50fhi}
  Integrated \HI\ flux density, \FHI, versus the 50\% velocity width, \wfi. 
  The lower limit to the detectable integrated \HI\ flux density in the HIPASS 
  BGC is approximately given by \FHI\ = $0.7 \times 0.116$ \wfi\ (indicated by
  the solid line). It applies quite well to galaxies with relatively narrow 
  profiles (\wfi\ $<$ 250\kms), but is a factor $\sim$2 lower for wide 
  double-horn profiles. Not displayed here are 60 galaxies with \FHI\ $\ga$ 
  100 Jy\kms.}
\end{figure}

\begin{figure} % Figure 11
% \centering{\psfig{figure=Koribalski.fig11.ps,width=7.5cm,angle=0}}
\caption{\label{fig:recomb}
  Example of strong Galactic recombinations lines in HIPASS, here shown at 
  the position of the \HII\ region complex IC\,4701. In this case three narrow
  lines are visible at velocities of $-890$, 4465 and 9690\kms. The Galactic
  emission around 0\kms\ has mostly been blanked out.}
\end{figure}

\begin{figure} % Figure 12
 \centering{\psfig{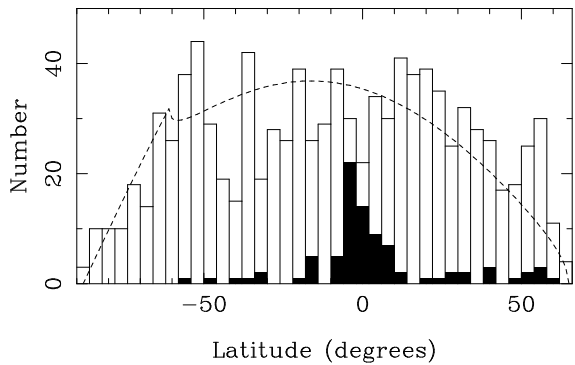}}
\caption{\label{fig:bhist}
  Histogram of the Galactic latitudes of all sources in the HIPASS BGC. Newly 
  cataloged galaxies are marked in black. The dotted line shows the relative 
  area of the southern sky covered at each Galactic latitude. Deviations from 
  this line indicate the presence of large-scale structure.} 
% NOTE. program BGClat.f, the kink occurs at b = -62.88 degr
%       The latitude distribution is linear from -90 until the kink.
%       Area = 57.2943
\end{figure}

\clearpage

\begin{figure} % Figure 13  
 \centering{\psfig{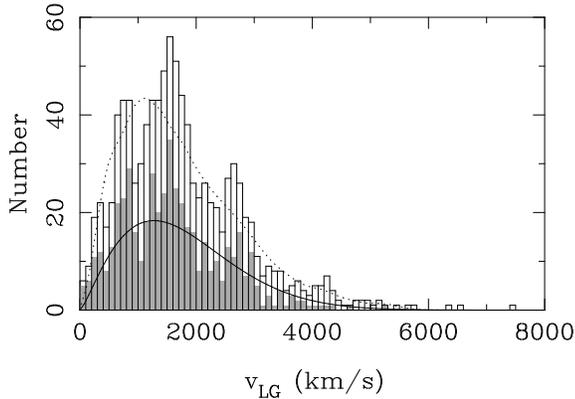}}
\caption{\label{fig:velhist1}
  Histogram of the \HI\ Local Group velocities, \vLG, in the HIPASS BGC.
  We show the full sample of 1000 sources (white) and the complete 
  \FHI-limited sub-sample (grey) with \FHI\ $>$ 25 Jy\kms\ (see 
  Section~\ref{sec:complete}). Overlaid are the selection functions for 
  both samples as derived by Zwaan et al. (2003).}
\end{figure}

\begin{figure}
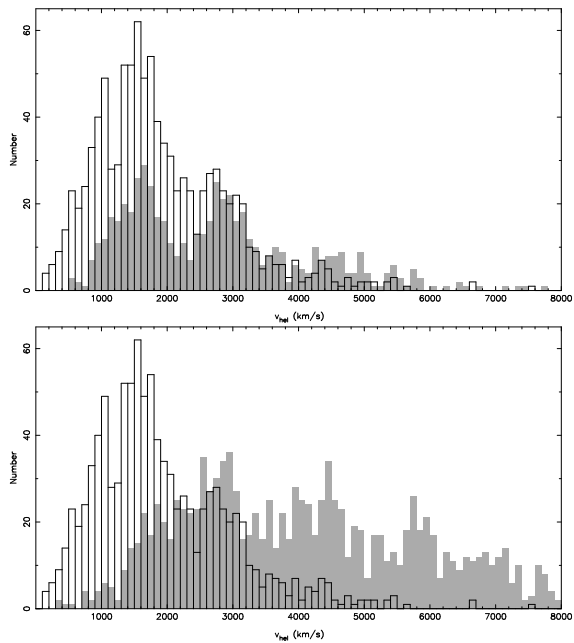
 % Figure 14
 \centering{\psfig{figure=Koribalski.fig14a.ps,width=7.5cm,angle=-90}} 
 \centering{\psfig{figure=Koribalski.fig14b.ps,width=7.5cm,angle=-90}}
\caption{\label{fig:velhist2}
  Histogram of the \HI\ systemic velocities, \vsys, of the HIPASS BGC 
  galaxies (white). For comparison we overlaid (in grey) the \HI\ velocity 
  histograms from the targeted samples of Mathewson et al. (1992), top, and 
  Theureau et al. (1998), bottom, selecting only southern galaxies with \HI\ 
  velocities less than 8000\kms. The samples contain 569 and 1188 galaxies, 
  respectively.}
\end{figure}

\begin{figure} % Figure 15
  \centering{\psfig{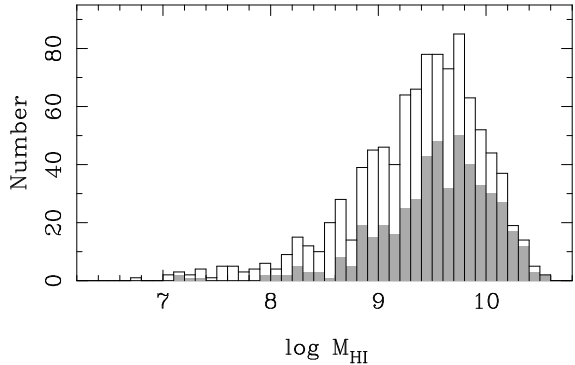}}
\caption{\label{fig:mhihist}
  Histogram of the \HI\ mass, \MHI, distribution in the HIPASS BGC.
  We show the full sample of 1000 sources (white) and the complete
  \FHI-limited sub-sample (grey) with \FHI\ $>$ 25 Jy\kms\ (see
  Section~\ref{sec:complete}).}
\end{figure}

\begin{figure} % Figure 16
  \centering{\psfig{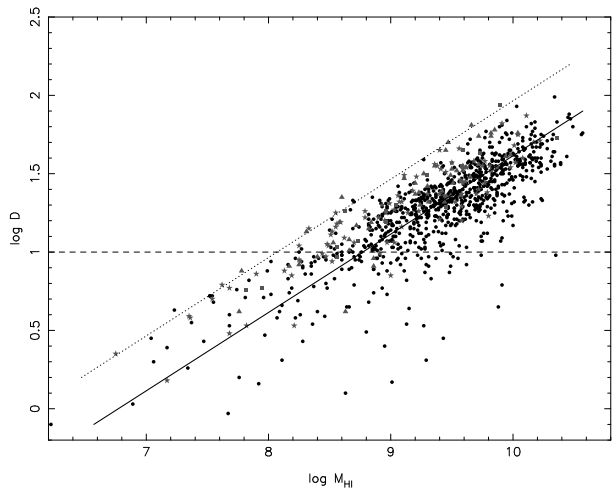}}
\caption{\label{fig:logdmhi}
  Derived distances, $D$, of all HIPASS BGC sources against derived \HI\ mass, 
  \MHI.
  The solid black line marks the \FHI-completeness limit (\FHI\ = 25 Jy\kms), 
  while the dotted line denotes the approximate detection limit, \FHI\ = 5 
  Jy\kms, of the BGC. The horizontal dashed line indicates a distance of $D$ 
  = 10 Mpc. 
% NOTE. MHI = 2.36 10^5 D^2 FHI; 
%       FHI = 25 Jy\kms\ => log(MHI) = 6.771 + 2 log(D)
%       FHI =  5 Jy\kms\ => log(MHI) = 6.072 + 2 log(D)
  Newly cataloged galaxies as well as \HI\ clouds are marked with stars, 
  galaxies with previously unknown velocities with triangles, and galaxies
  with previously incorrect velocities with squares. In total there are 
  157 (248) galaxies with \vLG\ $<$ 750\kms\ (1000\kms) of which 20 (27) 
  are newly cataloged. For comparison see Tully \& Fisher (1981a) and Briggs 
  (1997).}
\end{figure}

\clearpage

\begin{figure} % Figure 17
  \centering{\psfig{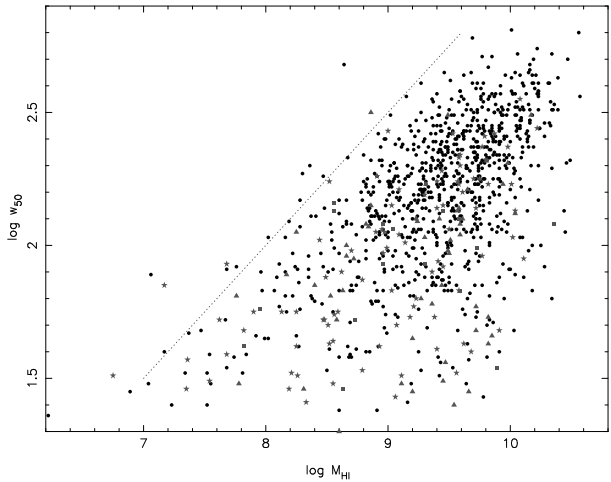}}
\caption{\label{fig:logw50mhi}
  Measured velocity width, \wfi, of all HIPASS BGC sources against \HI\ mass, 
  \MHI. While we find a wide range of \HI\ masses for galaxies with narrow 
  line widths (representing mostly low-mass dwarf irregular galaxies and some
  more massive, nearly face-on spirals), there is a much narrower range for 
  galaxies with large velocity widths (typically massive edge-on spiral 
  galaxies). This trend results from the fact that dwarf irregular galaxies 
  have typically low rotational velocities and masses, both of which gradually 
  increase for type Sd, Sc and Sb galaxies (see Roberts \& Haynes 1994). This 
  morphology segregation will be studied in detail in our optical follow-up 
  paper where the inclinations and morphological types of the majority of 
  galaxies in the BGC are available. The dotted line indicates the approximate 
  detection boundary, \wfi\ $\propto$ \MHI$^{0.5}$, for the BGC. For comparison
  and an explanation of the symbols see Fig.~\ref{fig:logdmhi}.}
% NOTE. Dotted line: log(MHI) = 4 + 2 log(w50)
\end{figure}

\begin{figure}
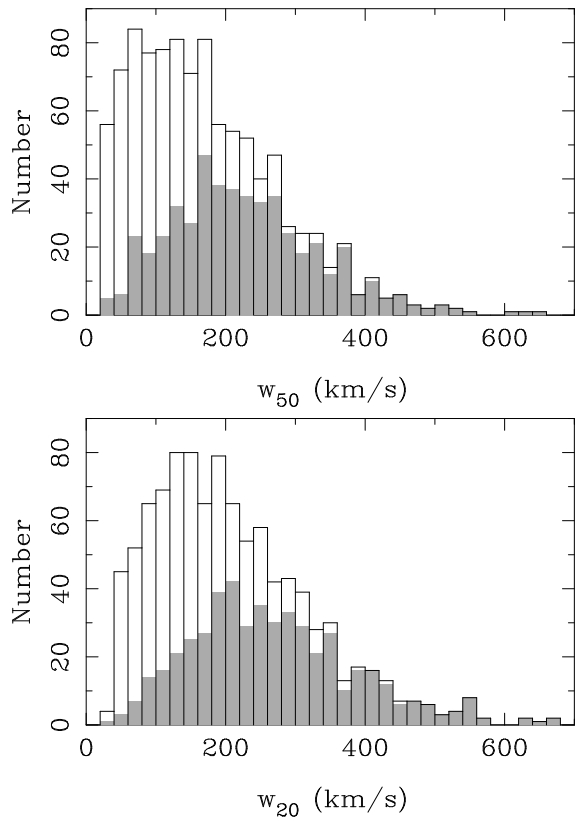
 % Figure 18
\begin{tabular}{c}
  \mbox{\psfig{figure=Koribalski.fig18a.ps,width=7.5cm,angle=-90}} \\
  \mbox{\psfig{figure=Koribalski.fig18b.ps,width=7.5cm,angle=-90}}
\end{tabular}
\caption{\label{fig:whist} 
  Histograms of the measured 50\% and 20\% \HI\ line widths in the HIPASS 
  BGC. We show the full sample of 1000 sources (white) and the complete 
  \FHI-limited sub-sample (grey) with \FHI\ $>$ 25 Jy\kms\ (see
  Section~\ref{sec:complete}).}
\end{figure}

\clearpage

\begin{figure} % Figure 19
 \centering{\psfig{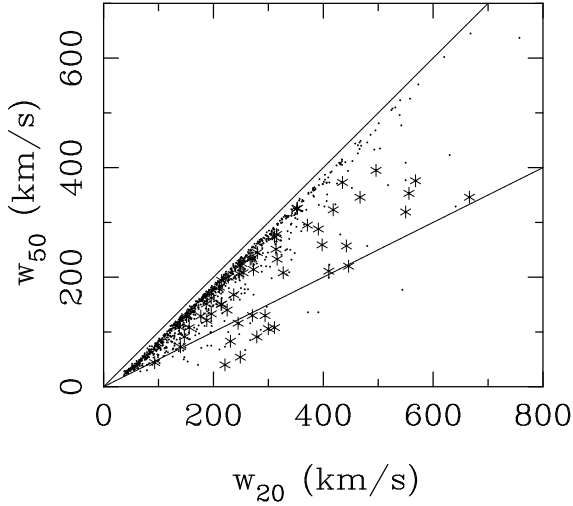}}
\caption{\label{fig:w20w50}
  A comparison of the 50\% and 20\% \HI\ velocity line widths. Galaxy pairs 
  and groups, which often show large \wtw\ / \wfi\ ratios, are shown as 
  stars. The two lines mark the ratios of 1.0 and 2.0.}
\end{figure}

\begin{figure} % Figure 20
 \centering{\psfig{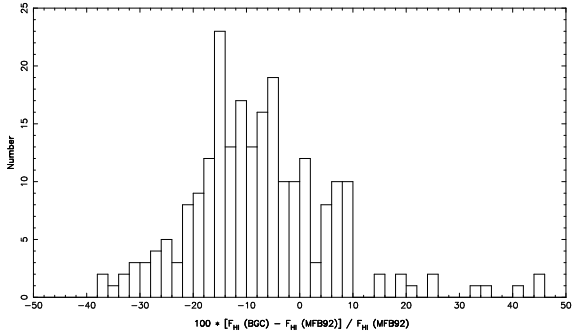}}
\caption{\label{fig:mfbfhi}
  Histogram of the percentage difference in \HI\ flux density (\FHI) 
  for the 228 galaxies common in the HIPASS BGC and the targeted sample 
  of Mathewson et al. (1992). The median offset from zero is --9\%, 
  consistent with the difference in absolute flux calibration (see 
  Section~\ref{sec:fhical}). Not displayed here are 6 galaxies with 
  percentage differences larger than 50\%.}
\end{figure}

\begin{figure} % Figure 21
 \centering{\psfig{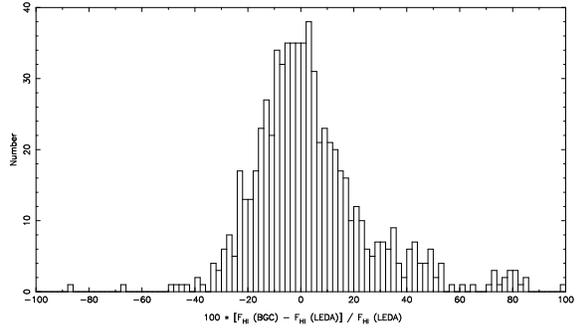}}
\caption{\label{fig:pgcflux}
  Comparison of the integrated \HI\ flux densities, \FHI, in the 
  HIPASS BGC with those given in LEDA for 692 galaxies (see 
  Section~\ref{sec:pgcflux}). Not displayed here are 23 galaxies 
  with percentage differences larger than 100\%.}
\end{figure}

\clearpage

\begin{figure*} % Figure 22
%% \centering{\psfig{figure=Koribalski.efig22.color.ps,width=16cm}} 
%%  in electronic edition
% \centering{\psfig{figure=Koribalski.fig22.ps,width=16cm}}
\caption{\label{fig:BGCdis} {\small
   The ``HIPASS Bright Galaxy Catalog'' contains the 1000 \HI-brightest 
   galaxies in the southern sky. Here we show their sky distribution, plus 
   the Magellanic Clouds (unfilled markers), using a zenithal equal area 
   projection of the south celestial hemisphere. Marker size and greyscale 
   represent the integrated \HI\ flux density, \FHI, and local group velocity,
   \vLG, respectively. Galaxies which are $\times 10$ stronger in \FHI\ are 
   displayed $\times 3$ greater in area. The greyscale ranges from \vLG\ =
   0 (black) to 4000\kms\ (white). Only 41 galaxies in the BGC exceed this 
   velocity, with the maximum at 7413\kms. Nearly 10\% of the HIPASS BGC 
   galaxies were previously uncataloged. The most prominent filament across 
   the southern sky is the Supergalactic Plane (SGP), roughly perpendicular 
   to the Galactic Plane. Parallel to the SGP we find at least two further 
   filaments which have previously been described as the toes of a dinosaur's 
   foot (Lynden-Bell 1994); the toes are clearly connected to the dinosaur's 
   heel near the Great Attractor (GA). Also prominent is the Local Void 
   ($v_{\rm LG} \la\ 1500$ km\,s$^{-1}$).
   --- The projection is not tangent to the south celestial pole (SCP); its 
   obliquity has instead been chosen so that the supergalactic equator is 
   projected as a straight line. The SCP is marked by a cross just below the
   center with arms at RA = $0^{\mathrm h}$ (left), $6^{\mathrm h}$ (bottom), 
   $12^{\mathrm h}$, and $18^{\mathrm h}$; the celestial equator on the 
   perimeter is distorted slightly from a circle. Supergalactic and Galactic
   coordinates are represented by the black and grey graticules, respectively.}}
\end{figure*}

\clearpage

\begin{figure*} % Figure 23
%% \centering{\psfig{figure=Koribalski.efig23.color.ps,width=16cm}} 
%% in electronic edition
% \centering{\psfig{figure=Koribalski.fig23.ps,width=16cm}}
\caption{\label{fig:LEDAdis}{\small
   For comparison with the HIPASS BGC (see Fig.~\ref{fig:BGCdis}) we show the 
   distribution of the 1000 {\em optically}-brightest galaxies in the southern 
   sky as obtained from LEDA on the basis of their mean apparent total blue
   magnitudes, $B_{\rm T}$. Marker size has again been set so that galaxies 
   with apparent blue luminosity $\times 10$ greater are $\times 3$ larger in 
   area, with the scale factor chosen to make them roughly comparable in size 
   with the HIPASS BGC. Thirty-eight galaxies for which no local group velocity
   was available in LEDA have been indicated with a cross and white marker. 
   Including these, 163 galaxies exceed \vLG\ = 4000\kms. ---
   The HIPASS BGC and our LEDA sample have only $\sim$400 galaxies in common. 
   The main differences to Fig.~\ref{fig:BGCdis} are: 
   (1) the Zone of Avoidance is clearly seen as near horizontal gap in this
   chart, and (2) galaxies appear more strongly clustered.  --- 
   The projection, coordinate grids and velocity ranges are the same as in 
   Fig.~\ref{fig:BGCdis}.}}
\end{figure*}

\clearpage

\begin{figure*} % Figure 24
\begin{tabular}{cc}
% \mbox{\psfig{figure=Koribalski.fig24a.ps,width=6.5cm}} &
% \mbox{\psfig{figure=Koribalski.fig24b.ps,width=6.5cm}} \\
% \mbox{\psfig{figure=Koribalski.fig24c.ps,width=6.5cm}} &
% \mbox{\psfig{figure=Koribalski.fig24d.ps,width=6.5cm}} \\
% \mbox{\psfig{figure=Koribalski.fig24e.ps,width=6.5cm}} &
% \mbox{\psfig{figure=Koribalski.fig24f.ps,width=6.5cm}} \\
\end{tabular}
\caption{\label{fig:BGCvel}
    The HIPASS Bright Galaxy Catalog (left; see also Fig.~\ref{fig:BGCdis})
    and the LEDA comparison sample (right; see also Fig.~\ref{fig:LEDAdis})
    reproduced in discrete velocity ranges (the number of sources in each 
    range is given in brackets):
    (a) \vLG\ $<$ 1400\kms\ (BGC + MC: 336 galaxies, LEDA: 244),
    (b) \vLG\ =   1400 -- 2200\kms\ (BGC: 343, LEDA: 248), and
    (c) \vLG\ $>$ 2200\kms\ (BGC: 323, LEDA: 508).
    Black saturation in (a) is at 0\kms, and white saturation in (c)
    is at 5200\kms. --- The projection and coordinate grids are the same 
    as in Figs.~\ref{fig:BGCdis} and \ref{fig:LEDAdis}.}
\end{figure*}

\clearpage

\begin{figure} % Figure 25
 \centering{\psfig{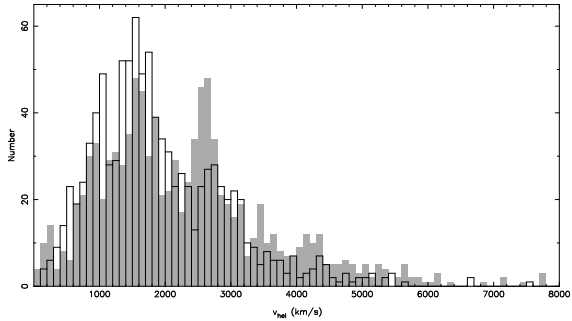}} 
\caption{\label{fig:velhist3}
  Histogram of the \HI\ systemic velocities, \vsys, of the HIPASS BGC 
  galaxies (white). For comparison we overlaid (in grey) the optical velocity 
  histogram of the 1000 optically brightest galaxies obtained from LEDA (see 
  Section~\ref{sec:lss}).}
\end{figure}

\begin{figure*} % Figure 26
% \centering{\psfig{figure=Koribalski.fig26.ps,width=16cm}}
\caption{\label{fig:lvoid}
  Distribution of galaxies ($v < 5000$\kms) in and around the southern Local 
  Void. Galaxies selected from NED (incl. HIZSS galaxies) are marked with open 
  circles, whereas BGC sources are indicated by stars. The newly cataloged 
  galaxies, several of which lie in the Local Void, are easily noted as stars 
  without a surrounding circle. Galactic coordinates are indicated with solid 
  lines (in steps of 10\degr); $\alpha, \delta$(J2000) coordinates are marked 
  with dotted lines.}
\end{figure*}

\begin{figure} % Figure 27
% \centering{\psfig{figure=Koribalski.fig27.ps,width=7.5cm,angle=-90}}
\caption{\label{fig:eso376-g022}
   ATCA \HI\ image (contours) of the galaxy ESO376-G022 (HIPASS J1051--34)
   overlaid onto an optical DSS image (greyscale). The contour levels are
   0.5, 1, 2, 3, and 4 Jy\,beam$^{-1}$\kms. The synthesized beam is displayed
   at the bottom left.}
\end{figure}

\begin{figure} % Figure 28
% \centering{\psfig{figure=Koribalski.fig28.ps,width=7.5cm,angle=-90}}
\caption{\label{fig:m83}
  Mean \HI\ velocity field in the nearby galaxy NGC~5236 (HIPASS J0317--41) 
  and neighbouring galaxies as obtained from the HIPASS data. The contour 
  levels go from 400 to 620\kms, in steps of 20\kms. We measure an 
  integrated \HI\ flux density of 1630 Jy\kms.}
\end{figure}

\clearpage

% Tables

\begin{table*} % Table 1
\caption{\label{tab:hisurveys}
   Some parameters of recent blind \HI\ surveys.}
\begin{flushleft}
\begin{tabular}{lcccccc}
\tableline
\tableline
\HI\ Survey&Galaxies (new)&Sky area&r.m.s.&$\delta v$&velocity range & Ref.\\
           &              &[deg$^2$]&[mJy]&  [\kkms] &   [\kkms]     &     \\
   (1)     &     (2)      &   (3)   & (4) &  (5)     &     (6)       & (7) \\
\tableline
Arecibo \\
-- Slice Survey                &  75 (40) &   55&1.7 &16&   100 to 8340 & (1)\\
-- \HI\ Strip Survey (AHISS)   &  66 (30) &   65&0.75&16& --700 to 7400 & (2)\\
-- Dual-Beam Survey (ADBS)     & 265 (81) &  430&3.5 &32& --650 to 7980 & (3)\\
HIPASS / ZOA \\
-- Shallow Survey (HIZSS)      & 110 (67) & 1840& 15 &27&--1200 to 12700& (4)\\
-- South Celestial Cap (SCC)   & 536 (114)& 2414& 13 &18&--1200 to 12700& (5)\\
-- Bright Galaxy Catalog (BGC) &1000~~(91)&20626& 13 &18&--1200 to  8000& (6)\\
-- HIPASS Catalog (HICAT)      &4315~~~~~~&21346& 13 &18&$\sim$300 to 12700& (7)\\
% NOTE. Curved surface of a spherical segment: A = 2 pi r h ; r = 180/pi
% SCC:  h = r (1 - sin(62degr)) => A = 2414  degr^2
% BGC:  h = r => A = 2 pi r^2   => A = 20626 degr^2
\tableline
\end{tabular}
\end{flushleft}
\tablecomments{The BGC selection criteria are given in 
               Section~\ref{sec:selection}. Cols.\,(4) and (5) show the 
	       1\,$\sigma$ sensitivity and velocity resolution, $\delta v$, 
	       of each survey.}
\tablerefs{(1) Spitzak \& Schneider 1998, (2) Zwaan et al. 1997, (3) Rosenberg
           \& Schneider 2000, (4) Henning et al. 2000, (5) Kilborn et al. 2002,
	   (6) this paper, (7) Meyer et al. (2004).}
\end{table*}

\newpage
\clearpage

\begin{deluxetable}{lccrrlrrrrrrrrrcc} % Table 2
\tabletypesize{\tiny}
\rotate
\tablecolumns{17}
\tablewidth{0pt}
\tablecaption{\label{tab:bgctable} The HIPASS Bright Galaxy Catalog}
\tablehead{
    \colhead{HIPASS Name} 
  & \multicolumn{2}{c}{RA (J2000) DEC}
  & \colhead{$l$}      & \colhead{$b$}      & \colhead{NED ID}
  & \multicolumn{2}{c}{\Speak $\pm \sigma$} 
  & \multicolumn{2}{c}{\FHI   $\pm \sigma$} 
  & \multicolumn{2}{c}{\vsys  $\pm \sigma$} & \colhead{\wfi}  & \colhead{\wtw}
  & \colhead{\vLG}     & \colhead{log \MHI} & \colhead{flag} \\
    \colhead{} 
  & \colhead{[$^{\rm h m s}$]} & \colhead{[\degr\,\arcmin\,\arcsec]}
  & \colhead{[\degr]}  & \colhead{[\degr]}  & \colhead{}
  & \multicolumn{2}{c}{[Jy]}    & \multicolumn{2}{c}{[Jy\kms]}
  & \multicolumn{2}{c}{[\kkms]} & \multicolumn{2}{c}{[\kkms]}
  & \colhead{[\kkms]} & \colhead{[\MMsun]} & \colhead{} \\
    \colhead{(1)}  & \colhead{(2)}  & \colhead{(3)}  & \colhead{(4)}
  & \colhead{(5)}  & \colhead{(6)}  & \colhead{(7)}  & \colhead{(8)}
  & \colhead{(9)}  & \colhead{(10)} & \colhead{(11)} & \colhead{(12)}
  & \colhead{(13)} & \colhead{(14)} & \colhead{(15)} & \colhead{(16)} 
  & \colhead{(17)} }
\startdata
HIPASS J0001--15 & 00:01:56 & --15:28:38 &  75.8 &--73.6 & WLM            &  4.769 & 0.240 &  231.8 &  28.4 &--122 &  2 &  53 &  74 & --40 &  7.67*&   \\      
HIPASS J0002--80 & 00:02:44 & --80:20:48 & 305.5 &--36.5 & ESO012-G014    &  0.255 & 0.019 &   23.9 &   3.1 & 1958 &  4 &  98 & 130 & 1762 &  9.49 &   \\      
HIPASS J0005--28 & 00:05:38 & --28:06:09 &  24.8 &--79.8 & ESO409-IG015   &  0.124 & 0.014 &    7.3 &   1.8 &  736 &  7 &  46 &  99 &  758 &  8.25 &   \\      
HIPASS J0006--41 & 00:06:18 & --41:29:06 & 332.9 &--72.9 & ESO293-IG034   &  0.127 & 0.018 &   23.3 &   4.2 & 1513 &  8 & 236 & 276 & 1473 &  9.33 &   \\      
HIPASS J0008--34 & 00:08:06 & --34:33:31 & 351.6 &--78.1 & ESO349-G031    &  0.166 & 0.016 &    5.8 &   1.6 &  221 &  6 &  30 &  79 &  212 &  7.04 &   \\      
HIPASS J0009--24 & 00:09:54 & --24:57:42 &  43.7 &--80.4 & NGC0024        &  0.340 & 0.025 &   50.3 &   5.1 &  554 &  2 & 210 & 223 &  588 &  8.86 &   \\      
HIPASS J0010--18 & 00:10:18 & --18:15:47 &  73.9 &--77.0 & UGCA003        &  0.304 & 0.023 &    7.4 &   1.9 & 1545 &  3 &  24 &  41 & 1610 &  8.91 &   \\      
HIPASS J0014--23 & 00:14:01 & --23:11:29 &  55.8 &--80.7 & NGC0045        &  1.374 & 0.071 &  195.8 &  14.3 &  467 &  2 & 167 & 185 &  507 &  9.33 &   \\      
HIPASS J0014--70 & 00:14:11 & --69:58:33 & 307.6 &--46.8 & ESO050-G006    &  0.124 & 0.016 &    8.6 &   2.3 & 4187 &  5 &  71 &  92 & 4024 &  9.77 &   \\      
HIPASS J0015--39 & 00:15:09 & --39:13:08 & 332.7 &--75.7 & NGC0055        & 14.595 & 0.732 & 1990.2 & 145.1 &  129 &  2 & 169 & 197 &   95 &  9.01*& (e)  \\   
HIPASS J0019--22 & 00:19:07 & --22:40:30 &  62.5 &--81.4 & MCG-04-02-003  &  0.191 & 0.016 &   16.0 &   2.5 &  669 &  3 & 119 & 131 &  709 &  8.53 &   \\      
HIPASS J0019--57 & 00:19:08 & --57:38:04 & 311.3 &--59.0 & AM0016-575     &  0.169 & 0.018 &   19.8 &   3.3 & 1745 &  4 & 150 & 166 & 1629 &  9.34 &   \\      
HIPASS J0019--77 & 00:19:48 & --77:06:47 & 305.2 &--39.9 & ESO028-G014    &  0.162 & 0.017 &   14.9 &   2.8 & 1791 &  4 & 114 & 132 & 1603 &  9.21 &   \\      
HIPASS J0022--53 & 00:22:30 & --53:38:55 & 312.4 &--62.9 & ESO150-G005    &  0.143 & 0.012 &   13.5 &   2.0 & 1436 &  3 & 111 & 125 & 1335 &  9.00 &   \\      
HIPASS J0026--33 & 00:26:36 & --33:40:38 & 340.7 &--81.5 & NGC0115        &  0.122 & 0.017 &   18.5 &   3.6 & 1824 &  9 & 200 & 246 & 1809 &  9.41 &   \\      
HIPASS J0030--33 & 00:30:22 & --33:13:41 & 338.4 &--82.4 & NGC0134        &  0.455 & 0.027 &  139.5 &   7.9 & 1582 &  3 & 460 & 486 & 1567 & 10.16~&   \\      
HIPASS J0031--22 & 00:31:25 & --22:45:59 &  75.7 &--83.7 & ESO473-G024    &  0.160 & 0.016 &    7.2 &   1.8 &  540 &  4 &  45 &  61 &  572 &  7.99 &   \\      
HIPASS J0031--10 & 00:31:30 & --10:28:37 & 106.2 &--72.7 & MCG-02-02-040  &  0.127 & 0.015 &   31.0 &   4.1 & 3573 & 13 & 278 & 431 & 3659 & 10.24~&   \\      
HIPASS J0031--05 & 00:31:43 & --05:10:15 & 110.0 &--67.5 & NGC0145        &  0.157 & 0.016 &   14.4 &   2.6 & 4139 &  8 &  78 & 146 & 4247 & 10.04~&   \\      
HIPASS J0033--01 & 00:33:23 & --01:07:45 & 112.7 &--63.6 & UGC00328       &  0.183 & 0.036 &   19.4 &   6.3 & 1985 &  7 & 141 & 158 & 2108 &  9.56 &   \\      
HIPASS J0034--30a& 00:34:11 & --30:46:00 & 348.0 &--84.8 & UGCA006        &  0.310 & 0.022 &   16.9 &   2.8 & 1582 &  3 &  57 &  78 & 1576 &  9.25 &   \\      
HIPASS J0034--27 & 00:34:12 & --27:48:39 &  21.8 &--86.1 & NGC0150        &  0.220 & 0.019 &   46.7 &   4.6 & 1584 &  4 & 313 & 342 & 1592 &  9.70 &   \\      
HIPASS J0034--08 & 00:34:48 & --08:24:23 & 110.3 &--70.9 & NGC0157        &  0.286 & 0.019 &   63.0 &   4.9 & 1652 &  7 & 209 & 338 & 1744 &  9.91 &   \\      
HIPASS J0035--20 & 00:35:40 & --20:08:51 &  94.8 &--82.1 & ESO540-G003    &  0.118 & 0.016 &   20.7 &   3.6 & 3292 &  8 & 321 & 360 & 3333 &  9.98 &   \\      
HIPASS J0035--25 & 00:35:45 & --25:22:57 &  58.5 &--86.1 & IC1558         &  0.261 & 0.018 &   27.7 &   3.1 & 1551 &  3 & 118 & 138 & 1569 &  9.46 &   \\      
HIPASS J0037--22 & 00:37:20 & --22:37:11 &  86.8 &--84.5 & NGC0172        &  0.151 & 0.039 &   22.3 &   8.0 & 3048 & 12 & 260 & 287 & 3077 &  9.95 & (r)  \\   
HIPASS J0038--46 & 00:37:43 & --46:34:34 & 310.0 &--70.4 & ESO242-G020    &  0.158 & 0.014 &   32.9 &   3.5 & 3367 & 12 & 199 & 410 & 3290 & 10.17~&   \\      
HIPASS J0038--24 & 00:38:29 & --24:17:55 &  76.3 &--85.9 & IC1562         &  0.184 & 0.018 &   10.8 &   2.3 & 3764 &  4 &  50 &  77 & 3785 &  9.81 &   \\      
HIPASS J0039--14a& 00:39:04 & --14:10:19 & 109.8 &--76.7 & NGC0178        &  0.328 & 0.020 &   25.5 &   3.0 & 1452 &  4 &  71 & 112 & 1517 &  9.39 &   \\      
HIPASS J0040--13 & 00:40:34 & --13:52:25 & 111.6 &--76.5 & NGC0210        &  0.372 & 0.022 &   69.6 &   5.0 & 1636 &  3 & 278 & 305 & 1701 &  9.93 &   \\      
HIPASS J0040--63 & 00:40:57 & --63:27:23 & 304.9 &--53.6 & ESO079-G005    &  0.176 & 0.020 &   22.5 &   3.8 & 1712 &  6 & 153 & 183 & 1566 &  9.36 &   \\      
HIPASS J0042--18 & 00:42:14 & --18:09:44 & 109.1 &--80.8 & ESO540-G016    &  0.138 & 0.016 &   17.7 &   3.0 & 1553 &  4 & 195 & 212 & 1598 &  9.28 &   \\      
HIPASS J0043--22 & 00:42:59 & --22:12:26 & 101.2 &--84.7 & IC1574         &  0.146 & 0.015 &    5.4 &   1.5 &  363 &  4 &  39 &  55 &  390 &  7.54 &   \\      
HIPASS J0046--11 & 00:46:03 & --11:30:14 & 118.0 &--74.3 & DDO005         &  0.187 & 0.017 &   16.6 &   2.7 & 1622 &  3 &  93 & 111 & 1694 &  9.30 &   \\      
HIPASS J0047--20 & 00:47:06 & --20:44:48 & 113.9 &--83.5 & NGC0247        &  3.972 & 0.201 &  608.2 &  42.1 &  156 &  2 & 198 & 224 &  187 &  8.95 & (e)  \\   
HIPASS J0047--25 & 00:47:31 & --25:17:22 &  97.2 &--88.0 & NGC0253        &  3.376 & 0.173 &  692.9 &  42.2 &  243 &  2 & 407 & 431 &  254 &  9.27 & (e)  \\   
HIPASS J0047--09 & 00:47:45 & --09:54:58 & 119.9 &--72.8 & UGCA014        &  0.135 & 0.016 &   18.5 &   3.3 & 1342 &  5 & 144 & 166 & 1419 &  9.19 &   \\      
HIPASS J0047--11 & 00:47:46 & --11:27:14 & 119.6 &--74.3 & NGC0255        &  0.339 & 0.023 &   42.8 &   4.4 & 1585 &  4 & 148 & 193 & 1656 &  9.69 &   \\      
HIPASS J0049--20 & 00:49:35 & --21:01:07 & 118.9 &--83.9 & UGCA015        &  0.144 & 0.018 &    3.9 &   1.5 &  294 &  4 &  25 &  41 &  322 &  7.23 &   \\      
HIPASS J0050--07 & 00:50:59 & --07:03:22 & 122.6 &--69.9 & NGC0274/5      &  0.185 & 0.014 &   36.6 &   3.4 & 1730 &  6 & 233 & 316 & 1817 &  9.70 & (pair)  \\
HIPASS J0051--00 & 00:51:57 & --00:28:21 & 123.2 &--62.4 & MCG+00-03-018  &  0.138 & 0.016 &   14.7 &   2.7 & 1621 &  4 & 169 & 188 & 1737 &  9.27 &   \\      
HIPASS J0052--31 & 00:52:43 & --31:12:17 & 299.1 &--85.9 & NGC0289        &  0.877 & 0.046 &  159.1 &  10.6 & 1629 &  2 & 274 & 300 & 1610 & 10.24~&   \\      
HIPASS J0054--37 & 00:54:52 & --37:40:25 & 299.2 &--79.4 & NGC0300        & 17.078 & 0.856 & 1972.6 & 156.1 &  146 &  2 & 147 & 166 &   98 &  9.29*& (e)  \\   
HIPASS J0054--07 & 00:54:57 & --07:20:15 & 125.5 &--70.2 & NGC0298        &  0.233 & 0.017 &   36.6 &   3.7 & 1756 &  3 & 209 & 230 & 1839 &  9.72 &   \\      
HIPASS J0056--09 & 00:56:39 & --09:55:33 & 127.3 &--72.8 & NGC0309        &  0.128 & 0.017 &   21.1 &   3.8 & 5661 &  7 & 206 & 236 & 5732 & 10.46~&   \\      
HIPASS J0059--07 & 00:59:50 & --07:35:08 & 129.1 &--70.4 & NGC0337        &  0.248 & 0.019 &   51.2 &   4.6 & 1648 &  4 & 233 & 272 & 1726 &  9.81 &   \\      
HIPASS J0101--07 & 01:01:34 & --07:35:22 & 130.4 &--70.3 & NGC0337A       &  1.231 & 0.064 &   98.1 &   9.7 & 1074 &  2 &  87 & 106 & 1151 &  9.74 &   \\      
HIPASS J0103--03 & 01:03:01 & --03:36:30 & 130.1 &--66.3 & MCG-01-03-079  &  0.122 & 0.015 &    8.9 &   2.2 & 2587 &  5 &  85 & 103 & 2679 &  9.43 &   \\      
HIPASS J0105--06 & 01:05:01 & --06:13:06 & 132.3 &--68.8 & MCG-01-03-085  &  0.475 & 0.028 &   62.7 &   5.5 & 1096 &  2 & 171 & 189 & 1176 &  9.56 &   \\      
HIPASS J0107--69 & 01:07:30 & --69:52:30 & 300.9 &--47.2 & NGC0406        &  0.241 & 0.018 &   48.3 &   4.4 & 1508 &  3 & 240 & 261 & 1333 &  9.56 &   \\      
HIPASS J0109--02 & 01:09:41 & --02:15:52 & 133.7 &--64.8 & MCG-01-04-005  &  0.140 & 0.016 &   16.4 &   2.9 & 1864 &  5 & 149 & 173 & 1957 &  9.42 &   \\      
\enddata
\tablecomments{The full version of this table is in the electronic 
               edition of the Journal.}
\end{deluxetable}

\begin{table*} % Table 3
\caption{\label{tab:closest}
         Galaxies for which we adopted independent distances.}
\begin{flushleft}
\begin{tabular}{llrrrrrrcl}
\tableline
\tableline
 Name & NED-ID & \vLG    & log \MHI & $D$   & Reference \\
      &        & [\kkms] & [\MMsun] & [Mpc] &           \\
 (1)  &  (2)   & (3)     & (4)      & (5)   &           \\
\tableline
HIPASS J0001--15  & WLM          &--40 & 7.67 & 0.925 & (1) \\
HIPASS J0015--39  & NGC~0055     &  95 & 9.01 & 1.480 & (1) \\
HIPASS J0054--37  & NGC~0300     &  98 & 9.29 & 2.04~ & (2) \\
HIPASS J1003--26A & NGC~3109     & 130 & 8.63 & 1.250 & (1) \\
HIPASS J1010--04  & Sextans\,A   & 113 & 7.92 & 1.440 & (1) \\
HIPASS J1929--17  & ESO594-G004  &  24 & 6.89 & 1.060 & (1) \\
HIPASS J1944--14  & NGC~6822     &  65 & 8.16 & 0.490 & (1) \\
HIPASS J2046--12  & DDO\,210     &   3 & 6.22 & 0.800 & (1) \\
HIPASS J2202--51  & IC\,5152     &  69 & 7.76 & 1.590 & (1) \\
\tableline
\end{tabular}
\end{flushleft}
\tablecomments{Col.\,(5) lists the adopted independent distances and Col.\,(6) 
   the corresponding reference. For the LMC and SMC, which are not contained 
   in the HIPASS BGC (see Section~\ref{sec:selection}), we adopted distances 
   of 50 and 60 kpc, respectively.}
\tablerefs{(1) Mateo 1998, (2) Willick \& Batra 2001.}
\end{table*}

\begin{table*} % Table 4
\caption{\label{tab:hiclouds}
         \HI\ properties of the definite \HI\ clouds in the HIPASS BGC.}
\begin{flushleft}
\begin{tabular}{llrrrrrrc}
\tableline
\tableline
 Name & NED-ID(s) + flag & \Speak & \FHI 
      & \vsys & \wfi & \wtw & \vLG  & log \MHI \\
      &                  & [Jy]   & [Jy\kms]
      & \multicolumn{4}{c}{[\kkms]} & [\MMsun] \\
 (1)&(2)&(3)&(4)&(5)&(6)&(7)&(8)&(9)\\
\tableline
HIPASS J0731--69&new     & 0.119 &  14.9 &1469&116&267&1195& ~8.95~~~   \\
HIPASS J1616--55&new (e) & 0.455 &  23.7 & 409& 48&100& 256& (7.82)$^+$ \\
HIPASS J1712--64&new     & 0.193 &   6.2 & 455& 29& 49& 295& (7.35)$^+$ \\
HIPASS J1718--59&new (e) & 1.688 &  58.2 & 396& 33& 51& 256& (8.21)$^+$ \\
\tableline
\end{tabular}
\end{flushleft}
\tablecomments{$^+$If these sources are Magellanic debris, as suggested, then 
               the adopted distances ($D$ = \vLG / \Ho) and resulting \HI\ 
	       masses are incorrect. Assuming $D$ = 50 kpc we obtain \HI\ 
	       masses of $\sim 10^5$\Msun.}
\end{table*}

\begin{table*} % Table 5
\caption{\label{tab:ext}
  \HI\ properties for the extended sources in the HIPASS BGC; 
  these are flagged with "e" in Table~\ref{tab:bgctable}. }
\begin{flushleft}
\begin{tabular}{llrrcccr}
\tableline
\tableline
      & & & & & \multicolumn{2}{c}{Deconvolved Gaussian} \\
 Name & NED-ID(s) & \FHI     & \vsys   & log \MHI 
      & \multicolumn{2}{c}{\HI\ diameter} & $PA$ \\
      & \multicolumn{2}{r}{[Jy\kms]}& [\kkms] & [\MMsun]  
      & [arcmin] & [kpc] & [deg] \\
 (1)&(2)&(3)&(4)&(5)&(6)&(7)&(8)\\
\tableline  
HIPASS J0015--39 &NGC~0055      & 1990.2 & 129 &~9.01* 
       & $23.2 \times  4.9$ & $10 \times ~2$ & --71.4 \\
HIPASS J0047--20 &NGC~0247      &  608.2 & 156 & 8.95
       & 14.4               & 10             & --10.1 \\
HIPASS J0047--25 &NGC~0253      &  692.9 & 243 & 9.27
       & 15.6               & 15             &   47.3 \\
HIPASS J0054--37 &NGC~0300      & 1972.6 & 146 &~9.29*
       & $27.9 \times 18.0$ & $17 \times 11$ & --48.9 \\
HIPASS J0317--41 &NGC~1291      &   90.2 & 838 & 9.25
       & $12.9 \times 10.9$ & $34 \times 29$ &   87.7 \\
HIPASS J0317--66 &NGC~1313      &  462.7 & 470 & 9.12
       & 10.1               & 10             &   20.0 \\
HIPASS J0403--43 &NGC~1512/0    &  259.3 & 898 & 9.74
       & $17.4 \times 10.7$ & $48 \times 30$ &   59.4 \\
HIPASS J0409--56 &NGC~1533      &   67.6 & 785 & 8.97
       & $12.6 \times ~2.4$ & $28 \times ~5$ & --47.8 \\
HIPASS J0411--32 &NGC~1532 (c)  &  248.8 &1037 & 9.90
       & 10.9               & 37             &   36.6 \\
HIPASS J0926--76 &NGC~2915      &  108.4 & 468 & 8.28
       & ~9.0               & ~7             & --63.5  \\
HIPASS J1003--26A&NGC~3109      & 1147.9 & 403 &~8.63*
       & $22.7 \times ~6.2$ & $~8 \times ~2$ &   83.4 \\
HIPASS J1118--32 &NGC~3621      &  884.3 & 730 & 9.91 
       & $19.0 \times ~3.6$ & $34 \times ~7$ &  --9.4 \\
HIPASS J1305--49 &NGC~4945      &  319.1 & 563 & 9.15 
       & $20.2 \times ~2.5$ & $25 \times ~3$ &   47.9 \\
HIPASS J1324--42 &NGC~5128 (r)  &   91.8$^{\rm a}$ & 556 & 8.64$^{\rm a}$
       & $44.9 \times 21.0$ & $59 \times 28$ &   28.8 \\ 
HIPASS J1337--29 &NGC~5236 (c)  & 1630.3 & 513 & 9.88 
       & $31.2 \times 21.4$ & $40 \times 28$ &    6.0 \\
HIPASS J1413--65 &Circinus      & 1450.5 & 434 & 9.43 
       & $33.3 \times 19.8$ & $27 \times 16$ &   31.2 \\
HIPASS J1532--56 &new           &   64.2 &1363 & 9.58
       & $28.4 \times 18.2$ & $131 \times 84$ & --89.3 \\
HIPASS J1616--55 &new           &   23.7 & 409 & (7.82)$^+$
       & $58.5 \times 26.3$ & ($58 \times 26$)$^+$ &   41.6 \\
HIPASS J1718--59 &new           &   58.2 & 396 & (8.21)$^+$
       & $74.4 \times 26.0$ & ($74 \times 26$)$^+$ & --88.1 \\
HIPASS J1909--63a&NGC~6744      & 1031.0 & 841 &10.35 
       & $19.3 \times 14.7$ & $54 \times 41$ &    2.5 \\
HIPASS J1943--06 &MCG-01-50-001/2& 44.8 &1515 & 9.72 
       & 26.3               & 170            &   63.3 \\
HIPASS J1944--14 &NGC~6822      & 2524.6 &--57 &~8.16* 
       & $39.7 \times 13.9$ & $~6 \times ~2$ & --54.3  \\
HIPASS J2318--42 &NGC~7582/90/99&   96.8 &1571 & 9.99
       & 17.9               & 107            &   67.1 \\
HIPASS J2357--32 &NGC~7793      &  278.5 & 227 & 8.80
       & $~7.6 \times ~3.3$ & $~7 \times ~3$ & --59.1 \\
\tableline
\end{tabular}
\end{flushleft}
\tablecomments{* The star after log \MHI\ in Col.\,(5) marks galaxies for 
   which we adopted independent distances (see Table~\ref{tab:closest}).\\
   --- $^+$If these sources are Magellanic debris, as suggested, then the 
   adopted distances ($D$ = \vLG / \Ho) and resulting \HI\ masses and 
   diameters are incorrect. Assuming $D$ = 50 kpc we obtain \HI\ masses 
   of $\sim 10^5$\Msun\ and \HI\ diameters less than 1 kpc. \\
   --- $^{\rm a}$A Gaussian fit to the extended \HI\ emission in NGC~5128 
   results in \FHI\ = 186 Jy\kms\ (see Section~\ref{sec:hiabs}).}
\end{table*}

\begin{table*} % Table 6 
\caption{\label{tab:lss} Nearby large-scale structures, clusters and galaxy
   groups as discussed in Section~\ref{sec:lss}.}
\begin{flushleft}
\begin{tabular}{ccccccc}
\tableline
\tableline
 Group & Alternate & $\alpha,\delta$(J2000) & $l, b$ & velocity & Comments \\
 Name  & Name      &[$^{\rm h m}$, \degr]   & [\degr, \degr] & [\kkms] &   \\
\tableline
Fornax Wall \\
\tableline
NGC~1672 & LGG 119 & 04\,45, --59 &  269, --39   & 900--1200 & Dorado Cloud \\
NGC~1566 & LGG 114 & 04\,20, --55 &  264, --43   &     "     &   "    \\
NGC~1433 &LGG 102/6& 03\,40, --47 &  256, --51   &     "     &   "    \\
NGC~1399 & LGG 096 & 03\,40, --36 &  237, --54   & 1300      & Fornax Cluster\\
NGC~1332 & LGG 097 & 03\,24, --22 &  213, --55   & 1800      & Eridanus Cloud\\
\tableline
Puppis filament \\
\tableline
Antlia G1& new     & 09\,50, --30 &  262, 18     &  900      & Antlia Cluster\\
Antlia G2& new     & 10\,40, --37 &  276, 19     &   "       &   "         \\
         &         & 13\,30, --30 &  313, 32     &           & SGP crossing\\
         &         & 12\,30, --05 &  293, 57     & 1100      & Southern Virgo
                                                               Extension \\
\tableline
NGC~4038 & LGG 263 & 12\,00, --20 &  287, 42     & 1750      & Crater  Cloud\\
NGC~5084 & LGG 345 & 13\,20, --21 &  312, 41     & 1750      &   "     \\
\tableline
SGP, Centaurus Wall \\
\tableline
A\,3742  &         & 21\,00, --47 & 353, --42    & 4900      & Indus Cluster \\
IC\,4765 & LGG 422 & 18\,50, --63 & 332, --24    & 4500      & Pavo  Cluster \\
IC~3370  & LGG 298 & 12\,50, --40 & 303, 20      & 3000      & Centaurus Cluster\\
A\,1060  &         & 10\,40, --27 & 270, 27      & 3800      & Hydra Cluster \\
NGC~3258 & LGG 196 & 10\,30, --35 & 273, 19      & 2800      & Antlia Cluster \\
Puppis G2& new     & 07\,50, --35 & 250, --04    & 2800      & Puppis Cluster\\
\tableline
\end{tabular}
\end{flushleft}
% \tablecomments{...}
\end{table*}

\begin{table*} % Table 7 
\caption{\label{tab:wrong}
   \HI\ properties of HIPASS BGC galaxies with potentially incorrect (optical)
   velocities in NED (at the time of the optical identification); for further 
   details see Appendix~\ref{app:wrong}. These galaxies are flagged with "w" 
   in Table~\ref{tab:bgctable}.}
\begin{flushleft}
\begin{tabular}{ccccccc}
\tableline
\tableline
 Name    & HIPASS \vsys & NED-ID     & \vopt   & Ref. & \vHI    & Ref. \\
         & [\kkms]      &            & [\kkms] &      & [\kkms] &      \\
 (1)     &   (2)        &  (3)       &  (4)    &  (5) &  (6)    &  (7) \\
\tableline
HIPASS J0312--21 & 6614 & ESO547-G019     &15711$\pm$56 & (1)& \\
HIPASS J0340--45 & 1547 & IC\,1986\ta     & 1554$\pm$35 & (2)&1566 &(17)\\
                 &      &                 & 1274$\pm$30 & (3)& \\
                 &      &                 & 1451$\pm$50 & (1)& \\
HIPASS J0341--01 & 3421 & [ISI96]\,0339--0209& 8459$\pm$275& (4)& \\
% NOTE The nucleus of ISI96-0339-0209 is very close to a bright star. 
%      The offset between the HIPASS and optical positions is 1.0 arcmin.
HIPASS J0457--42 &  657 & ESO252-IG001    & 3220        & (5)& \\
                 &      &                 & 3430        & (6)& \\
                 &      &                 & 3501$\pm$63 & (7)& \\ 
HIPASS J0732--77 & 1632 & ESO017-G002\tb  & 1380$\pm$200& (8)& \\
HIPASS J1043--47 & 1095 & ESO264-G035\tc  &  770$\pm$60 & (9)& \\
HIPASS J1047--38 &  711 & ESO318-G013\td  &   17$\pm$19 & (7)& \\ 
                 &      &                 &   17$\pm$26 &(10)& \\
HIPASS J1051--34 &  972 & ESO376-G022     & 1364$\pm$53 & (7)& 972$\pm$2 &(18)\\
                 &      &                 & 1410$\pm$90 &(11)& \\
HIPASS J1231--55 & 1717 & ESO172-G004     & 3939$\pm$150&(12)& \\
HIPASS J1319--33 & 1683 & ESO382-G040     & 1587$\pm$50 &(13)&1682       &(19)\\
                 &      &                 &13800?       & (9)& \\
HIPASS J1329--55 & 2573 & ESO173-G016\te  & 3608$\pm$50 &(13)& \\
HIPASS J1344--47 & 1386 & ESO270-G028     & 3410$\pm$70 &(14)& \\
HIPASS J1350--48b& 4253 & ESO221-G006     & 4510$\pm$100& (9)&4013       &(20)\\
                 &      &                 & 4285$\pm$50 &(13)& \\
                 &      &                 & 4400        & (6)& \\
HIPASS J1657--60 & 1035 & ESO138-G009     &  ---        & ---& 881       &(20)\\
HIPASS J1832--57 & 2746 & ESO140-G023     & 2698$\pm$50 &(13)&2730$\pm$2 &(17)\\
                 &      &                 & 4800$\pm$5  &(15)& \\
HIPASS J2257--02 & 3009 & PGC070104\tf    & 2770$\pm$275& (4)& \\
                 &      &                 & 4605$\pm$10 &(16)& \\
\tableline
\end{tabular}
\end{flushleft}
\tablecomments{ 
\ta\ The optical velocities listed for IC\,1986 are likely to be from 
     different parts of this irregular galaxy.
\tb\ The face-on appearance of the Sa galaxy ESO017-G002 agrees with its 
     narrow \HI\ profile (\wfi\ = 29\kms).
\tc\ There is an uncataloged neighbour galaxy $\sim$4\arcmin\ SW of 
     ESO264-G035. 
% NOTE. ESO264-G035 is confused by ESO214-G017; ATCA \HI\ data are needed.
\td\ The ATCA \HI\ image of ESO318-G013 shows a rather asymmetric \HI\ 
     distribution, centered on the western part of the galaxy. The HIPASS 
     \HI\ profile is very narrow (\wfi\ = 42\kms) for what appears to be a 
     peculiar, nearly edge-on galaxy. 
\te\ ESO173-G016 is a spiral galaxy with an extended low-surface brightness 
     disk, barely visible in the DSS\,1 and 2.
\tf\ Bothun et al. (1993) refer to a preprint by Impey and collaborators.
     Note that Impey et al. (1996; ISI96) list two LSB galaxies:
     2254(36)--0245 (\vopt\ = 2770\kms) and
     2254(58)--0246 (\vopt\ = 4606\kms).}
\tablerefs{(1) Da Costa et al. 1991, (2) Peterson et al. 1986, (3) Lauberts et 
al. 1989 (ESO-LV), (4) Impey et al. 1996 (ISI96), (5) Lauberts \& Valentijn 
1989, (6)
Fairall \& Jones 1991, (7) de Vaucouleurs et al. 1991 (RC3), (8) Fairall 1980,
(9) Fairall et al. 1989, (10) Hopp \& Materne (1985), (11) Penston et al. 1977,
(12) Fairall et al. 1998, (13) Dressler 1991, (14) Visvanathan \& Yamada 1996,
(15) Peterson 1986, (16) Bothun et al. (1993), (17) Tully 1988 (Nearby Galaxy 
Catalog), (18) Fouqu\'e et al.  1990, (19) Cot\'e et al. 1997, (20) Huchtmeier 
\& Richter 1989.}
\end{table*}

\end{document}